\documentclass[useAMS,usenatbib,onecolumn,letterpaper]{mn2e}
\usepackage{natbib}
\usepackage{graphicx}
\usepackage{times}
\newcommand{\bmf}[1]{\mathbf {#1}}
\newcommand{\simgt}{\lower.5ex\hbox{$\; \buildrel > \over \sim \;$}}
\newcommand{\simlt}{\lower.5ex\hbox{$\; \buildrel < \over \sim \;$}}

\begin{document}
\title[Using galaxy-galaxy weak lensing measurements to correct the Finger-of-God]
{Using galaxy-galaxy weak lensing measurements to correct the Finger-of-God}

\author[Hikage, Takada \& Spergel]{Chiaki Hikage$^{1,2}$, Masahiro Takada$^3$, David N. Spergel$^{1,3}$ \\
$^1$ Department of Astrophysical Sciences, Princeton University, Peyton Hall, Princeton NJ 08544, USA \\
$^2$ Kobayashi-Maskawa Institute for the Origin of Particles and the Universe (KMI), Nagoya University, Aichi 464-8602, Japan \\
$^3$ Institute for the Physics and Mathematics of the Universe (IPMU), 
The University of Tokyo, Chiba 277-8582, Japan}
\maketitle

%%%%%%%%%%%%%%%%%%%%%%%%%%%%%%%%%%%%%%%%%%%%%%%%%%%%%%%%
\begin{abstract}
For decades, cosmologists have been using galaxies to trace the
large-scale distribution of matter.  At present, the largest source of
systematic uncertainty in this analysis is the challenge of modeling
the complex relationship between galaxy redshift and the distribution
of dark matter. If all galaxies sat in the centers of halos, there
would be minimal Finger-of-God (FoG) effects and a simple relationship
between the galaxy and matter distributions. However, many galaxies,
even some of the luminous red galaxies (LRGs), do not lie in the
centers of halos. Because the galaxy-galaxy lensing is also sensitive
to the off-centered galaxies, we show that we can use the lensing
measurements to determine the amplitude of this effect and to
determine the expected amplitude of FoG effects.  We develop an
approach for using the lensing data to model how the FoG suppresses
the power spectrum amplitudes and show that the current data implies a
30\% suppression at wavenumber $k=0.2~h{\rm Mpc}^{-1}$.  Our analysis
implies that it is important to complement a spectroscopic survey with
an imaging survey with sufficient depth and wide field coverage. Joint
imaging and spectroscopic surveys allow a robust, unbiased use of the
power spectrum amplitude information: it improves the marginalized
error of growth rate $f_g\equiv d\ln D/d\ln a$ by up to a factor of 2
over a wide range of redshifts $z<1.4$. We also find that the dark
energy equation-of-state parameter, $w_0$, and the neutrino mass,
$f_\nu$, can be unbiasedly constrained by combining the lensing
information, with an improvement of 10--25\% compared to a
spectroscopic survey without lensing calibration.
\end{abstract}
%%%%%%%%%%%%%%%%%%%%%%%%%%%%%%%%%%%%%%%%%%%%%%%%%%%%%%%%


\section{Introduction}
\label{sec:intro}

Over the past three decades, astronomers have been conducting ever
larger redshift surveys in their efforts to probe the large-scale
structure of the universe
\citep{DavisHuchra:82,deLapparentetal:86,Kirshneretal:87,SDSS,Peacocketal:01}.
In the coming decade, we are embarking on even larger surveys:
BOSS\footnote{http://cosmology.lbl.gov/BOSS/},
WiggleZ\footnote{http://wigglez.swin.edu.au/site/}
\citep{Blakeetal:11}, Vipers\footnote{http://vipers.inaf.it/},
FMOS\footnote{http://www.naoj.org/Observing/Instruments/FMOS/},
HETDEX\footnote{http://hetdex.org/},
BigBOSS\footnote{http://bigboss.lbl.gov/} \citep{BigBOSS},
LAMOST\footnote{http://www.lamost.org/website/en}, Subaru
PFS\footnote{http://sumire.ipmu.jp/en/},
Euclid\footnote{http://sci.esa.int/euclid}, and
WFIRST\footnote{http://wfirst.gsfc.nasa.gov/}.  This upcoming
generation of surveys are motivated by our desire to understand cosmic
acceleration and to measure the composition of the universe by
simultaneously measuring geometry and dynamics.  The combination of
cosmic microwave background (CMB) data and large redshift surveys
trace the growth of structure formation from the last-scattering
surface ($z\simeq 1100$) to low redshifts and determine cosmological
parameters to high precision
\citep[][]{WangSpergelStrauss:99,Eisensteinetal:99,Tegmarketal:04,Coleetal:05}.
Measurements of the baryon acoustic oscillation (BAO) scale provide us
with a robust geometrical probe of the angular diameter distance and
the Hubble expansion rate
\citep{Eisensteinetal:05,Percivaletal:07BAO}.  Observations of
redshift-space distortion measure the growth rate of structure
formation \citep[][]{Zhangetal:07,Guzzoetal:08,Wang:08,Guziketal:10,
  Whiteetal:09,PercivalWhite:09,SongPercival:09,SongKayo:10,Yamamotoetal:10,Tangetal:11}.
%\citep[also see][for the first attempt on a model-independent gravity test]{Reyesetal:10}. 
Combining measurements of the growth of structure formation and the
geometry of the universe provides a key clue to understanding the
nature of dark energy, properties of gravity on cosmological scales,
or the nature of cosmic acceleration \citep[][]{DETF,Peacocketal:06}.

The galaxy power spectrum in redshift space, a direct observable from
a redshift survey, is a two-dimensional function of wavelengths
perpendicular and parallel to the line-of-sight direction
\citep{Peacocketal:01,Okumuraetal:08,Guzzoetal:08}. While galaxy
clustering in real space is statistically isotropic in an isotropic
and homogeneous universe, the line-of-sight components of galaxies'
peculiar velocities alter galaxy clustering in redshift space
\citep[][]{Kaiser:87}. For review, see \cite{Hamilton:98}. The
amplitude of the distortion depends both on geometry and dynamics
\citep[][]{AlcockPaczynski:79,SeoEisenstein:03}.

For the surveys to achieve their ambitious goals for precision
cosmology, we will need a detailed understanding of the underlying
systematics.  One of the major systematic uncertainties in
redshift-space power spectrum measurements is non-linear redshift
distortion due to the internal motion of galaxies within halos, the
so-called Finger-of-God (FoG) effect
\citep[][]{Jackson:72,Scoccimarro:04}. Since it is sensitive to highly
non-linear physics as well as difficult to model galaxy
formation/assembly histories, the FoG effect is the dominant
systematic in redshift surveys.

\cite{Reidetal:09} advocated using halos rather than Luminous Red
Galaxies \citep[LRGs;][]{Eisensteinetal:01} to trace large-scale
structure.  In an analysis of LRGs sampled with the Sloan Digital Sky
Survey (SDSS)\footnote{http://www.sdss.org/},
%\citep[][]{Eisensteinetal:01}. 
\cite{Reidetal:10} implemented this scheme by removing satellite LRGs
from the same halo with the aid of the mock catalog and the halo model
prescription. From the SDSS LRG dataset, \cite{Reidetal:10} found that
about 6\% of LRGs are satellite galaxies, while the remaining 94\% are
central galaxies of halos with masses $\simgt 10^{13}M_\odot$. Once
such a halo catalog is constructed, clustering properties of halos are
easier to model, because halos have only bulk motions in large-scale
structure, and therefore have the reduced FoG effect.  Despite this
effort, the remaining FoG effect is a dominant systematic uncertainty.

FoG effects are just one of the non-linear systematics.  Future
analysis of the redshift-space power spectrum of halos will need to
model non-linear clustering, non-linear bias, and non-linear redshift
distortion effect due to their bulk motions. Recent simulations and
refined perturbation theory suggest that halo clustering based
approach seems a very promising probe of cosmology
%a halo approach for cosmological analysis
\citep[][]{Scoccimarro:04,CrocceScoccimarro:06,Matsubara:08,Saitoetal:11,Taruyaetal:10,Tangetal:11,ReidWhite:11,SatoMatsubara:11}.

For a halo-based catalog, a significant source of uncertainty is the
position of the galaxies in the halos. \cite{Hoetal:09} compared LRG
positions with the X-ray surface-brightness peak, reporting a sizable
positional difference. For the LRG analysis, this is the dominant
uncertainty \citep[see][for the useful discussion in
  Appendix~C]{Reidetal:10}.  The virial theorem implies that
off-centered LRGs are moving relative to the halo center thus
producing an FoG effect.

In this paper, we propose a novel method of using a cross-correlation
of spectroscopic galaxies (e.g., LRGs) with background galaxy images
to correct the FoG contamination to the redshift-space power
spectrum. Dark matter halos hosting spectroscopic galaxies induce a
coherent lensing distortion effect on background galaxy images, and
the signals are measurable using the cross-correlation method -- the
so-called galaxy-galaxy or cluster-galaxy weak lensing. The lensing
signals have been now measured at a high significance by various
groups
\citep{Mandelbaumetal:06,Sheldonetal:09,Leauthaudetal:10,Okabeetal:10}. If
we include off-centered galaxies and use the galaxy position as a halo
center proxy of each halo in the lensing analysis, the lensing signals
at angular scales smaller than the typical offset scale are diluted
\citep[see][for a useful formulation of the off-centering effect on
  cluster-galaxy weak lensing]{OguriTakada:10}.  Thus the
galaxy-galaxy lensing signals can be used to infer the amount of the
off-centered galaxy contamination
\citep[][]{Johnstonetal:07,Leauthaudetal:10,Okabeetal:10}. Furthermore,
since lensing is a unique means of reconstructing the dark matter
distribution, it may allow us to infer the halo center on individual
halo basis if a sufficiently high signal-to-noise ratio is available
\citep[][]{Ogurietal:10}. Hence, a weak-lensing based calibration of
the FoG effect in redshift-space power spectrum measurements may be
feasible if spectroscopic and imaging surveys observe the same region
of the sky. Fortunately, many upcoming surveys will survey the same
region of the sky: the BOSS and Subaru Hyper SuprimeCam (HSC) Survey
\citep[][]{Miyazakietal:06}, the Subaru PFS and HSC surveys (Subaru
Measurements of Images and Redshifts: the SuMIRe project), Euclid and
WFIRST or a combination of LSST \citep{LSST} with spectroscopic
surveys.

In Section~\ref{sec:basic}, we will first develop a model of computing
the redshift-space power spectrum of LRGs based on the halo model
approach \citep[see][for a thorough review]{CooraySheth:02}.
Extending \cite{White:01} and \cite{Seljak:01}, we model the
distribution of off-centered LRGs as a source of FoG distortions.
Following the method in \cite{OguriTakada:10}, we also model the
distribution of off-centered LRGs as a source of smoothing of the
LRG-galaxy lensing signal. Assuming survey parameters of the Subaru
HSC imaging survey combined with the BOSS and/or Subaru PFS
spectroscopic surveys as well as the Euclid imaging and spectroscopic
surveys, we study the impact of the FoG effect on parameter
estimations.  We also study the ability of the combined imaging and
spectroscopic surveys for correcting for the FoG effect contamination
based on the off-centering information inferred from the LRG-galaxy
lensing measurements. For the parameter forecast, we pay particular
attention to the dark energy equation-of-state parameter, $w_0$, the
neutrino mass parameter, $f_\nu$, and the growth rate at each redshift
slice.  Unless explicitly stated we will throughout this paper assume
a WMAP-normalized $\Lambda$CDM model as our fiducial cosmological
model \citep{Komatsu:09}: $\Omega_{\rm b}h^2=0.0226$, $\Omega_{\rm
  cdm}h^2=0.1109$, $\Omega_\Lambda=0.734$, respectively, $\tau=0.088$,
$n_s=0.963$, $A(k=0.002{\rm Mpc}^{-1})=2.43\times 10^{-9}$, where
$\Omega_{\rm b}$, $\Omega_{\rm cdm}$ and $\Omega_{\Lambda}$ are the
energy density parameters of baryon, CDM and dark energy (the
cosmological constant with $w_0=-1$ here), $\tau$ is the optical depth
to the last scattering surface, and $n_s$ and $A$ are the tilt and
amplitude of the primordial curvature power spectrum.

\section{Formulation: Redshift-space power spectrum}
\label{sec:basic}

In this section, we give a formulation for modeling the redshift-space
power spectrum of luminous red galaxies (LRGs) based on the halo model
approach \citep[][]{White:01,Seljak:01}.

\subsection{Dominant Luminous Red Galaxies (DLRGs)}
Weak lensing studies \citep[][]{Mandelbaumetal:06,Johnstonetal:07}
%,Leauthaudetal:10,Okabeetal:10}
and clustering analyses
\citep[][]{Rossetal:07,Rossetal:08,Wakeetal:08,Zhengetal:09,ReidSpergel:09,Whiteetal:11}
find that most LRGs reside in massive halos.  While the typical
massive halo contains only one LRG, roughly 5-10\% of all LRGs are
satellite galaxies in a halo containing multiple LRGs.  These
satellite galaxies contribute a large one halo term that is an
additional source of shot noise and non-linearity in power spectrum
estimation. \citet{Reidetal:09} outline a procedure of identifying
these satellite LRGs through finding multiple pairs that lie in common
halos (or the small spatial region) and then using only the brightest
luminous red galaxies in each halo as a tracer.  We call these
galaxies dominant luminous red galaxies (DLRGs). These DLRGs are more
linear tracers of the underlying matter field than the LRGs are.
\cite{Reidetal:10} and \cite{Percivaletal:10} adopt this procedure to
determine the SDSS LRG power spectrum. In this paper, we focus on
these DLRGs so that each halo contains either zero or one DLRG.

\subsection{Halo Model Approach for DLRGs}

%%%%%%%%%%%%%%%%%%%%%%%%%%%%%%%%%%%%%%%%%%%%%%%%%%%%%%%%
\begin{figure}
\begin{center}
\includegraphics[width=13cm]{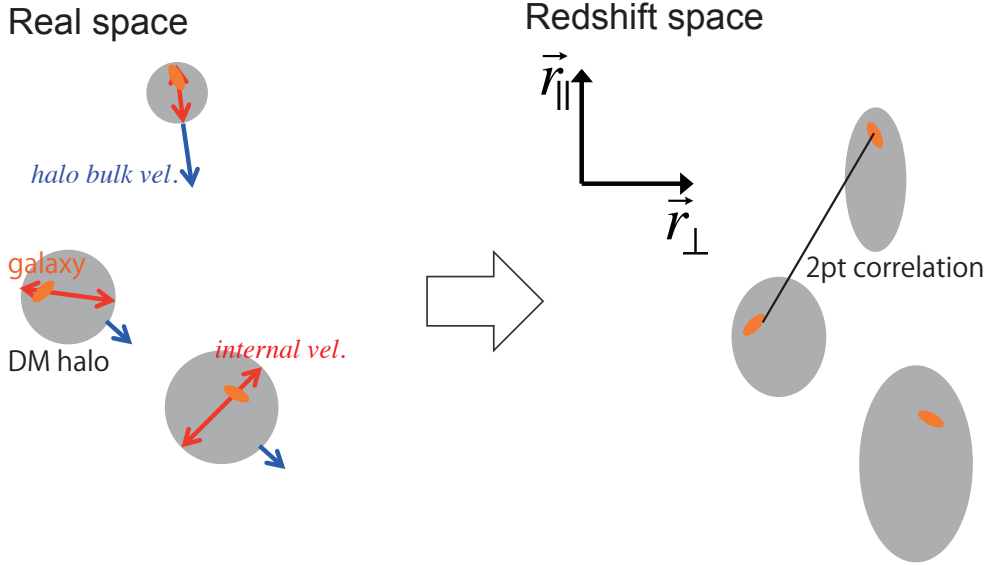} 
\caption{A schematic illustration of the redshift-distortion effect on
  redshift-space clustering of dominant luminous red galaxies (DLRGs;
  see text for details). We assume that a catalog of DLRGs is
  constructed so that each halo contains one DLRG.  The redshift
  distortion effect on the redshift-space power spectrum of DLRGs
  arises from two contributions: the bulk motion of halos that host
  each DLRG, and the internal motion of DLRG within a halo, the
  Finger-of-God (FoG) effect.  If some of the LRGs are not in the
  center of their halos, then their motions produce significant FoG
  effects.  The halo bulk motion causes a displacement of halo
  position in redshift space, while the internal motion stretches the
  distribution region of DLRGs within a halo along the line-of-sight
  direction.}
\label{fig:illust}
\end{center}
\end{figure}
%%%%%%%%%%%%%%%%%%%%%%%%%%%%%%%%%%%%%%%%%%%%%%%%%%%%%%%

Since there is only one DLRG per halo, the two-halo term determines
the clustering of these galaxies in the halo model picture
\citep[][]{CooraySheth:02,TakadaJain:03a}. If the DLRGs sat in the
center of each halo, then the DLRG power spectrum would be linearly
related to the halo power spectrum.  However, since the DLRGs do not
always lie in the center of the halo, the power spectrum is given as
%%%%%%%%%%%%%%%%%%%%%%%%%%%%%%%%%%%%%%%%%%%%%%%%%%%%%%%%%%%%%%%%%%%%%%%%%%
\begin{equation}
\label{eq:pk_dlrg}
%CH add ^2
P_{\rm DLRG}(k)=\frac{1}{\bar{n}_{\rm DLRG}^2}
%P_{\rm DLRG}(k)=\frac{1}{\bar{n}_{\rm DLRG}}
\int\!dM\int\!dM'~ \frac{dn}{dM}N_{\rm HOD}(M)
\tilde{p}_{\rm off}(k;M)
\frac{dn}{dM'}
N_{\rm HOD}(M')\tilde{p}_{\rm off}(k;M')
P_{\rm hh}(k;M,M'),
\end{equation}
%%%%%%%%%%%%%%%%%%%%%%%%%%%%%%%%%%%%%%%%%%%%%%%%%%%%%%%%%%%%%%%%%%%%%%%%%%
where $dn/dM$ is the halo mass function, $N_{\rm HOD}(M)$ is the halo
occupation number (note $N_{\rm HOD}\le 1$ as described below),
%which specifies how many LRGs reside in each halo of a given mass $M$ on average
and $P_{\rm hh}(k;M,M')$ is the cross-power spectrum of halos of
masses $M$ and $M'$. Numerical simulations show that the halo
cross-power spectrum is approximately a linearly biased version of the
matter power spectrum \citep{Reidetal:09}: $P_{\rm hh}(k;M,M')\simeq
b(M)b(M')P_{\rm m}^{\rm NL}(k)$, where $b(M)$ is the halo bias, and
$P_{\rm m}^{\rm NL}(k)$ is the non-linear matter power spectrum.
% (note the text says linear power spectrum is incorrect). 
This approximation simplifies the relationship between the DLRG and
matter power spectrum:
%%%%%%%%%%%%%%%%%%%%%%%%%%%%%%%%%%%%%%%%%%%%%%%%%%%%%%%%%%%%%%%%%%%%%%%%%%
\begin{equation}
\label{eq:pk_real}
P_{\rm DLRG}(k)=\left[\frac{1}{\bar{n}_{\rm DLRG}}
\int\!dM~\frac{dn}{dM}b(M)N_{\rm HOD}(M)
\tilde{p}_{\rm off}(k;M)\right]^2 P_{\rm m}^{\rm NL}(k).
\end{equation}
%%%%%%%%%%%%%%%%%%%%%%%%%%%%%%%%%%%%%%%%%%%%%%%%%%%%%%%%%%%%%%%%%%%%%%%%%%
The quantity $\bar{n}_{\rm DLRG}$ is the mean number density of DLRGs
defined as
%%%%%%%%%%%%%%%%%%%%%%%%%%%%%%%%%%%%%%%%%%%%%%%%%%%%%%%%%%%%%%%%%%%%%%%%%%x
\begin{equation}
\label{eq:nlrg}
\bar{n}_{\rm DLRG}\equiv \int\!dM~ \frac{dn}{dM}N_{\rm HOD}(M). 
\end{equation}
%%%%%%%%%%%%%%%%%%%%%%%%%%%%%%%%%%%%%%%%%%%%%%%%%%%%%%%%%%%%%%%%%%%%%%%%%%
The mean bias of halos hosting DLRGs is defined as
%%%%%%%%%%%%%%%%%%%%%%%%%%%%%%%%%%%%%%%%%%%%%%%%%%%%%%%%%%%%%%%%%%%%%%%%%%
\begin{equation}
\label{eq:b1}
\bar{b}\equiv 
\frac{1}{\bar{n}_{\rm DLRG}}\int\!dM~ b(M)\frac{dn}{dM}N_{\rm HOD}(M). 
\end{equation}
%%%%%%%%%%%%%%%%%%%%%%%%%%%%%%%%%%%%%%%%%%%%%%%%%%%%%%%%%%%%%%%%%%%%%%%%%%
The mean mass of halos hosting DLRGs is similarly estimated as
$\bar{M}_h\equiv (1/\bar{n}_{\rm DLRG})\int\!dM~ M (dn/dM)N_{\rm
  HOD}(M)$.

The coefficient $\tilde{p}_{\rm off}(k; M)$ in Eq.~(\ref{eq:pk_real})
is the Fourier transform of the average radial profile of DLRGs within
a halo with mass $M$:
%%%%%%%%%%%%%%%%%%%%%%%%%%%%%%%%%%%%%%%%%%%%%%%%%%%%%%%%%%%%%%%%%%%%%%%%%%
\begin{equation}
\label{eq:poff_fourier}
\tilde{p}_{\rm off}(k;M)=4\pi\int_{0}^{r_{\rm vir}}\!r^2 
dr~ p_{\rm off}(r)\frac{\sin(kr)}{kr},
\end{equation}
%%%%%%%%%%%%%%%%%%%%%%%%%%%%%%%%%%%%%%%%%%%%%%%%%%%%%%%%%%%%%%%%%%%%%%%%%%
where $p_{\rm off}(r)$ is normalized so as to satisfy $\int^{r_{\rm
    vir}}_0 4\pi r^2 dr~ p_{\rm off}(r)=1$, and $r_{\rm vir}$ is the
virial radius of a halo with mass $M$, which can be defined once the
virial overdensity and the background cosmology are specified. Note
that, since the power spectrum is a statistical quantity, we just need
the averaged DLRG distribution within a halo, which is therefore a
one-dimensional function of radius $r$ with respect to the halo center
in a statistically homogeneous and isotropic universe.  In the
following, quantities with tilde symbol denote their
Fourier-transformed coefficients for our notational convention.

The term in the square bracket in Eq.~(\ref{eq:pk_real}) describes the
halo exclusion effect.  Because halos have finite sizes, roughly their
virial radius, there is only one dominant galaxy in this region (see
Fig.~\ref{fig:illust}). If we are implementing an algorithm that
eliminates multiple galaxies in the finite region, then we impose an
exclusion region around each galaxy. The two halo term describes the
correlations between two DLRGs in {\em two} different halos.
%we cannot select two DLRGs within the same halo -- this is the halo
%exclusion effect.
The DLRG power spectrum at small scales (large $k$'s) is thus
suppressed compared to the matter power spectrum multiplied with
$\bar{b}^2$ \citep[see Fig.~11 in][]{CooraySheth:02}.

If each DLRG resides at the center of each halo (e.g., the center of
mass), $\tilde{p}_{\rm off}(k)=1$ (or $p_{\rm off}\propto
\delta_D(r)$). However some fraction of DLRGs in the sample are
expected to have an offset from the halo center
\citep[][]{Skibbaetal:11}.  Due to the collision-less nature of dark
matter, dark matter halos lack clear boundary with surrounding
structures and do not have a spherically symmetric mass distribution.
Thus the halo center is not a well-defined quantity.  While DLRGs, the
most massive galaxy in the halo, will eventually sink toward the
center of the halo through dynamical friction, many clusters are
dynamically young and have experienced recent interactions.  Thus, we
expect that DLRGs are not all in the centers of halos and that the
distribution of the their positions in the halos evolve with redshift.

How does this halo model picture need to be changed in redshift space?
To model the redshift-space power spectrum, we need to properly take
into account the redshift distortion effect due to peculiar velocities
of DLRGs.  If all DLRGs are located at the center in their host halos,
DLRGs move together with their host halos having coherent, bulk
velocities in large-scale structure, and the redshift-space clustering
is not affected by the FoG effect.  However, as illustrated in
Fig.~\ref{fig:illust}, if some DLRGs are offset from the center, they
will have internal motions within their host halos, which causes the
FoG effect.  The virial theorem implies that the amplitude of the
displacement of the DLRG from the center of its halo is directly
related to the DLRG velocity dispersion within the halo.

In the halo model picture, the FoG effect can be incorporated by
stretching the average radial profile of DLRGs along the line-of-sight
direction by the amount of the internal motion, as illustrated in the
right panel of Fig.~\ref{fig:illust}.  This stretch enhances the halo
exclusion effect, which suppresses the power spectrum amplitudes.
Thus the redshift-space distribution of DLRGs within a halo becomes
two-dimensional, given as a function of two radii, $r_\perp$ and
$r_\parallel$, perpendicular and parallel to the line-of-sight
direction with respect to the halo center.  Also note that the
internal velocity distribution of DLRGs within a halo is considered to
be statistically isotropic and therefore it depends on the radius $r$
from the halo center, halo mass $M$ and redshift $z$ (see
Section~\ref{sec:sigv} for details). The averaged redshift-space
distribution of DLRGs within a halo, denoted as $p_{s,{\rm
    off}}(r_\perp,r_\parallel)$, can be given as a smearing of the
real-space distribution with the displacement function:
%%%%%%%%%%%%%%%%%%%%%%%%%%%%%%%%%%%%%%%%%%%%%%%%%%%%%%%%%%%%%%%%%%%%%%%%%%
\begin{equation}
p_{s,{\rm off}}(r_\perp,r_{\parallel};M)=\int_{-\infty}^{\infty}\!dr_{\parallel}'~
R(r_{\parallel}-r_{\parallel}'; r', M)~ p_{\rm off}\!\left(
\sqrt{r_\perp^2+r'_\parallel{}^2}\right),
\label{eq:pdf_off_real}
\end{equation}
%%%%%%%%%%%%%%%%%%%%%%%%%%%%%%%%%%%%%%%%%%%%%%%%%%%%%%%%%%%%%%%%%%%%%%%%%%
where $R(\Delta r_\parallel;r,M)$ is the displacement function of
DLRGs due to the velocity distribution inside a halo and satisfies the
normalization condition: $\int\!d(\Delta r_\parallel)~ R(\Delta
r_\parallel)=1$. Assuming that the internal motion of DLRGs is much
smaller than the speed of light, the displacement of the radial
position of a given DLRG is directly related to the line-of-sight
component of the internal velocity $v_\parallel$ as
%%%%%%%%%%%%%%%%%%%%%%%%%%%%%%%%%%%%%%%%%%%%%%%%%%%%%%%%%%%%%%%%%%%
\begin{equation}
\Delta\!r_{\parallel}=\frac{v_{\parallel}}{aH(z)},
\end{equation}
%%%%%%%%%%%%%%%%%%%%%%%%%%%%%%%%%%%%%%%%%%%%%%%%%%%%%%%%%%%%%%%%%%%
where $H(z)$ is the Hubble expansion rate at the redshift of DLRG.

Hence, assuming a distant observer approximation, the redshift-space
power spectrum of DLRGs can be given in terms of the Fourier-transform
of $p_{s, {\rm off}}(r_\perp,r_\parallel;M)$ as
%%%%%%%%%%%%%%%%%%%%%%%%%%%%%%%%%%%%%%%%%%%%%%%%%%%%%%%%%%%%%%%%%%%
\begin{equation}
P_{s,{\rm DLRG}}(k,\mu)=\left[\frac{1}{\bar{n}_{\rm DLRG}}\int\!dM~\frac{dn}{dM}b(M)N_{\rm HOD}(M)
\tilde{p}_{s, {\rm off}}(k,\mu; M)
\right]^2P^{\rm NL}_{s,{\rm m}}(k,\mu),
\label{eq:ps_lrg}
\end{equation}
%%%%%%%%%%%%%%%%%%%%%%%%%%%%%%%%%%%%%%%%%%%%%%%%%%%%%%%%%%%%%%%%%%
where $\mu$ is the cosine angle between the line-of-sight direction
and the wavevector $\bmf{k}$, i.e.  $\mu\equiv k_{\parallel}/k$, and
$P^{\rm NL}_{s,{\rm m}}(k_\perp,k_\parallel)$ is the non-linear
redshift-space power spectrum. In this paper, we simply assume that
the redshift distortion effect due to the coherent bulk motion of
halos is described by linear theory \citep{Kaiser:87}:
%%%%%%%%%%%%%%%%%%%%%%%%%%%%%%%%%%%%%%%%%%%%%%%%%%%%%%%%%%%%%%%%%%
\begin{equation}
P^{\rm NL}_{s,{\rm m}}(k,\mu)=P^{\rm NL}_{\rm m}(k)[1+2\beta\mu^2+\beta^2\mu^4],
\label{eq:Kaiser}
\end{equation}
%%%%%%%%%%%%%%%%%%%%%%%%%%%%%%%%%%%%%%%%%%%%%%%%%%%%%%%%%%%%%%%%%%
where $\beta\equiv f_g/\bar{b}$, $f_g$ is the linear growth rate,
$f_g\equiv d\ln D/d\ln a$, and $\bar{b}$ is the effective bias of
halos hosting the DLRGs (Eq.~[\ref{eq:b1}]). As given by the term in
the square bracket in Eq.~(\ref{eq:ps_lrg}), the FoG effect due to
off-centered DLRGs causes scale-dependent, angular anisotropies in the
redshift power spectrum amplitudes.

In the limit that all DLRGs are at the true center of each halo, the
redshift-space power spectrum (Eq.~[\ref{eq:ps_lrg}]) is reduced to
the halo power spectrum in redshift space:
%%%%%%%%%%%%%%%%%%%%%%%%%%%%%%%%%%%%%%%%%%%%%%%%%%%%%%%%%%%%%%%%%%%
\begin{equation}
p_{\rm off}(r)=\frac{1}{4\pi r^2}\delta_D(r)
\rightarrow 
P_{s,{\rm DLRG}}(k,\mu)=\bar{b}^2 P^{\rm NL}_{s,{\rm m}}(k,\mu) \simeq
P_{s,{\rm  halo}}(k,\mu), 
\label{eq:ps_limit}
\end{equation}
%%%%%%%%%%%%%%%%%%%%%%%%%%%%%%%%%%%%%%%%%%%%%%%%%%%%%%%%%%%%%%%%%%%
where $P_{s, {\rm halo}}(k,\mu)$ is the halo power spectrum in
redshift space. More rigorously speaking, halo clustering is affected
by non-linearities in gravitational clustering, redshift distortion and
biasing at scales even in the weakly non-linear regime we are
interested in
\citep{Scoccimarro:04,Taruyaetal:09,Taruyaetal:10,Saitoetal:11,Tangetal:11,ReidWhite:11,SatoMatsubara:11}.
For these we can use an accurate model of the redshift-space spectrum
of halos by using refined perturbation theory and/or N-body and mock
simulations \citep[][]{Taruyaetal:09,Taruyaetal:10,SatoMatsubara:11}.
Hence we can extend the formulation above in order to include these
non-linear effects, simply by using a model of non-linear,
redshift-space halo power spectrum for $P_{s,{\rm hh}}(k,\mu; M, M')$,
instead of $b(M)b(M')P^{\rm NL}_{s, {\rm m}}(k,\mu)$ in
Eq.~(\ref{eq:ps_lrg}).  However, this is beyond the scope of this
paper, and we here focus on the FoG effect by assuming the Kaiser
formula (\ref{eq:Kaiser}) for the sake of clarity of our discussion.

\subsection{Model ingredients}

To compute the redshift-space power spectrum (Eq.~[\ref{eq:ps_lrg}]),
we need to specify the model ingredients: halo occupation distribution
of DLRGs, the off-centered distribution of DLRGs and the velocity
distribution inside halos. In this subsection, we will give these
model ingredients adopted in this paper.

\subsubsection{HOD and the halo model ingredients}

First we need to specify the halo mass function and the halo bias.  We
use the fitting formula developed in \cite{ShethTormen:99} to compute
the halo mass function and the halo bias in our fiducial cosmological
model \citep[also see][]{TakadaJain:03a}.

A useful, empirical method for describing clustering properties of
galaxies is the halo occupation distribution (HOD) \citep[][also see
  references therein]{Scoccimarroetal:01,Zhengetal:05}. The HOD gives
the average number of galaxies residing in halos of mass $M$ and at
redshift $z$. The previous works have shown that the halo model
prediction using the HOD modeling can well reproduce the observed
properties of LRG clustering over wide ranges of length scales and
redshifts ($0\simlt z\simlt 0.5$)
\citep[][]{Zhengetal:09,ReidSpergel:09,Whiteetal:11}. Since we assume
that satellite LRGs can be removed based on the method of
\cite{Reidetal:10}, we use the HOD for central LRGs that is found in
\cite{ReidSpergel:09}:
%%%%%%%%%%%%%%%%%%%%%%%%%%%%%%%%%%%%%%%%%%%%%%%%%%%%%%%%
\begin{equation}
N_{\rm HOD}(M) \simeq N_{\rm cen}(M)
=\frac{1}{2}\left[1+{\rm erf}\left(\frac{\log_{10}(M)-\log_{10}
(M_{\rm min})}{\sigma_{\log M}}\right)\right],
\label{eq:HOD}
\end{equation}
%%%%%%%%%%%%%%%%%%%%%%%%%%%%%%%%%%%%%%%%%%%%%%%%%%%%%%%%
where ${\rm erf}(x)$ is the error function, and we adopt $M_{\rm
  min}=8.05\times 10^{13}M_\odot$ and $\sigma_{\log M}=0.7$.  We do
not consider a possible redshift evolution of the HOD, because any
strong redshift dependence has not been found from actual data.  Note
$N_{\rm HOD}(M)\le 1$.  Also note that we use the HOD model for
``{\em central}'' galaxies, but this does not mean that all DLRGs
under consideration are central galaxies, 
%CH
%and we meant by this model
but each halo has {\em one} DLRG at most.

\subsubsection{Radial profile of DLRGs}

The radial profile of DLRGs is not well known, as the true center of a
halo is not easy to estimate observationally.  Several studies, both
observational and numerical, suggest that the DLRG are more centrally
concentrated than the dark matter, but do not all lie in the bottom of
the dark matter potentials
\citep{LinMohr:04,Koesteretal:07,Johnstonetal:07,Hoetal:09,HilbertWhite:10,Okabeetal:10,Ogurietal:10,
Skibbaetal:11}.  \cite{Hoetal:09} compared the LRG positions with
X-ray peak positions for known X-ray clusters, and found that the LRG
radial distribution can be fitted with an NFW profile with high
concentration parameter ($c\sim 20$).  Using the Subaru weak lensing
observations for about 20 X-ray luminous clusters, \cite{Ogurietal:10}
fit an elliptical NFW model to the dark matter distribution.  They
found that, for most clusters, the positional difference between the
lensing-inferred mass center and the brightest cluster galaxy is well
fitted by a Gaussian distribution with width of $\sim 100~h^{-1}{\rm
kpc}$, a scale comparable to the positional uncertainties in the
lensing analysis.  For these clusters, the lensing data is consistent
with the DLRGs lying in the center of mass of their host halos.
However, in a few clusters, the DLRGs are clearly offset from the
center of the potential with characteristic displacements of $\sim
400~h^{-1}{\rm kpc}$.  \cite{Johnstonetal:07} reached a similar
conclusion: most DLRGs are in the centers of their halo; however, a
handful are significant displaced.  However, the results are not yet
conclusive due to the limited statistics. In this paper, we employ the
following two empirical models for a radial profile of DLRGs based on
these observational implications:
%%%%%%%%%%%%%%%%%%%%%%%%%%%%%%%%%%%%%%%%%%%%%%%%%%%%%%%%
\begin{equation}
p_{\rm off}(r; M)=
\left\{
\begin{array}{ll}
{\displaystyle \frac{1}{(2\pi)^{3/2}r_{\rm off}^3(M)}
\exp\left(-\frac{r^2}{2r_{\rm off}^2(M)}\right)}, & \mbox{(Gaussian
offset model)}, \nonumber\\
{\displaystyle 
\frac{c_{\rm off}^3}{4\pi r_{\rm vir}^3}f
\frac{1}{(c_{\rm off} r/r_{\rm vir})(1+c_{\rm off} r/r_{\rm vir})^2}
}, & \mbox{(NFW offset model)}, 
\end{array}
\right.
\label{eq:model_off}
\end{equation}
%%%%%%%%%%%%%%%%%%%%%%%%%%%%%%%%%%%%%%%%%%%%%%%%%%%%%%%%
where $f\equiv 1/[\ln(1+c_{\rm off})-c_{\rm off}/(1+c_{\rm off})]$ and
the prefactor of each model is determined so as to satisfy the
normalization condition $\int_0^{r_{\rm vir}}\!4\pi r^2dr~p_{\rm
  off}(r)=1$.  These profiles are specified by one parameter ($r_{\rm
  off}$ or $c_{\rm off}$), but differ in the shape. Note that $r_{\rm
  vir}$ is specified as a function of halo mass and redshift, and
$c_{\rm off}$ differs from the concentration parameter of dark matter
profile.

As a working example, we will employ $r_{\rm off}(M)=0.3 r_{s}(M)=0.3
r_{\rm vir}(M)/c_{\rm vir }(M)$ for Gaussian DLRG radial distribution
and $c_{\rm off}=20$ for NFW DLRG radial distribution as our fiducial
models. Here $r_s$ is the scale radius of dark matter NFW profile, and
$c_{\rm vir}$ is the concentration parameter. For the following
results we will use the simulation results in \citet{Duffyetal:08} to
specify $c_{\rm vir}$ as a function of halo mass and redshift. Our
fiducial model of $r_{\rm off}=0.3r_s$ gives $r_{\rm off}\simeq
100~{\rm kpc}$ for halos of $10^{14}M_\odot$, consistent with the
results in \cite{Ogurietal:10}. Note that the typical offset of the
DLRG from the center of the halo potential varies with halo mass.

\begin{figure}
\begin{center}
\includegraphics[width=8cm]{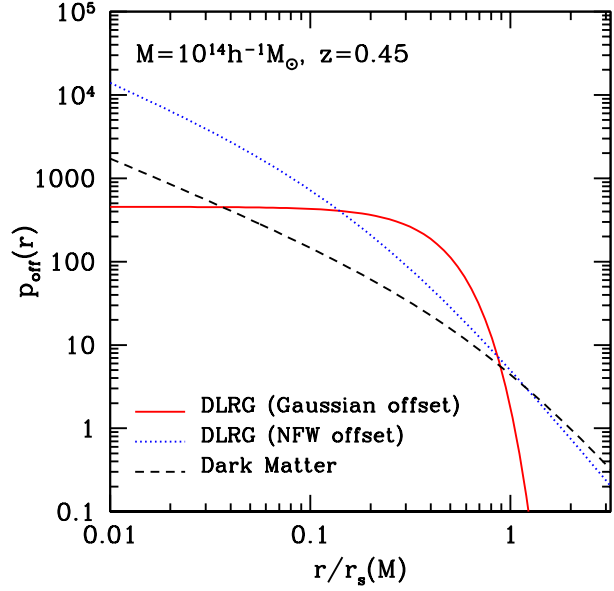} 
\caption{This figure shows the statistically averaged radial profile
  of dark matter and DLRGs in halos with mass $M=10^{14}h^{-1}M_\odot$
  and halo concentration $c_{\rm vir}=4.3$, and at redshift
  $z=0.45$. The solid curve denotes a Gaussian radial profile (or
  off-centered) model with the width that is taken to be $r_{\rm
    off}=0.3r_s (\sim 100~{\rm kpc})$, where $r_s$ is the scale radius
  of the dark matter NFW profile. The dotted curve is an NFW radial
  profile model, which is given by the concentration parameter of
  $c_{\rm off}=20$.}
\label{fig:profile}
\end{center}
\end{figure}

Fig.~\ref{fig:profile} shows the Gaussian and NFW models for the
radial profile of DLRGs inside a halo of mass $M=10^{14}h^{-1}M_\odot$
and at $z=0.45$. Note that for our fiducial model, $r_s\simeq
250~h^{-1}{\rm kpc}$.  These profiles are much more centrally
concentrated than a typical dark matter halo of this mass scale,
represented by an NFW profile with $c=4.3$.
%%%%%%%%%%%%%%%%%%%%%%%%%%%%%%%%%%%%%%%%%%%%%%%%%%%%%%%%
%%%%%%%%%%%%%%%%%%%%%%%%%%%%%%%%%%%%%%%%%%%%%%%%%%%%%%%

The Fourier transforms of these radial profiles are analytic
functions:
%%%%%%%%%%%%%%%%%%%%%%%%%%%%%%%%%%%%%%%%%%%%%%%%%%%%%%%%
\begin{equation}
\tilde{p}_{\rm off}(k;M)=
\left\{
\begin{array}{ll}
\exp[-r_{\rm off}^2(M)k^2/2], & \mbox{(Gaussian model)}, \nonumber \\
{\displaystyle f\left[
\sin \eta\left\{
{\rm Si}(\eta(1+c_{\rm off}))-{\rm Si}(\eta)
\right\}
+\cos\eta\left\{
{\rm Ci}(\eta(1+\eta))-{\rm Ci}(\eta)
\right\}-\frac{\sin(\eta c_{\rm off})}{\eta(1+c_{\rm off})}
\right]}, & \mbox{(NFW model)}, 
\end{array}
\right.
\label{eq:tilde_poff}
\end{equation}
%%%%%%%%%%%%%%%%%%%%%%%%%%%%%%%%%%%%%%%%%%%%%%%%%%%%%%%%
where $\eta=k r_{\rm vir}/c_{\rm off}$, and ${\rm Si}(x)$ and ${\rm
  Ci}(x)$ are the sine and cosine integral functions.  The Fourier
transform $\tilde{p}_{\rm off}(k; M)\simeq 1 $ on relevant scales.

\subsubsection{Velocity dispersion of DLRGs}
\label{sec:sigv}

In this paper, we simply assume that galaxies at a given radius $r$
have the following 1D velocity dispersion that is determined by the
mass enclosed within the sphere:
%%%%%%%%%%%%%%%%%%%%%%%%%%%%%%%%%%%%%%%%%%%%%%%%%%%%%%%
\begin{equation}
\sigma_{v}^2(r;M)=\frac{GM(<r)}{2r}.
\label{eq:sigv_rM}
\end{equation}
%%%%%%%%%%%%%%%%%%%%%%%%%%%%%%%%%%%%%%%%%%%%%%%%%%%%%%%
At virial radius $r_{\rm vir}$, the velocity dispersion is
determined by virial mass:
%%%%%%%%%%%%%%%%%%%%%%%%%%%%%%%%%%%%%%%%%%%%%%%%%%%%%%%
\begin{equation}
\label{eq:sigvir}
\sigma_{v}(r=r_{\rm vir};M)=472{\rm km/s}
\left(\frac{M}{10^{14}h^{-1}M_\odot}\right)^{1/3}
\left(\frac{\Delta(z)}{18\pi^2}\right)^{1/6}\left(1+z\right)^{1/2},
\end{equation}
%%%%%%%%%%%%%%%%%%%%%%%%%%%%%%%%%%%%%%%%%%%%%%%%%%%%%%%
where we use the definition of virial mass given in terms of the
overdensity $\Delta(z)$ at redshift $z$ \citep[we use the fitting
  formulae given in][]{NakamuraSuto:97}.  Since an NFW profile has an
asymptotic behavior of $M(<r)\propto r^2$ as $r\rightarrow 0$, the
velocity dispersion has the limit $\sigma_{v}(r;M)\rightarrow 0$ as
$r\rightarrow 0$.

The velocity dispersion of LRGs is poorly known \citep[see][for the
  first attempt]{Skibbaetal:11}.  In Appendix \ref{app:smallv}, we
give an alternative model of computing the velocity dispersion by
assuming an isothermal distribution for the phase space density of
DLRGs within a halo, where we properly take into account the different
radial profiles of DLRGs and dark matter.

\begin{figure}
\begin{center}
\includegraphics[width=8cm]{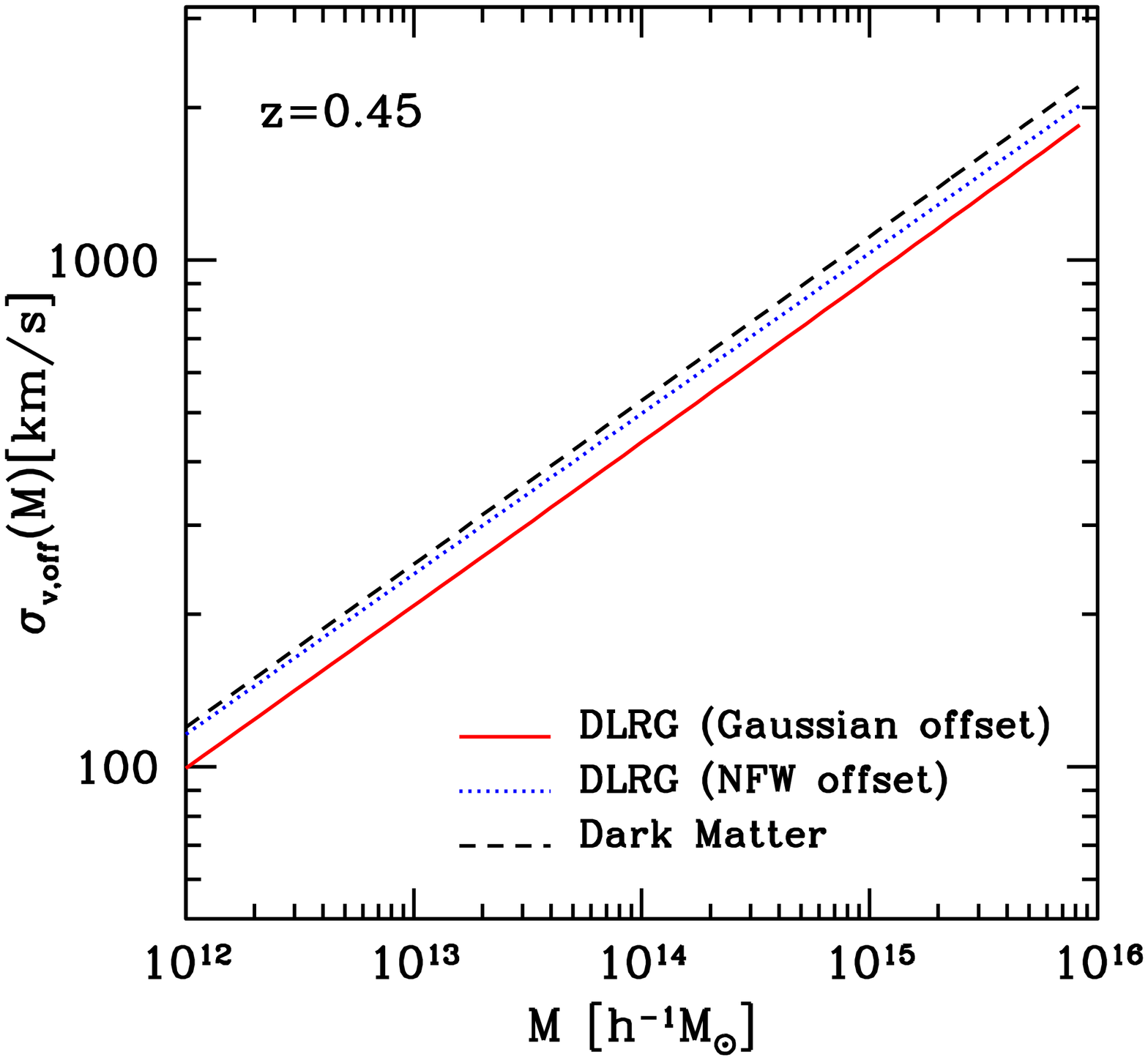}
\includegraphics[width=8cm]{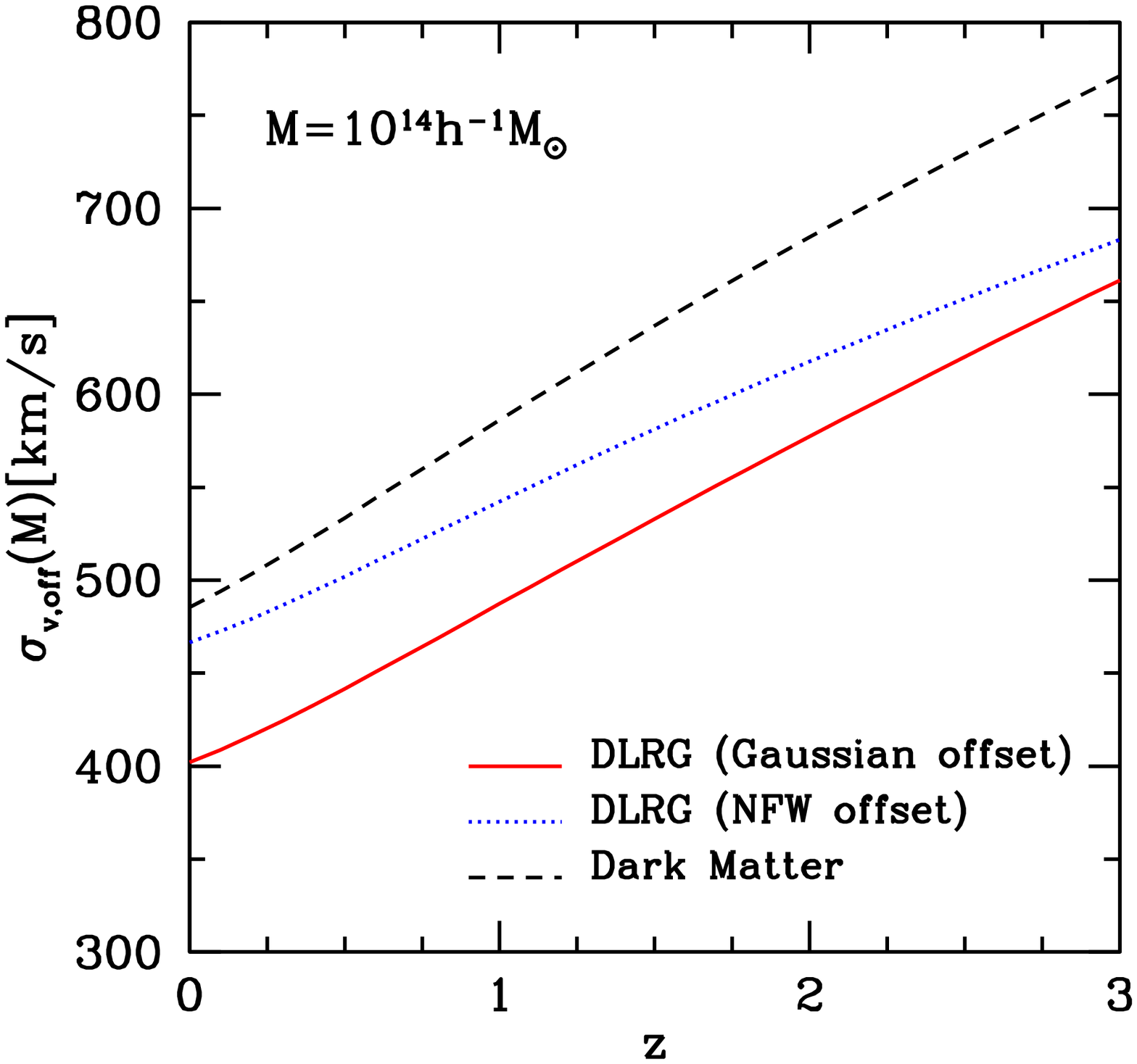} 
\caption{{\it Left panel:} Velocity dispersion of DLRGs averaged over
  DLRG radial profile, as a function of host halo mass at $z=0.45$,
  assuming the Gaussian (solid curve) and NFW (dotted) radial profile
  profiles as in Fig.~\ref{fig:profile}. For comparison the dashed
  curve shows the velocity dispersion of dark matter. {\it Right
    panel:} Redshift dependence of the velocity dispersions of DLRGs
  and DM for a halo with mass
  $M=10^{14}h^{-1}M_\odot$.}  \label{fig:sigv}
\end{center}
\end{figure}

The averaged velocity dispersion of DLRGs within halos of a given mass
scale $M$ can be obtained by averaging the velocity dispersion
(Eq.~[\ref{eq:sigv_rM}]) with the radial profile of DLRGs:
%%%%%%%%%%%%%%%%%%%%%%%%%%%%%%%%%%%%%%%%%%%%%%%%%%%%%%%%
\begin{equation}
\label{eq:sigvm}
\sigma^2_{v, {\rm off}}(M)\equiv 
\int_0^{r_{\rm vir}}\! 4\pi r^2dr~ p_{\rm off}(r; M)\sigma_{v}^2(r; M). 
\end{equation}
%%%%%%%%%%%%%%%%%%%%%%%%%%%%%%%%%%%%%%%%%%%%%%%%%%%%%%%%
This velocity dispersion has an asymptotic limit when all the DLRGs
are at the center of each halo: $\sigma_{v,{\rm off}}\rightarrow 0$
when $p_{\rm off}(r)\propto \delta_D(r)$.  Fig.~\ref{fig:sigv} plots
the velocity dispersion of DLRGs, $\sigma_{v,{\rm off}}(M)$, as a
function of halo mass $M$ for a fixed redshift (left panel), and as a
function of redshift $z$ for a fixed halo mass (right), respectively.
The velocity dispersion of DLRGs is larger in more massive halos and
at higher redshifts. Within the same halo, the velocity dispersion of
DLRGs is smaller than that of dark matter by 10-20\% in the
amplitudes, because DLRGs are more centrally concentrated.
Eq.~(\ref{eq:sigvir}) implies that the velocity dispersion of both the
DLRGs and the dark matter scales as ${\sigma}_{v,{\rm off}}(M)\propto
M^{1/3}$.
%%%%%%%%%%%%%%%%%%%%%%%%%%%%%%%%%%%%%%%%%%%%%%%%%%%%%%%%
%%%%%%%%%%%%%%%%%%%%%%%%%%%%%%%%%%%%%%%%%%%%%%%%%%%%%%%

\subsection{Redshift-space power spectrum of DLRG and the covariance matrix}

We use our model for the radial distribution of the DLRGs and their
velocity distribution to estimate the effect of the offset on the
power spectrum.  We assume the velocity distribution of DLRGs within
halos is Gaussian, where the width of the distribution is given by the
velocity dispersion (Eq.~[\ref{eq:sigv_rM}]):
%%%%%%%%%%%%%%%%%%%%%%%%%%%%%%%%%%%%%%%%%%%%%%%%%%%%%%%%
\begin{equation}
R(\Delta r_\parallel; r,M)d(\Delta r_\parallel)=
\frac{1}{\sqrt{2\pi}\sigma_{v,{\rm off}}(r,M)}
\exp\left[-\frac{v_{\parallel}^2}{2\sigma_{v, {\rm off}}^2(r,M)}\right]
dv_\parallel,
\label{eq:vel_gauss}
\end{equation}
%%%%%%%%%%%%%%%%%%%%%%%%%%%%%%%%%%%%%%%%%%%%%%%%%%%%%%%%
where $\Delta r_\parallel=v_\parallel/aH(z)$ and the prefactor is
determined so as to satisfy $\int_{-\infty}^{\infty}\!d \Delta
r'_\parallel~R(\Delta r_\parallel; M)=1$.

The Fourier transform of the redshift-space radial profile (see
Eq.~[\ref{eq:pdf_off_real}]) can be expressed as
%%%%%%%%%%%%%%%%%%%%%%%%%%%%%%%%%%%%%%%%%%%%%%%%%%%%%%%%
\begin{equation}
\label{eq:psoff}
%\tilde{p}_{s,{\rm off}}(k_{\perp},k_\parallel; M)=\tilde{p}_{\rm off}(k) 
%\exp\left[-\frac{\sigma_{v,{\rm off}}^2(M)k^2\mu^2}{2a^2H^2(z)}\right],
\tilde{p}_{s,{\rm off}}(k_{\perp},k_\parallel; M)\simeq
\int_0^{r_{\rm vir}}4\pi r^2 dr p_{\rm off}(r;M)
\exp\left[-\frac{\sigma_{v,{\rm off}}^2(r,M)k^2\mu^2}{2a^2H^2(z)}\right], 
\end{equation}
%%%%%%%%%%%%%%%%%%%%%%%%%%%%%%%%%%%%%%%%%%%%%%%%%%%%%%%%
where $k=\sqrt{k_\perp^2+k_\parallel^2}$ and the exponential function
above is the Fourier transform of Eq.~(\ref{eq:vel_gauss}).  We again
note that, exactly speaking, $\tilde{p}_{s,{\rm off}}$ also depends on
the Fourier transform of the real-space radial profile $\tilde{p}_{\rm
  off}$ as implied by Eq~(\ref{eq:pdf_off_real}), but we use
$\tilde{p}_{\rm off}\simeq 1$ at large length scales of interest, much
larger than the characteristic offset of the DLRG from the halo
center.  Also note that, for the limit $p_{\rm off}\rightarrow
\delta_D(r)$, $\tilde{p}_{s,{\rm off}}\rightarrow 1$ as
$\sigma_{v,{\rm off}}\rightarrow 0$ at $r\rightarrow 0$.

Hence the redshift-space power spectrum of DLRGs (see
Eq.~[\ref{eq:ps_lrg}]) can be computed for a given cosmological model
by inserting Eq.~(\ref{eq:psoff}) into
%%%%%%%%%%%%%%%%%%%%%%%%%%%%%%%%%%%%%%%%%%%%%%%%%%%%%%%%
\begin{eqnarray}
P_{s,{\rm DLRG}}(k,\mu)&=&\left[\frac{1}{\bar{n}_{\rm
DLRG}}\int\!dM~\frac{dn}{dM}b(M)N_{\rm HOD}(M)\tilde{p}_{
s, {\rm off}}(k,\mu; M) \right]^2P^{\rm NL}_{s,{\rm m}}(k,\mu).
\label{eq:ps_lrg_2}
\end{eqnarray}
%%%%%%%%%%%%%%%%%%%%%%%%%%%%%%%%%%%%%%%%%%%%%%%%%%%%%%%%
At very large length scales (or very small $k$'s) the redshift-space
power spectrum can be approximated as
%%%%%%%%%%%%%%%%%%%%%%%%%%%%%%%%%%%%%%%%%%%%%%%%%%%%%%%%
\begin{equation}
P_{s, {\rm DLRG}}(k,\mu)\approx \bar{b}^2 \left[1-\frac{
    \overline{\sigma}_{v, {\rm off}}^2 k^2\mu^2}{a^2H^2(z)}
  \right]P^{\rm NL}_{s, {\rm m}}(k,\mu),
\label{eq:ps_lrg_app}
\end{equation}
%%%%%%%%%%%%%%%%%%%%%%%%%%%%%%%%%%%%%%%%%%%%%%%%%%%%%%%%
where $\overline{\sigma}_{v,{\rm off}}^2$ is the velocity dispersion
averaged over the halo mass function weighted with the DLRG HOD:
%%%%%%%%%%%%%%%%%%%%%%%%%%%%%%%%%%%%%%%%%%%%%%%%%%%%%%%%%
\begin{equation}
\overline{\sigma}_{v,{\rm off}}^2\equiv 
\frac{1}{\bar{b}\bar{n}_{\rm DLRG}}
\int\!dM\frac{dn}{dM}b(M)N_{\rm HOD}(M) \sigma^2_{v,{\rm off}}(M).
\label{eq:sigv}
\end{equation}
%%%%%%%%%%%%%%%%%%%%%%%%%%%%%%%%%%%%%%%%%%%%%%%%%%%%%%%%%
As we will show below, the approximation (\ref{eq:ps_lrg_app}) is not
accurate at $k\simgt 0.15h$/Mpc.

Thus the key quantity characterizing the FoG effect on DLRG power
spectrum is the halo-mass averaged velocity dispersion,
$\overline{\sigma}_{v,{\rm off}}^2$. Table~\ref{tab:survey} gives the
values for different redshifts assuming our fiducial model
parameters. The DLRG velocity dispersion is also compared with that of
dark matter within the same halos hosting DLRGs. It can be found that
the typical FoG suppression scale, estimated as $\overline{\sigma}_{v,
  {\rm off}}/(aH)$, is of scales of $5~h^{-1}{\rm Mpc}$.  Therefore,
even at large length scales $k\simeq 0.1~h{\rm Mpc}^{-1}$, which is
employed in the literature in order to extract cosmological
information, the FoG effect suppresses the power spectrum amplitudes
by a factor of 0.75 ($1-[0.1\times 5]^2\simeq 0.75$) according to
Eq.~(\ref{eq:ps_lrg_app}), a systematic correction that is much larger
than the reported statistical errors in many surveys \citep[see
  Appendix C in][for discussion]{Reidetal:10}.
%%%%%%%%%%%%%%%%%%%%%%%%%%%%%%%%%%%%%%%%%%%%%%%%%%%%%%%%
\begin{figure}
\begin{center}
\includegraphics[width=8cm]{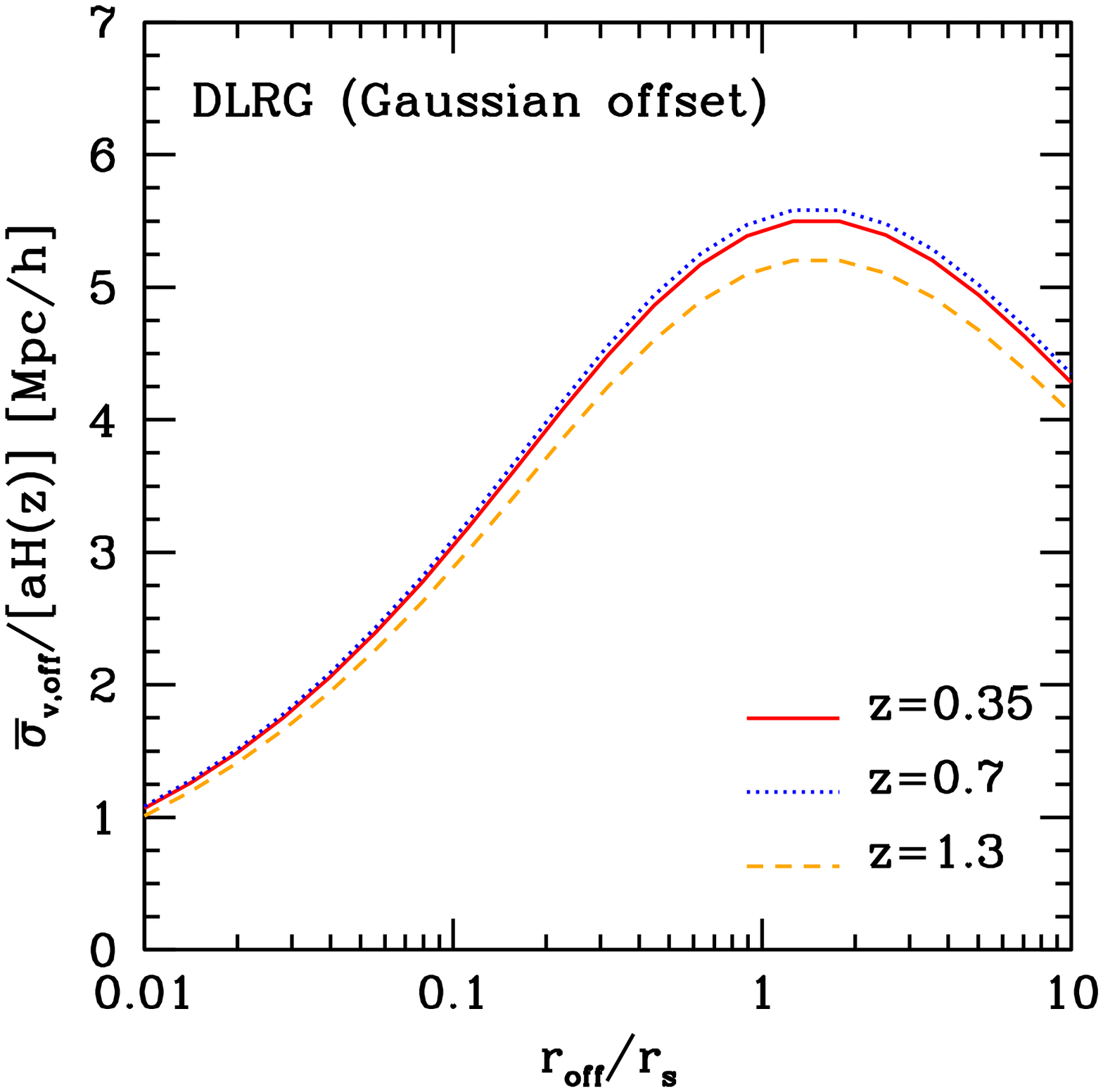}
\includegraphics[width=8cm]{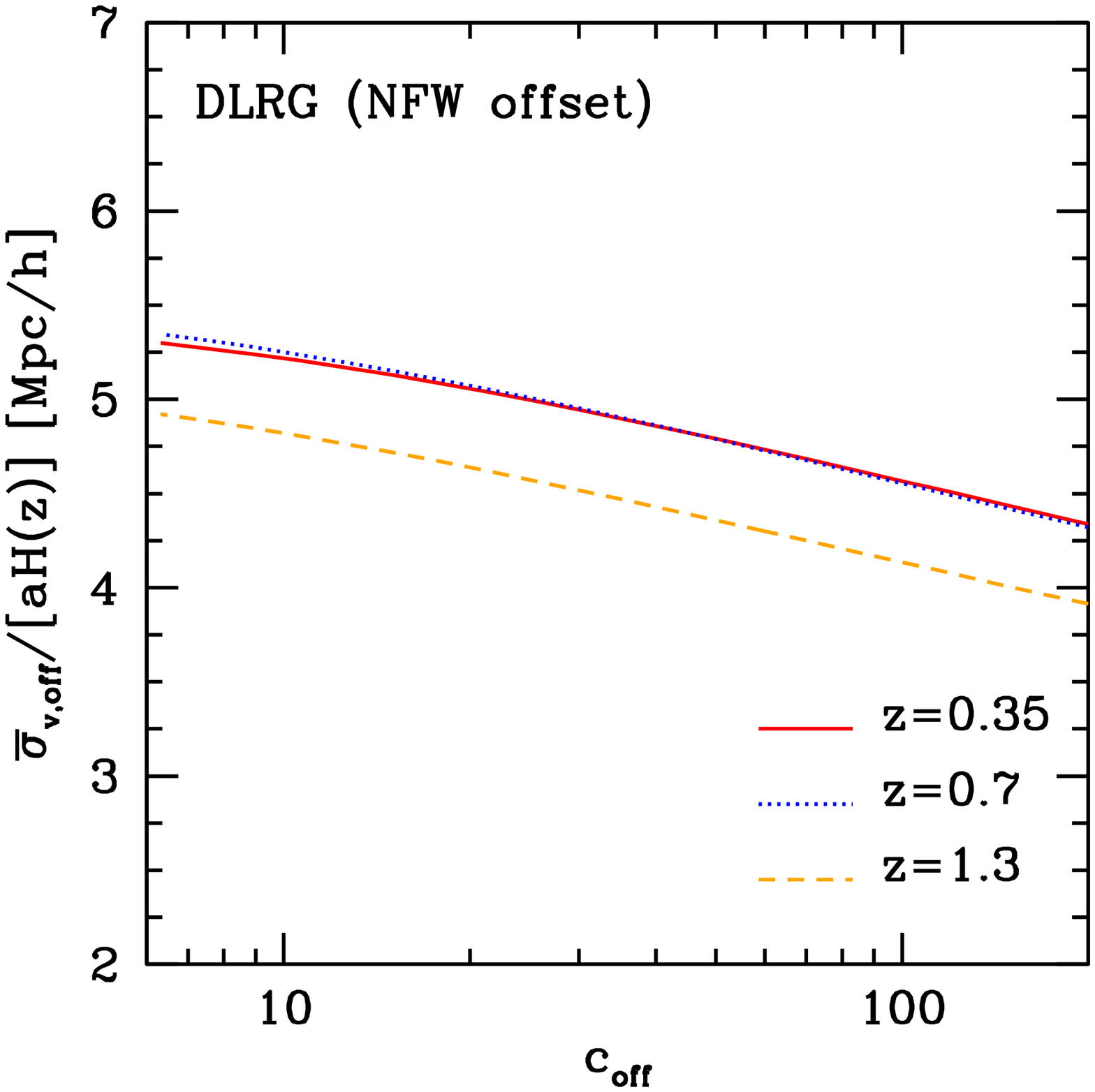}
\caption{The FoG suppression scale (see Eq.~[\ref{eq:sigv}]),
  $\overline{\sigma}_{v,{\rm off}}/aH(z)$, as a function of the model
  parameters of DLRG radial profiles, for the Gaussian radial profile
  model ({\em left panel}) and the NFW model ({\em right panel}). The
  FoG scale is computed by averaging the velocity dispersion of DLRGs
  over the halo mass function weighted by the DLRG halo occupation
  distribution (Eq.~[\ref{eq:HOD}]). The different curves are for
  different redshifts.
\label{fig:fogscale}
}
\end{center}
\end{figure}
%%%%%%%%%%%%%%%%%%%%%%%%%%%%%%%%%%%%%%%%%%%%%%%%%%%%%%%

%%%%%%%%%%%%%%%%%%%%%%%%%%%%%%%%%%%%%%%%%%%%%%%%%%%%%%%%
\begin{figure}
\begin{center}
\includegraphics[width=8cm]{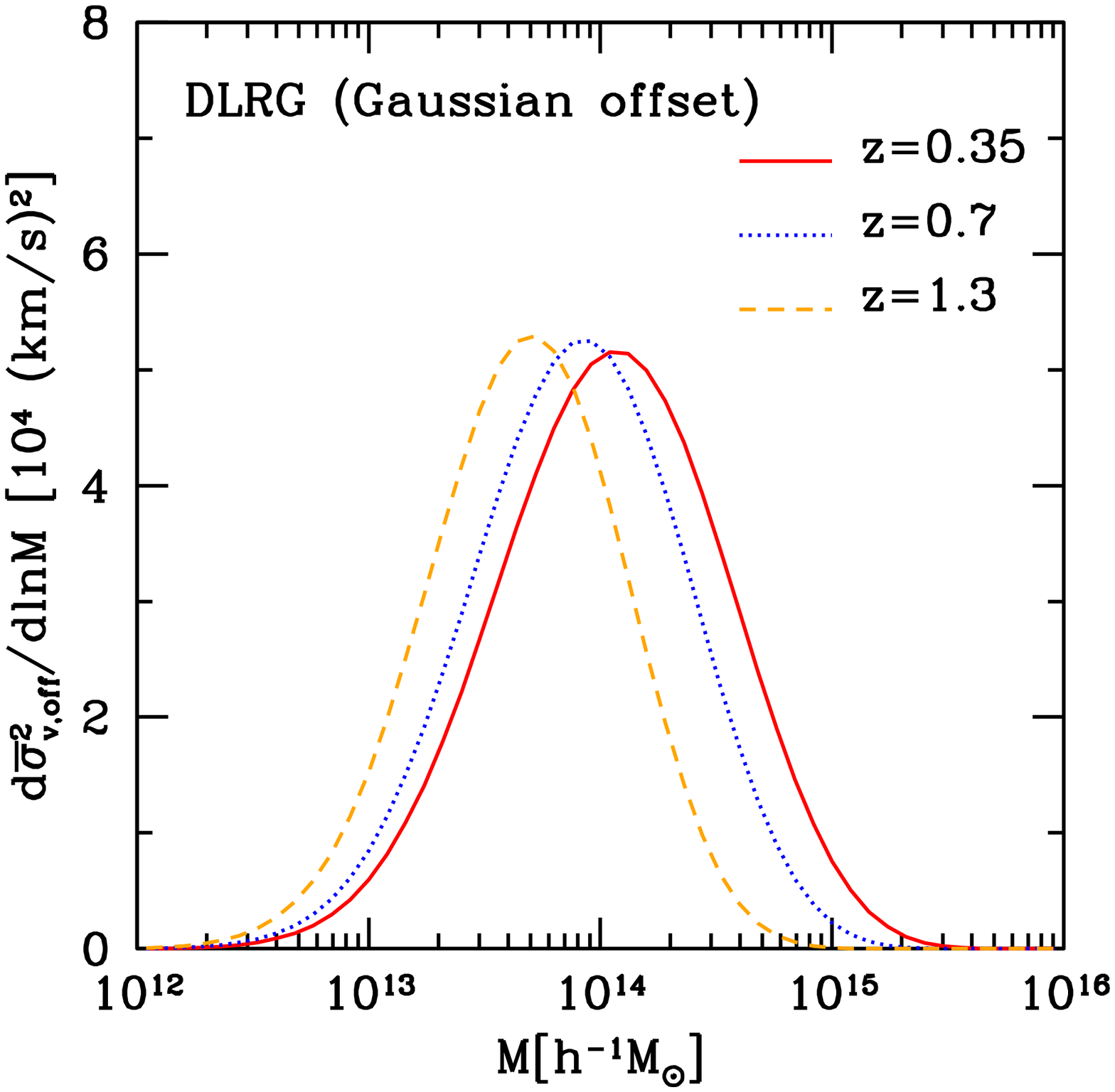}
\includegraphics[width=8cm]{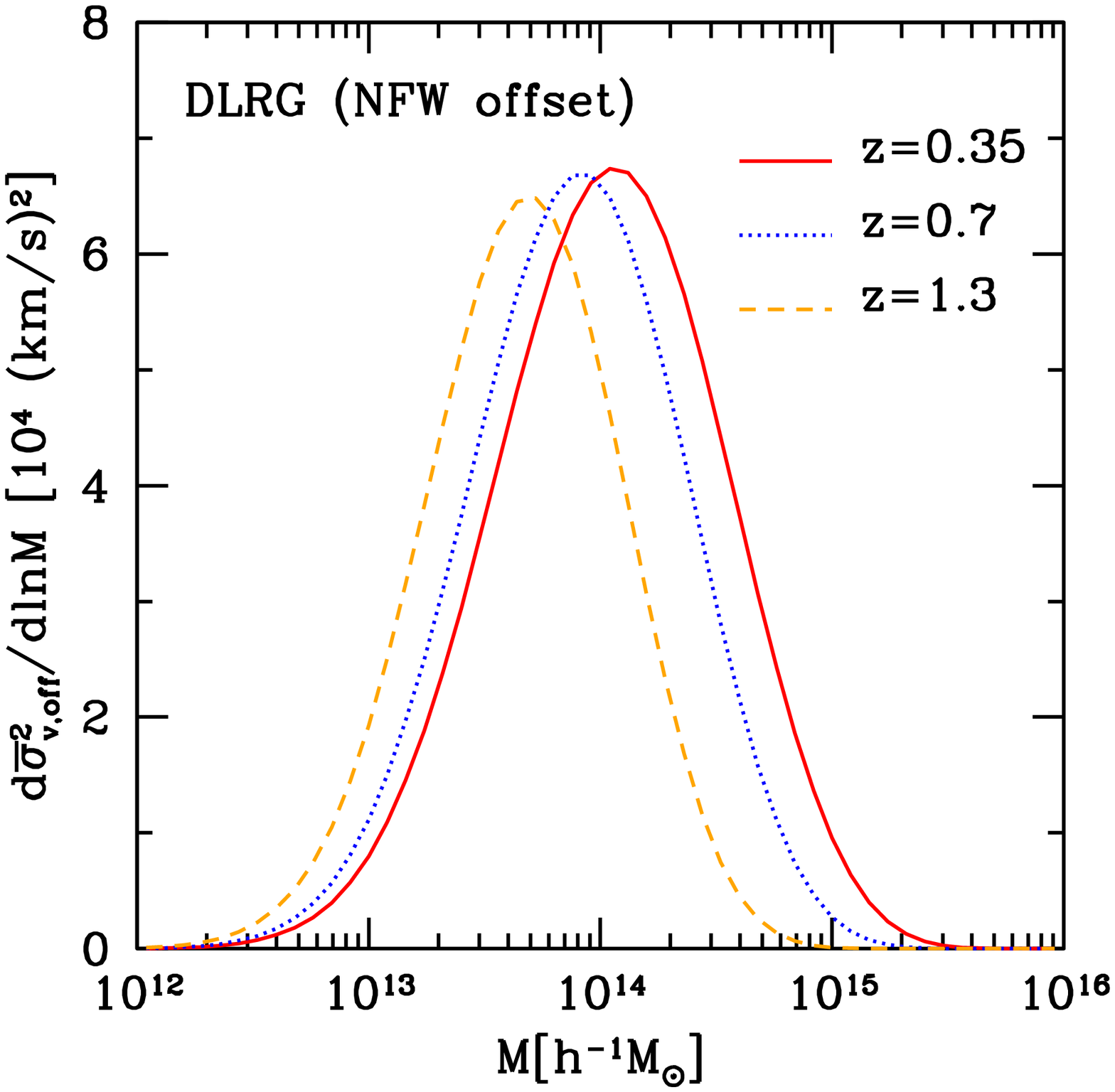}
\caption{This figure shows how different mass halos contribute to the
  power spectrum suppression shown in Fig.~\ref{fig:fogscale}.  While
  the typical DLRG sits in a halo with mass $\sim 10^{13}M_\odot$, the
  FoG in the more massive halos plays the most important role in
  suppressing the galaxy power spectrum.  The left and right panels
  are the results for the Gaussian and NFW radial profile profiles,
  respectively, where we assume the fiducial model parameters $r_{\rm
    off}=0.3r_s $ and $c_{\rm off}=20$, respectively, as in
  Fig.~\ref{fig:profile}.  }  \label{fig:dsigmav_dlnM}
\end{center}
\end{figure}
%%%%%%%%%%%%%%%%%%%%%%%%%%%%%%%%%%%%%%%%%%%%%%%%%%%%%%%
Fig.~\ref{fig:fogscale} plots how the typical FoG suppression scale
changes with changing parameters of the Gaussian and NFW DLRG radial
profiles.  For the Gaussian radial profile, the FoG displacement scale
has a maximum scale of $\overline{\sigma}_{v, {\rm off}}/(aH)\sim
5~h^{-1}{\rm Mpc}$ around $r_{\rm off} \sim r_s$, as the velocity
dispersion peaks at the scale radius $r_s$ for an NFW profile.  On the
other hand, for the NFW radial profile, the displacement scale has a
weak dependence on $c_{\rm off}$, and leads to $\sim 5~h^{-1}{\rm
  Mpc}$ over all the range of $c_{\rm off}$.  The redshift dependence
of the FoG scale is weak and changes by only $\sim 10\%$.

Fig.~\ref{fig:dsigmav_dlnM} shows that the more massive halos are
responsible for most of the FoG suppression.  While the typical DLRG
sits in halos of mass $5\times 10^{13}~h^{-1}M_\odot$, the FoG arises
primarily from more massive halos as these halos have larger velocity
dispersions.  Here, we again assume the fiducial model parameters of
the FoG effects as in Fig.~\ref{fig:profile}. The plot shows that
halos with masses $M\sim 10^{14}~h^{-1}M_\odot$ have a dominant
contribution to the FoG effect at redshifts $z=0.35$ and 0.7 for both
the Gaussian and NFW radial profiles, while less massive halos become
more important at higher redshifts.
%Fig.~\ref{fig:dsigmav_dlnM} also shows that
%halos in roughly two decades of mass scales contribute.  Compared to
%the mean mass scale of halos hosting DLRGs, which is about $0.5\times
%10^{14}h^{-1}M_\odot$ as given in Table~\ref{tab:survey}, the FoG
%effect arises mainly from more massive halos of $10^{14}M_\odot$ due
%to a more weight on the velocity dispersion of DLRGs in more massive
%halos as implied in Eq.~(\ref{eq:sigv}).

The covariance matrix describes statistical uncertainties in measuring
the redshift-space power spectrum from a given survey, and the
correlations between the power spectra of different wavenumbers.
\cite{Takahashietal:09} show that the assumption of Gaussian errors is
valid on scales of interest.  In this limit of Gaussian errors, the
covariance matrix has a simple form:
%%%%%%%%%%%%%%%%%%%%%%%%%%%%%%%%%%%%%%%%%%%%%%%%%%%%%%%%
\begin{equation}
{\rm Cov}[P_{s, {\rm DLRG}}(k_i,\mu_a)
P_{s, {\rm DLRG}}(k_j,\mu_b)]
=\frac{2 \delta^K_{ij}\delta^K_{ab}}{N^{\rm mode}(k_i,\mu_b)}
\left[P_{s,{\rm DLRG}}(k_i,\mu_b)+\frac{1}{\bar{n}_{\rm DLRG}}\right]^2, 
%\frac{2}{N_i^{\rm mode}}\left[1+\frac{1}{\bar{n}_{\rm DLRG}
%P_{s,DLRG}(k_i)}\right]^2,
\label{eq:cov_pk}
\end{equation}
%%%%%%%%%%%%%%%%%%%%%%%%%%%%%%%%%%%%%%%%%%%%%%%%%%%%%%%%
where $k_i$ and $\mu_a$ are the $i$-th and $a$-th bins of wavenumber
and cosine angle, respectively, and $\delta^K_{ij}$ and
$\delta^K_{ab}$ are the Kronecker delta function: $\delta^K_{ij}=1$ if
$i=j$ within the bin width, otherwise $\delta^k_{ij}=0$ and so on.
The Kronecker delta functions impose that the power spectrum of
different wavenumber bins are independent.  The quantity $N^{\rm
  mode}(k_i,\mu_a)$ is the number of independent Fourier modes around
the bin centered at $k_i$ and $\mu_a$ with widths $\Delta k$ and
$\Delta \mu$, which can be resolved for a given survey volume $V_s$:
$N^{\rm mode}(k_i,\mu_a)=2\pi k_i^2\Delta k\Delta \mu
V_s/(2\pi)^3$. Here we assume the fundamental Fourier mode is
determined by the survey volume as $k_f=2\pi/L$, a reasonable
approximation for a simple survey geometry.

\section{Angular power spectrum of DLRG-galaxy weak lensing}
\label{sec:lensing}

Observations of DLRG-galaxy lensing measure the radial distribution of
DLRGs in the halo.  In this subsection, we briefly review
\cite{OguriTakada:10} discussion of how the radial distribution of the
DLRGs in their halo affects the galaxy-galaxy lensing observables.

The halos hosting DLRGs distort background galaxy images.  By
cross-correlating positions of DLRGs on the sky with tangential
ellipticity component of background galaxy images with respect to the
line connecting DLRG and background galaxy, we can measure the
radially averaged mass distribution around a DLRG
\citep[][]{Mandelbaumetal:06}.  While this stacking analysis is
usually done in real space, we will describe the results in Fourier
space as the effect of the DLRG offsets are convolution in real space
and multiplication in Fourier space \citep[][] {OguriTakada:10}.

Since we are interested in small angular scales, we can use the
flat-sky approximation \citep{Limber:54} and use the halo model in
\cite{OguriTakada:10} to compute the angular power spectrum of
DLRG-galaxy lensing \citep[also see][]{TakadaBridle:07}:
%%%%%%%%%%%%%%%%%%%%%%%%%%%%%%%%%%%%%%%%%%%%%%%%%%%%%%%
\begin{equation}
C_{\gamma g}(l)=C_{\gamma g}^{1h}(l)+C_{\gamma g}^{2h}(l), 
\end{equation}
%%%%%%%%%%%%%%%%%%%%%%%%%%%%%%%%%%%%%%%%%%%%%%%%%%%%%%%
where $C_{\gamma g}^{1h}$ and $C_{\gamma g}^{2h}$ are the 1- and
2-halo term spectra defined as follows. For the full-sky expression of
the lensing power spectrum, see \cite{dePutterTakada:10}.  The 1-halo
term contribution to galaxy-galaxy lensing arises from the mass
distribution within one halo that hosts DLRGs and gives dominant
contribution to the signal on small angular separations:
%%%%%%%%%%%%%%%%%%%%%%%%%%%%%%%%%%%%%%%%%%%%%%%%%%%%%%%
\begin{eqnarray}
C_{\gamma g}^{1h}(l)\equiv \frac{1}{\bar{n}^{\rm 2D}_{\rm DLRG}}
\int\!d\chi\frac{d^2V}{d\chi d\Omega}S_{\rm DLRG}(z)W^{\rm GL}(\chi)
\chi^{-2}
\int\!dM~\frac{dn}{dM}N_{\rm HOD}(M) \frac{1}{\bar{\rho}_{m 0}}
\left[\left.
M\tilde{u}_{{\rm NFW}}\!\left(k; M,z
\right) \tilde{p}_{\rm off}(k; M)\right|_{k=l/\chi}
+ m_{\rm sh, DLRG}
\right],
\label{eq:lens_1h}
\end{eqnarray}
%%%%%%%%%%%%%%%%%%%%%%%%%%%%%%%%%%%%%%%%%%%%%%%%%%%%%%%%
where $\bar{\rho}_{m0}$ is the mean mass density today, $\chi$ is the
comoving angular diameter distance (which is given as a function of
redshift via the distance-redshift relation), $W^{\rm GL}(\chi)$ is
the lensing efficiency function for a given source galaxy population
\citep[see Eq.~19 in][]{OguriTakada:10}, and $d^2V/d\chi d\Omega$ is
the volume element in the unit comoving interval and the unit solid
angle; $d^2V/d\chi d\Omega=\chi^2$ for a flat universe. 
The function $S_{\rm DLRG}(z)$ is
the redshift selection function of DLRGs.  For simplicity, we assume a
complete selection function: $S_{\rm DLRG}(z)=1$ within the redshift
range of the survey, and otherwise $S_{\rm DLRG}=0$. The quantity
$\bar{n}_{\rm DLRG}^{\rm 2D}$ is the mean angular number density of
DLRG in the redshift slice: $\bar{n}_{\rm DLRG}^{\rm 2D}\equiv
\int\!d\chi (d^2V/d\chi d\Omega) \int\!dM (dn/dM) N_{\rm HOD}(M)S_{\rm
  DLRG}(z) $. The term denoted by $m_{\rm sh, DLRG}$ gives the
contribution arising from a subhalo hosting DLRG, and we will
throughout this paper assume the subhalo mass $m_{\rm sh,
  DLRG}=0.32\times 10^{12}h^{-1}M_\odot$ as our fiducial value,
implied from the results in \cite{Johnstonetal:07}.

There are two contributions to the 1-halo term in
Eq.~(\ref{eq:lens_1h}): the first term in the bracket describes the
contribution of the halo of mass $M$ and the second term describes the
contribution of a subhalo hosting the DLRG.  For the first term, we
assume
%In the 1-halo term above we consider the two contributions: one is
%from the main halo that hosts DLRG and has mass $M$ appearing in the
%halo mass function, and the other is from subhalo because DLRG very
%likely resides on its own subhalo of galactic mass scales. For the
%former we assume 
an NFW profile characterizing the dark matter distribution within a
halo, and $\tilde{u}_{{\rm NFW}}$ is the Fourier-transform \citep[see
  Eq.~29 in][]{OguriTakada:10}.  Including off-centered DLRGs in the
galaxy-galaxy lensing analysis dilutes the measured lensing signal
amplitudes at the small scales
\citep{Johnstonetal:07,OguriTakada:10}. This off-centering effect on
the lensing power spectrum can be included by simply replacing the
dark matter profile with $\tilde{u}\tilde{p}_{\rm off}$, where
$\tilde{p}_{\rm off}$ is the Fourier-transformed coefficients of the
DLRG radial profile (see Eq.~[\ref{eq:tilde_poff}]).  For the subhalo
contribution we simply assume the delta function for the mass profile,
a good approximation at the relevant angular scales. In this limit,
the power spectrum behaves like a white shot noise.

Similarly, the 2-halo term contribution, which dominates at large
scales, is given as
%%%%%%%%%%%%%%%%%%%%%%%%%%%%%%%%%%%%%%%%%%%%%%%%%%%%%%%%
\begin{equation}
C^{2h}_{\gamma g}(l)\equiv \frac{1}{\bar{n}_{\rm DLRG}^{\rm 2D}}
\int\!d\chi~ \frac{d^2V}{d\chi d\Omega}S_{\rm DLRG}(z)W^{\rm GL}(\chi)\chi^{-2}\left[
\int\!dM\frac{dn}{dM}b(M)N_{\rm HOD}(M)
\right] P_m^L\!\left(k=\frac{l}{\chi}; z\right),
\end{equation}
%%%%%%%%%%%%%%%%%%%%%%%%%%%%%%%%%%%%%%%%%%%%%%%%%%%%%%%%
where $P^L_m(k)$ is the linear mass power spectrum. This 2-halo term
scales with halo bias: if DLRGs are residing on more massive halos or
equivalently more biased halos, the 2-halo term has greater
amplitudes.

To perform parameter forecasts for planned lensing surveys, we also
need to model the lensing power spectrum covariance.  Following
\cite{TakadaJain:09}, we assume Gaussian errors so that
the covariance matrix of the lensing power spectrum is given by a
product of sampling variance and shot noise terms \citep[see][for the
  definition of the covariance matrix]{OguriTakada:10}.
%%%%%%%%%%%%%%%%%%%%%%%%%%%%%%%%%%%%%%%%%%%%%%%%%%%%%%%%%
%\begin{equation}
%{\rm Cov}[C_{\gamma g}(l),C_{\gamma g}(l')]=\frac{\delta^K_{ll'}}{(2l+1)\Delta l
% f_{\rm sky}}\left[
%\left(C_{gg}(l)+\frac{1}{\bar{n}_{\rm DLRG}^{2D}}\right)
%\left(
%%C_{\gamma\gamma}+\frac{\sigma_\epsilon^2}{\bar{n}_{g}^{\rm imaging}}\right)
%+C_{\gamma g}(l)^2
%\right].
%\label{eq:lens_cov}
%\end{equation}
%%%%%%%%%%%%%%%%%%%%%%%%%%%%%%%%%%%%%%%%%%%%%%%%%%%%%%%%%%
%where $\Delta l$ is the bin width around multipole bin $l$, and
%$f_{\rm sky}$ is the sky coverage of a given survey. The power
%spectrum $C_{gg}(l)$ is the auto-spectrum of DLRGs taken in the lensing
%analysis, while $C_{\gamma\gamma}(l)$ is the shear power spectrum of
%the source galaxy images. The first term in the square bracket of the
%equation above takes into account the shot noise contamination arising
%from the finite number of DLRGs as well as from the intrinsic
%ellipticity noise, where $\sigma_{\epsilon }$ is the rms intrinsic
%ellipticities per component and $\bar{n}^{\rm imaging}_g$ is the mean
%angular number density of source galaxies. 

%%%%%%%%%%%%%%%%%%%%%%%%%%%%%%%%%%%%%%%%%%%%%%%%%%%%%%%%
\begin{figure}
\begin{center}
\includegraphics[width=17.5cm]{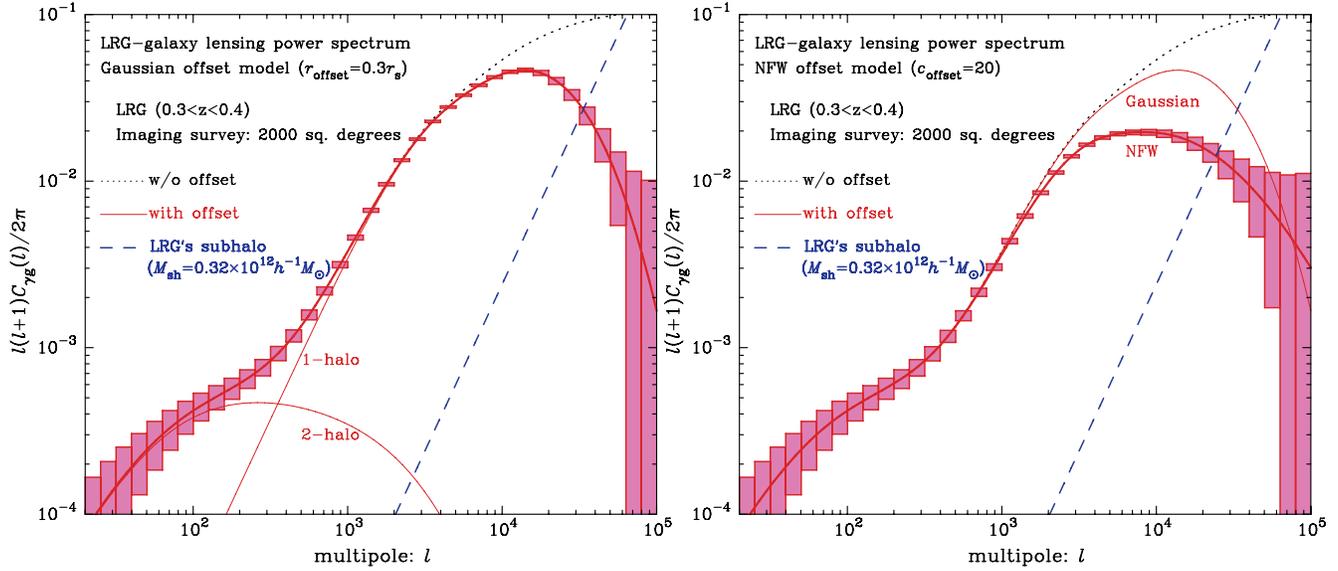}
\hspace{4cm} 
\caption{Angular power spectra of DLRG-galaxy lensing assuming that
  spectroscopic and imaging surveys are available for an overlapping
  area of $2000$ square degrees and the DLRGs are in the redshift
  range of $0.3<z<0.4 $ (see Section~\ref{sec:survey} for
  details). The bold solid curves are the angular power spectra
  including the off-centering effect for the Gaussian ({\em left
    panel}) and NFW ({\em right}) DLRG radial profiles. The thin
  curves in the left panel are the 1- and 2-halo term contributions.
  For comparison, the dotted curves are the spectra without the
  off-centering effect. The dashed curves are the contribution from a
  subhalo hosting DLRG, assuming the subhalo mass $m_{\rm sh,
    DLRG}=0.32\times 10^{12}h^{-1}M_\odot$. The boxes around the bold
  solid curve are the expected $1\sigma$ uncertainties in measuring
  band powers of the power spectrum at each $l$ bins, where we assume
  survey parameters given in Table~\ref{tab:survey}.}
\label{fig:lrglens}
\end{center}
\end{figure}
%%%%%%%%%%%%%%%%%%%%%%%%%%%%%%%%%%%%%%%%%%%%%%%%%%%%%%%

Fig.~\ref{fig:lrglens} shows the angular power spectrum of DLRG-galaxy
weak lensing expected when cross-correlating DLRGs in redshift slice
$0.3<z<0.4$ with background galaxy images that have a typical redshift
of $z\sim 1$ as expected for a Subaru-type imaging survey. The top
dotted curve shows the power spectrum when all DLRGs are at each
halo's center, while the bold solid curves show the spectrum with the
off-centering effect assuming our fiducial models of the Gaussian
({\em left panel}) and NFW ({\em right panel}) radial profiles as in
Fig.~\ref{fig:profile}. The figure shows that the off-centering effect
significantly dilutes the lensing power spectrum amplitudes at angular
separations smaller than the projected offset scale. The dashed curve
shows the contribution from DLRG subhalo assuming $m_{\rm sh,
  DLRG}=0.32\times 10^{12}h^{-1}M_\odot$.  The DLRG subhalo
contribution becomes significant at $l\simgt 2\times 10^{4}$, where an
angular scale of $l\sim 10^4$ corresponds to the projected scale $\sim
300~{\rm kpc}$ for the DLRG redshift of $z=0.35$ for a $\Lambda$CDM
model.

The boxes around the curve show statistical uncertainties in measuring
band powers at each multipole bins, expected when measuring the
DLRG-galaxy lensing for an overlapping area of $2000$ square degrees
between spectroscopic and imaging surveys. Here we assume the depth
expected for Subaru HSC survey that probes galaxies at typical
redshifts of $z_s\sim 1$ (see Section~\ref{sec:survey} for
details). The plot clearly implies that combining such imaging and
spectroscopic surveys allows us to infer the off-centering effect at a
high significance. We will below give a more quantitative estimate.

\section{Results}
\label{sec:results}

In this section, we estimate the ability of ongoing and planned
surveys for using the DLRG-galaxy weak lensing measurements to correct
the FoG effect on redshift-space power spectrum of DLRGs.

\subsection{Survey parameters}
\label{sec:survey}

To model the DLRG power spectrum available from ongoing and upcoming
spectroscopic surveys we assume survey parameters that resemble BOSS,
Subaru PFS, and Euclid surveys. The expected statistical uncertainties
in measuring the DLRG power spectrum in each redshift slice depend on
the area coverage (or equivalently the comoving volume) and the number
density and bias parameter of DLRGs.  For the sky coverage we assume
10,000, 2,000, and 20,000 square degrees for the BOSS, PFS, and Euclid
surveys, respectively. Table~\ref{tab:survey} summarizes the survey
parameters. For the redshift slices we consider 3 redshift bins with
width $\Delta z=0.1$ over the range $0.3<z<0.6$ for the BOSS survey, 4
slices with $\Delta z=0.2$ over $0.6<z<1.4$ for the PFS survey, and 10
slices with $\Delta z=0.1$ over $0.5<z<1.5$ for the Euclid
survey. Thus the BOSS and PFS surveys are complementary to each other
in their redshift coverages, while Euclid survey covers a wide range
of redshifts on its own.  Having a wider redshift coverage allows us
to trace the redshift evolution of mass clustering growth over a wider
range of redshifts and therefore improves cosmological constraints
\citep[][]{Takadaetal:06}. In Table~\ref{tab:survey} we also show the
comoving volume of each redshift slice.  We simply assume the same HOD
model given in Eq.~(\ref{eq:HOD}) over all the redshifts to estimate
the number density of DLRGs and the mean bias. The number densities
are somewhat smaller than those actually obtained from data
\citep[e.g.,][]{Whiteetal:11} or the target number densities for the
survey. This is because we use the HOD model for DLRGs: we assume that
we can select one DLRG per each halo based on the method of
\cite{Reidetal:10}.  Table~\ref{tab:survey} also shows the FoG
displacement scale $\overline{\sigma}_{v,{\rm off}}/(aH)$ in each
redshift slice, which is estimated using Eq.~(\ref{eq:sigv}).

The parameter forecast is also sensitive to the maximum wavenumber
$k_{\rm max}$, because complex non-linearities alter the power
spectrum at high wavenumber.  In our analysis, we use
\cite{Taruyaetal:09,Taruyaetal:10} to determine $k_{\rm max}$:
%%%%%%%%%%%%%%%%%%%%%%%%%%%%%%%%%%%%%%%%%%%%%%%%%%%%%%%%%
\begin{equation}
\frac{k_{\rm max}^2}{6\pi^2}\int_0^{k_{\rm max}}dk~ P^L_m(k)=C.
\label{eq:Ckmax}
\end{equation}
%%%%%%%%%%%%%%%%%%%%%%%%%%%%%%%%%%%%%%%%%%%%%%%%%%%%%%%%%
We use either $C=0.2$ or $0.7$, motivated by the fact that the model
predictions based on the standard and improved perturbation theory are
sufficiently accurate in a sense that the predictions well match
N-body simulation results to a few \% accuracies in the amplitudes up
to the determined $k_{\rm max}$. Table~\ref{tab:survey} gives the
$k_{\rm max}$ values for $C=0.2$ and 0.7, respectively, in each
redshift slice.  The approach in our analysis is to assume that
refined model prescriptions based on N-body simulations
\cite[][]{Taruyaetal:10} are used to estimate non-linearities in the
real space power spectrum and focus on the FoG effects on the DLRG
power spectrum.  Our goal is to show how the lensing analysis can
calibrate the FoG effects.

For our analysis, we assume a lensing survey with either the
properties of the planned Subaru HSC survey or the proposed EUCLID
imaging survey.  For the Subaru HSC survey we assume the survey area
$\Omega_{\rm s}=2,000$ square degrees, the mean number density of
imaging galaxies $\bar{n}_g^{\rm imaging }=30~$arcmin$^{-2}$, and the
redshift distribution is given by the functional form $n_g(z)\propto
z^2\exp(-z/z_0)$ where the parameter $z_0$ is determined so that the
mean redshift $\langle z \rangle=1$. For the Euclid survey we simply
assume the same survey parameters, except for the survey area of
$\Omega_{\rm s}=20,000$ square degrees.  When computing the
DLRG-galaxy lensing we use background galaxies at redshifts $z>z_{{\rm
    DLRG},i}+\Delta z/2+0.05$, where $z_{{\rm DLRG}, i}+\Delta z/2$ is
the upper bound on redshifts of DLRGs in the $i$-th redshift slice of
a given spectroscopic survey. That is, we include galaxies at
redshifts higher than the redshifts of any DLRGs in a given redshift
slice by $\delta z=0.05$. We assume that such background galaxies can
be selected based on their available photometric redshift estimates
\citep[][]{Nishizawaetal:10}. In addition we assume the rms intrinsic
ellipticities $\sigma_\epsilon=0.22$ per component, which determines
the intrinsic ellipticity noise contamination to the error covariance
matrix of the DLRG-galaxy lensing power spectrum
\citep{OguriTakada:10}.

%%%%%%%%%%%%%%%%%%%%%%%%%%%%%%%%%%%%%%%%%%%%%%%%%%%%%%%%
\begin{table*}
\begin{center}
Parameters of Spectroscopic Surveys\\
\begin{tabular}{ccccccccccc}
  \hline\hline
& & $V_s$ & $k_{{\rm max},C=0.2} $ & $k_{{\rm  max}, C=0.7} $ & $\bar{n}_{\rm DLRG}$ &
& $\bar{M}_{h}$ 
& \multicolumn{3}{c}{$\overline{\sigma}_{ v,{\rm off}}/aH(z)$ [Mpc/h]} \\
\cline{9-11}
Survey & $z$ & \raisebox{-1ex}{[$({\rm Gpc}/h)^3$]} & \raisebox{-1ex}{[$h/{\rm Mpc}$]} 
& \raisebox{-1ex}{[$h/{\rm Mpc}$]} & \raisebox{-1ex}{[$(h/{\rm Mpc})^{3}$]} & $\bar{b}$ 
& \raisebox{-1ex}{[$10^{14}M_\odot/h$]} & \raisebox{-1ex}{DM} & \raisebox{-1ex}{DLRG}  & \raisebox{-1ex}{DLRG} \\
& & & & & & & & & \raisebox{0ex}{(Gauss)} & \raisebox{0ex}{(NFW)} 
\\ \hline
BOSS             & 0.3 -- 0.4 & 0.76 & 0.10 & 0.17 & $1.25\times 10^{-4}$ & 1.80 & 0.63 & 5.35 & 4.42 & 5.06  \\
(10,000~deg$^2$) & 0.4 -- 0.5 & 1.14 & 0.10 & 0.18 & $1.15\times 10^{-4}$ & 1.85 & 0.58 & 5.41 & 4.48 & 5.10  \\
                 & 0.5 -- 0.6 & 1.53 & 0.11 & 0.19 & $1.05\times 10^{-4}$ & 1.89 & 0.54 & 5.43 & 4.50 & 5.11  \\
\hline
PFS              & 0.6 -- 0.8 & 0.84 & 0.11 & 0.20 & $0.91\times 10^{-4}$ & 1.95 & 0.48 & 5.42 & 4.49 & 5.07  \\
(2,000deg$^2$)   & 0.8 -- 1.0 & 1.13 & 0.12 & 0.22 & $0.72\times 10^{-4}$ & 2.02 & 0.42 & 5.33 & 4.43 & 4.97  \\
                 & 1.0 -- 1.2 & 1.37 & 0.13 & 0.23 & $0.57\times 10^{-4}$ & 2.08 & 0.36 & 5.20 & 4.32 & 4.81  \\
                 & 1.2 -- 1.4 & 1.56 & 0.14 & 0.25 & $0.44\times 10^{-4}$ & 2.14 & 0.31 & 5.03 & 4.19 & 4.64  \\
\hline
Euclid           & 0.5 -- 0.6 & 3.06 & 0.11 & 0.19 & $1.05\times 10^{-4}$ & 1.89 & 0.54 & 5.43 & 4.50 & 5.11 \\
(20,000deg$^2$)  & 0.6 -- 0.7 & 3.84 & 0.11 & 0.20 & $0.95\times 10^{-4}$ & 1.93 & 0.50 & 5.56 & 4.60 & 5.09 \\
                 & 0.7 -- 0.8 & 4.60 & 0.12 & 0.20 & $0.86\times 10^{-4}$ & 1.96 & 0.46 & 5.53 & 4.59 & 5.05 \\
                 & 0.8 -- 0.9 & 5.32 & 0.12 & 0.21 & $0.77\times 10^{-4}$ & 2.00 & 0.43 & 5.49 & 4.56 & 5.00 \\
                 & 0.9 -- 1.0 & 5.98 & 0.13 & 0.22 & $0.69\times 10^{-4}$ & 2.04 & 0.40 & 5.43 & 4.51 & 4.93 \\
                 & 1.0 -- 1.1 & 6.58 & 0.13 & 0.23 & $0.61\times 10^{-4}$ & 2.07 & 0.37 & 5.36 & 4.45 & 4.85 \\
                 & 1.1 -- 1.2 & 7.11 & 0.14 & 0.24 & $0.54\times 10^{-4}$ & 2.10 & 0.34 & 5.28 & 4.39 & 4.77 \\
                 & 1.2 -- 1.3 & 7.58 & 0.14 & 0.25 & $0.47\times 10^{-4}$ & 2.13 & 0.32 & 5.20 & 4.33 & 4.68 \\
                 & 1.3 -- 1.4 & 8.00 & 0.15 & 0.26 & $0.41\times 10^{-4}$ & 2.16 & 0.30 & 5.11 & 4.26 & 4.59 \\
                 & 1.4 -- 1.5 & 8.35 & 0.15 & 0.26 & $0.36\times 10^{-4}$ & 2.18 & 0.28 & 5.01 & 4.18 & 4.50 \\
\hline
\end{tabular}
\end{center}
\caption{Survey parameters considered in this paper, which are chosen
  to resemble the ongoing BOSS survey, and the planned Subaru PFS
  survey and Euclid survey.  The range of redshift $z$, the survey
  volume of each redshift slice $V_s$, the mean number density of
  DLRGs $\bar{n}_{\rm DLRG}$, the effective linear bias $\bar{b}$, and
  the mean halo mass $\bar{M}_h$ are given. The quantity $k_{{\rm
  mac},C=0.2}$ or $k_{{\rm mac}, C=0.7}$ is the maximum wavenumber of
  each redshift slice up to which the power spectrum information is
  included when studying parameter forecasts. The maximum wavenumbers
  are chosen by setting $C=0.2$ or $0.7 $ in Eq.~(\ref{eq:Ckmax}),
  which is motivated by the fact that there are accurate model
  predictions of halo clustering available up to such scales.  The
  last three columns denote the FoG suppression scale,
  $\overline{\sigma}_{v,{\rm off}}/aH(z)$ [Mpc/h], for dark matter and
  for DLRGs having the Gaussian and NFW radial profiles.  }
\label{tab:survey}
\end{table*}
%%%%%%%%%%%%%%%%%%%%%%%%%%%%%%%%%%%%%%%%%%%%%%%%%%%%%%%%

\subsection{Weak lensing information on the off-centered DLRGs}

In this subsection, we estimate the ability of future lensing survey
to determine the radial distribution of DLRGs in the halos.

As shown in Fig.~\ref{fig:lrglens}, the offset of the DLRGs from the
centers of their halos dilutes the lensing signals at small angular
scales.  Thus if we assume that the dark matter halos are
well-described by NFW halos on the $\sim$100~kpc scale, the lensing
observations can be used to infer the DLRG radial distribution.

There are two approaches that can be considered. The first one is a
method fully based on the halo model. That is, by fitting the measured
lensing profile to the halo model prediction (see
Eq.~[\ref{eq:lens_1h}]), one can constrain parameters including the
parameters of either Gaussian or NFW DLRG radial profile model:
%%%%%%%%%%%%%%%%%%%%%%%%%%%%%%%%%%%%%%%%%%%%%%%%%%%%%%%%%
\begin{eqnarray}
&& p_\alpha \equiv \left\{ 
r_{\rm off} ({\rm Gauss}) \mbox{ or } c_{\rm off} ({\rm NFW}), c_N,
\beta, \sigma_{\log M}, m_{\rm sh, DLRG}
\right\},
\label{eq:para_lens}
\end{eqnarray}
%%%%%%%%%%%%%%%%%%%%%%%%%%%%%%%%%%%%%%%%%%%%%%%%%%%%%%%%%
where $r_{\rm off}$ or $c_{\rm off}$ is the off-centering parameter
(see Eq.~[\ref{eq:model_off}]), $\sigma_{\log M}$ is the parameter of
DLRG HOD (see Eq.~[\ref{eq:HOD}]), and the mean mass scale of subhalo
hosting each DLRG (see Eq.~[\ref{eq:lens_1h}]). The parameters $c_N$
and $\beta$ are introduced to model the scaling relation of halo
concentration with halo mass, $c_{\rm vir}(M_{\rm vir})=c_N(M_{\rm
vir}/10^{12} h^{-1}M_\odot)^{-\beta} $, where $c_N=7.85$ and
$\beta=0.081$ are adopted for the fiducial values. Note that the HOD
models are given by the two parameters $\sigma_{\log M}$ and $M_{\rm
min}$, but one of the two is determined by the observed mean number
density of DLRGs. In this parameter estimation we assume that the halo
mass function, halo bias and NFW profile or more generally halo mass
profile are well calibrated based on a suite of simulations. We also
assume that, although the lensing strength depends on cosmology, the
background cosmology is well constrained from other cosmological
observables such as CMB and the galaxy redshift survey itself (e.g.,
via the BAO experiment). Hence we include only the 1-halo term power
spectrum of DLRG-galaxy lensing to perform the parameter forecast
based on the Fisher matrix formalism.

We can use external information as priors on the model parameters. For
the concentration-mass scaling relation, simulation-based studies
\citep[][]{Duffyetal:08} and/or cluster lensing studies
\cite[][]{Okabeetal:10} can be used as the priors. For the HOD
parameter, $\sigma_{\log M}$, the clustering analysis of DLRGs
\citep[][]{Whiteetal:11} can be used.  We employ the priors
$\sigma(\ln c_{N})=\sigma(\ln \beta) =\sigma(\ln \sigma_{\log M})=0.2$
in our Fisher analysis.

%%%%%%%%%%%%%%%%%%%%%%%%%%%%%%%%%%%%%%%%%%%%%%%%%%%%%%%%
\begin{table}
\begin{center}
Off-centering Parameter Determination from DLRG-Galaxy Lensing\\
\begin{tabular}{ll|cc|c|cc}
  \hline\hline
& &
 \multicolumn{2}{c}{Gaussian offset}
&& 
 \multicolumn{2}{c}{NFW offset}
\\
\cline{3-4}
\cline{6-7}
Survey & Redshift range 
& \raisebox{-1ex}{$\sigma(\ln r_{\rm off})$}
& \raisebox{-1ex}{$\sigma(\ln \overline{\sigma}_{v, {\rm off}})$}
&
& \raisebox{-1ex}{$\sigma(\ln c_{\rm off})$ }
& \raisebox{-1ex}{$\sigma(\ln \overline{\sigma}_{v, {\rm off}})$}
\\ \hline 
BOSS             & 0.3 -- 0.4 & 0.07  & 0.019 & & 0.31 & 0.016 \\
(10,000~deg$^2$) & 0.4 -- 0.5 & 0.08  & 0.022 & & 0.33 & 0.018 \\
                 & 0.5 -- 0.6 & 0.10  & 0.027 & & 0.34 & 0.019 \\ \hline
PFS              & 0.6 -- 0.8 & 0.12  & 0.032 & & 0.34 & 0.020 \\
(2,000~deg$^2$)  & 0.8 -- 1.0 & 0.24  & 0.066 & & 0.36 & 0.022 \\
                 & 1.0 -- 1.2 & 0.49  & 0.15  & & 0.39 & 0.024 \\
                 & 1.2 -- 1.4 & 0.85  & 0.33  & & 0.46 & 0.030 \\ \hline
Euclid           & 0.5 -- 0.6 & 0.058 & 0.016 & & 0.22 & 0.012 \\
(20,000~deg$^2$) & 0.6 -- 0.7 & 0.087 & 0.024 & & 0.24 & 0.014 \\
                 & 0.7 -- 0.8 & 0.13  & 0.036 & & 0.27 & 0.015 \\
                 & 0.8 -- 0.9 & 0.20  & 0.055 & & 0.29 & 0.017 \\
                 & 0.9 -- 1.0 & 0.31  & 0.086 & & 0.31 & 0.019 \\
                 & 1.0 -- 1.1 & 0.47  & 0.14  & & 0.33 & 0.020 \\
                 & 1.1 -- 1.2 & 0.71  & 0.24  & & 0.35 & 0.022 \\
                 & 1.2 -- 1.3 & 1.03  & 0.51  & & 0.37 & 0.023 \\
                 & 1.3 -- 1.4 & 1.44  & 0.53  & & 0.39 & 0.025 \\
                 & 1.4 -- 1.5 & 1.87  & 0.54  & & 0.41 & 0.027 \\ \hline
\end{tabular}
\end{center}
\caption{Marginalized uncertainties in the DLRG off-centering
 parameters in each redshift slice, which are expected to obtain from
 the DLRG-galaxy lensing measurements (see around
 Eq.~[\ref{eq:para_lens}] for details of our Fisher analysis).  For
 the BOSS and PFS surveys, we assume that the Subaru HSC-type imaging
 survey is overlapped with the spectroscopic surveys for an area of
 2000 square degrees.  For the Euclid survey we assume the
 overlapping area of $20000$ square degrees for the joint imaging and
 spectroscopic surveys.  As in Fig.~\ref{fig:profile} we consider the
 Gaussian or NFW DLRG radial profiles as a working example, and the
 the error bars show the relative errors: e.g., $\sigma(\ln r_{\rm
 off})\equiv \sigma(r_{\rm off})/r_{\rm off}$, where the denominator
 is the fiducial value.  The relative error on the FoG suppression
 scale, denoted by $\sigma(\ln \sigma_{v,{\rm off}})$, is estimated by
 propagating the error of $r_{\rm off}$ or $c_{\rm off}$ using
 Eq.~(\ref{eq:sigv}).
\label{tab:lens_error}
}
\end{table}
%%%%%%%%%%%%%%%%%%%%%%%%%%%%%%%%%%%%%%%%%%%%%%%%%%%%%%%%

Table~\ref{tab:lens_error} gives the marginalized errors on the
parameter, $r_{\rm off}$ or $c_{\rm off}$, for the Gaussian or NFW
radial profile model, respectively, in each redshift slice. Here we
assume the DLRG-galaxy lensing measurements for an area of 2,000 or
20,000 square degrees for the Subaru HSC or Euclid survey,
respectively, as discussed in \S~\ref{sec:survey}.  Note that we
include the lensing information up to $l_{\rm max}=5\times 10^4$ in
our Fisher analysis, but we find that the results in
Table~\ref{tab:lens_error} are not so sensitive to the choice of the
maximum multipole as such small-scale signals are dominated by the
lensing contribution of DLRG's subhalo. Most of the information on the
characteristic offsets comes from the power spectrum at $l\simlt
10^4$. Table~\ref{tab:lens_error} also shows the expected error on the
FoG suppression scale, which is obtained by propagating the error of
the characteristic offset based on Eq.~(\ref{eq:sigv}).

The method above is a model-dependent method. Since the Gaussian or
NFW radial profile models may differ from the genuine radial profile,
the estimated offset parameters may have systematic biases. Moreover
the radial profile may change with halo masses.  Hence an alternative
approach is based on a method that the measured lensing profile is
compared with an single NFW profile, including the off-centering
effect.  Previous measurements
\citep[][]{Johnstonetal:07,Okabeetal:10} mostly used this method.
However, we find that this method does not work well if we use the
full sample of DLRGs for the DLRG-galaxy lensing measurements. If the
lensing profile in Fig.~\ref{fig:lrglens} is fitted to a single NFW
profile, the extracted halo mass is found to be $\sim 0.5\times
10^{14}h^{-1}M_\odot$, i.e. the mean halo mass of halos hosting DLRGs
as listed in Table~\ref{tab:survey}. This mass scale is smaller than
the mass scale of halos, $10^{14}h^{-1}M_\odot$, that give a dominant
contribution to the FoG effect as shown in
Fig.~\ref{fig:dsigmav_dlnM}.  In turn the characteristic offset
inferred from such a single NFW profile-fitting is not that accurate.
Since the FoG effects are larger for the more massive halos, it would
be more optimal to focus on studies of the DLRG-galaxy lensing for
halos with a narrower mass range around $10^{14}h^{-1}M_\odot$, where
such a halo catalog can be constructed based on optical richness
available form multi-color imaging survey itself, X-ray and/or
Sunyaev-Zel'dovich (SZ) data.  

%%% MT (10/7/11)
%Nevertheless we would like to comment on our current on-going work.  
In fact we are now working on the SDSS DR7 LRG catalog in order to
explore the feasibility of the method above with real data. In this
on-going study, we have focused on the regions including multiple LRGs
in the small spatial region, which are likely to reside in the same halo
with masses more massive than other majority of halos hosting a single
LRG inside. We have then studied the galaxy-galaxy lensing signal
measured via cross-correlation of the LRG regions with background galaxy
shapes, where the background galaxies are taken from the photometric
SDSS galaxy catalog based on their photometric redshift estimates.  We
have so far found, preliminarily though, that the lensing signals at
small scales do change with different centers; the brightest LRG
position, the faintest LRG or the mean of their positions. Then we are
now trying to constrain the off-centered profile of LRGs from the varied
lensing signals including marginalizations over other parameters such as
the mean halo mass, the halo profile parameters and the sub-halo mass
scale. This is in working progress, and will be presented elsewhere
(Hikage et al. in prep.).

%This is beyond the scope of this paper,
%and 
In this paper, we will just assume that the first approach stated above,
i.e. the halo model based fitting method, is feasible for the following
analysis.

\subsection{The FoG effect on DLRG power spectrum}

\begin{figure}
\begin{center}
\includegraphics[width=8.7cm]{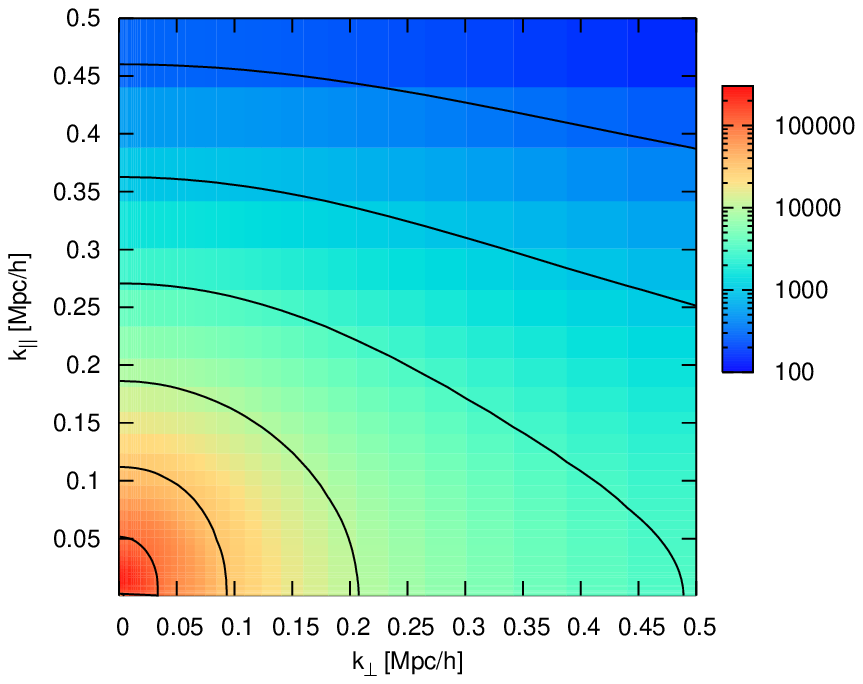}
\includegraphics[width=7.3cm]{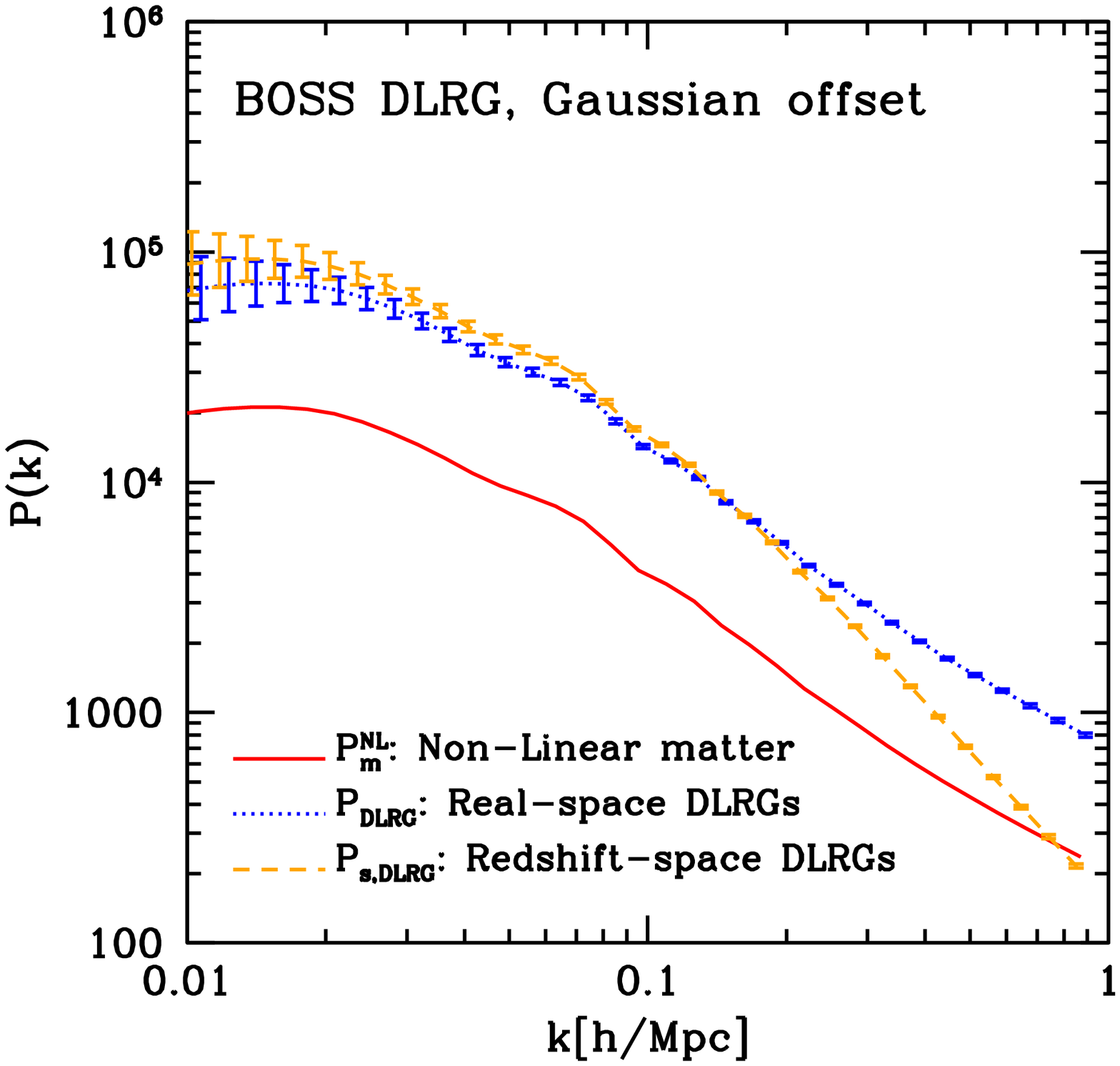}
\caption{{\it Left panel:} Redshift-space power spectrum of DLRGs,
$P_{s,{\rm DLRG}}(k_\parallel,k_\perp)$, in the two-dimensional
wavenumber space ($k_\parallel,k_\perp$). We use
Eq.~(\ref{eq:ps_lrg}) to compute the power spectrum assuming the
Gaussian DLRG radial profile in Fig.~\ref{fig:profile}, and assume
the parameters expected for BOSS survey in Table~\ref{tab:survey}. The
spectrum shown is the spectra averaged over the redshift range $0.3\le
z\le 0.6$.  The contours denote the power spectrum amplitudes of
$10^5, 3.2\times 10^4, 10^4, 3.2\times 10^3, 10^3, 3.2\times 10^2,
10^2~(h/{\rm Mpc})^3$, respectively.  {\it Right panel:} The monopole
power spectrum of DLRGs obtained by averaging the redshift-space power
spectrum over the angle $\mu(\equiv k_\parallel/k)$ for a fixed
$k(\equiv\sqrt{k_\perp^2+k_\parallel^2})$.  For comparison, the
non-linear matter power spectrum $P_m^{\rm NL}(k)$ and the real-space
power spectrum of DLRGs (Eq.~[\ref{eq:pk_real}]) are plotted.  The
error bars show the 1$\sigma$ statistical errors in measuring the
power spectrum at each $k $ bin for BOSS survey.}
    \label{fig:pk}
\end{center}
\end{figure}

In this subsection, we explore how the off-centered DLRGs suppress the
power spectrum through their FoG effects. The left panel of
Fig. \ref{fig:pk} shows the DLRG power spectrum $P_{s,{\rm
DLRG}}(k_\perp, k_\parallel)$ (Eq.~[\ref{eq:ps_lrg}]) in the
two-dimensional space of $(k_{\perp},k_\parallel)$ assuming the
Gaussian DLRG radial profile model.  The plotted power spectrum is the
spectra averaged over a range of redshifts, $0.3\le z \le 0.6$,
covered by the BOSS survey.  Apparent anisotropic features can be
found: at large length scales ($k\simlt 0.15h$/Mpc), the Kaiser effect
enhances the power along the line-of-sight direction, while the FoG
effect suppresses the power more significantly with increasing
wavenumbers and thus squashes the contours along $k_\parallel$.  The
right panel shows the monopole power spectrum which is obtained by
averaging the redshift-space power spectrum over the angle $\mu$
($\equiv k_\parallel/k$) for a fixed $k(\equiv
\sqrt{k_\parallel^2+k_\perp^2})$.  For comparison we also show the
real-space DLRG power spectrum $P_{\rm DLRG}(k)$
%DS L --> NL, add Smith et al. 2003
(Eq.~[\ref{eq:pk_real}]) and the input non-linear matter power
spectrum $P_m^{\rm NL}(k)$, where the latter is computed using
the fitting formula by \cite{Smithetal:03}. It is clear that the
redshift-space distortion effect causes a scale-dependent modification
in the amplitude as well as shape of monopole power spectrum.

%%%%%%%%%%%%%%%%%%%%%%%%%%%%%%%%%%%%%%%%%%%%%%%%%%%%%%%%
%%%%%%%%%%%%%%%%%%%%%%%%%%%%%%%%%%%%%%%%%%%%%%%%%%%%%%%

\begin{figure*}
\begin{center}
\includegraphics[width=8cm]{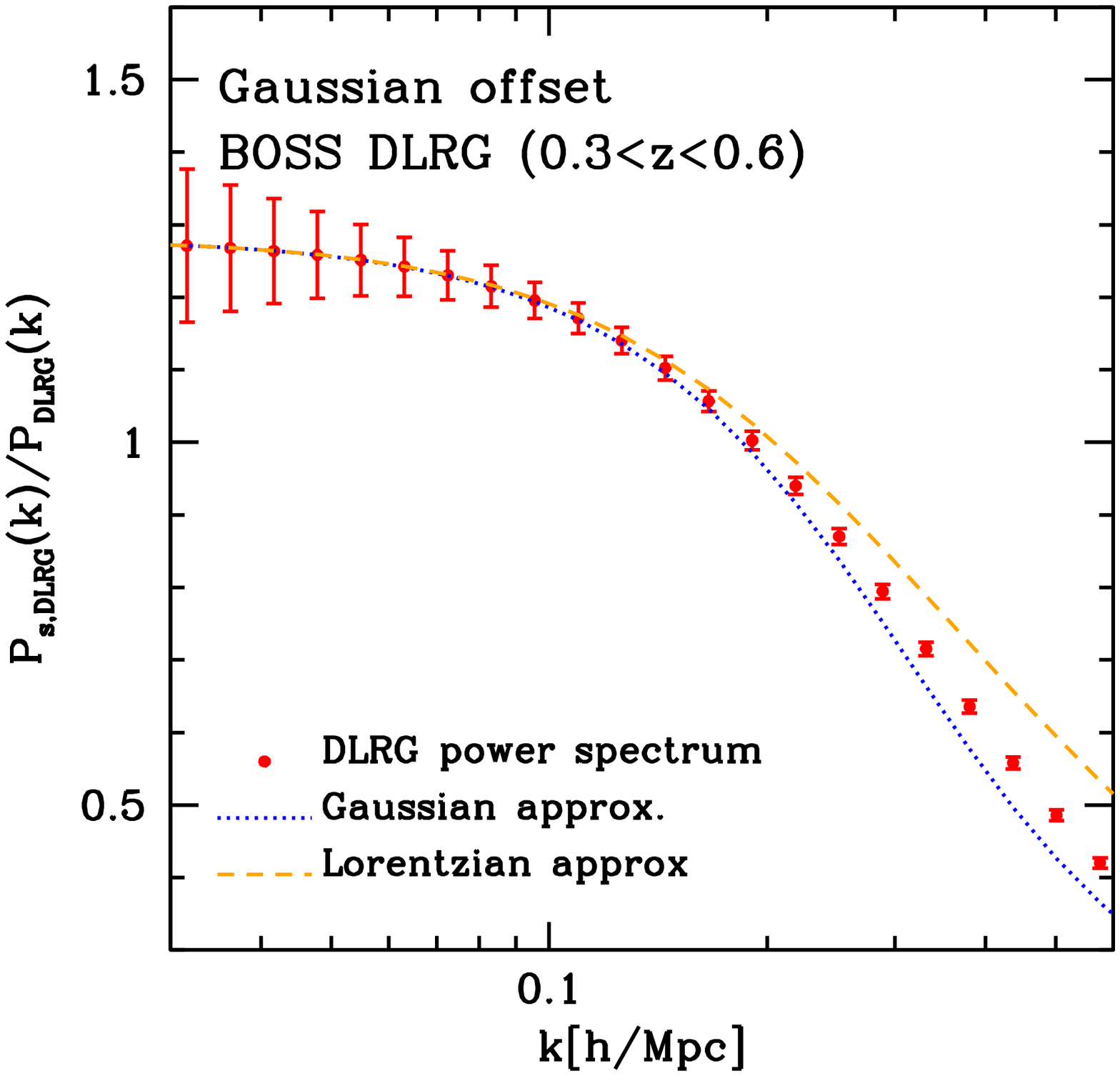}
\includegraphics[width=8cm]{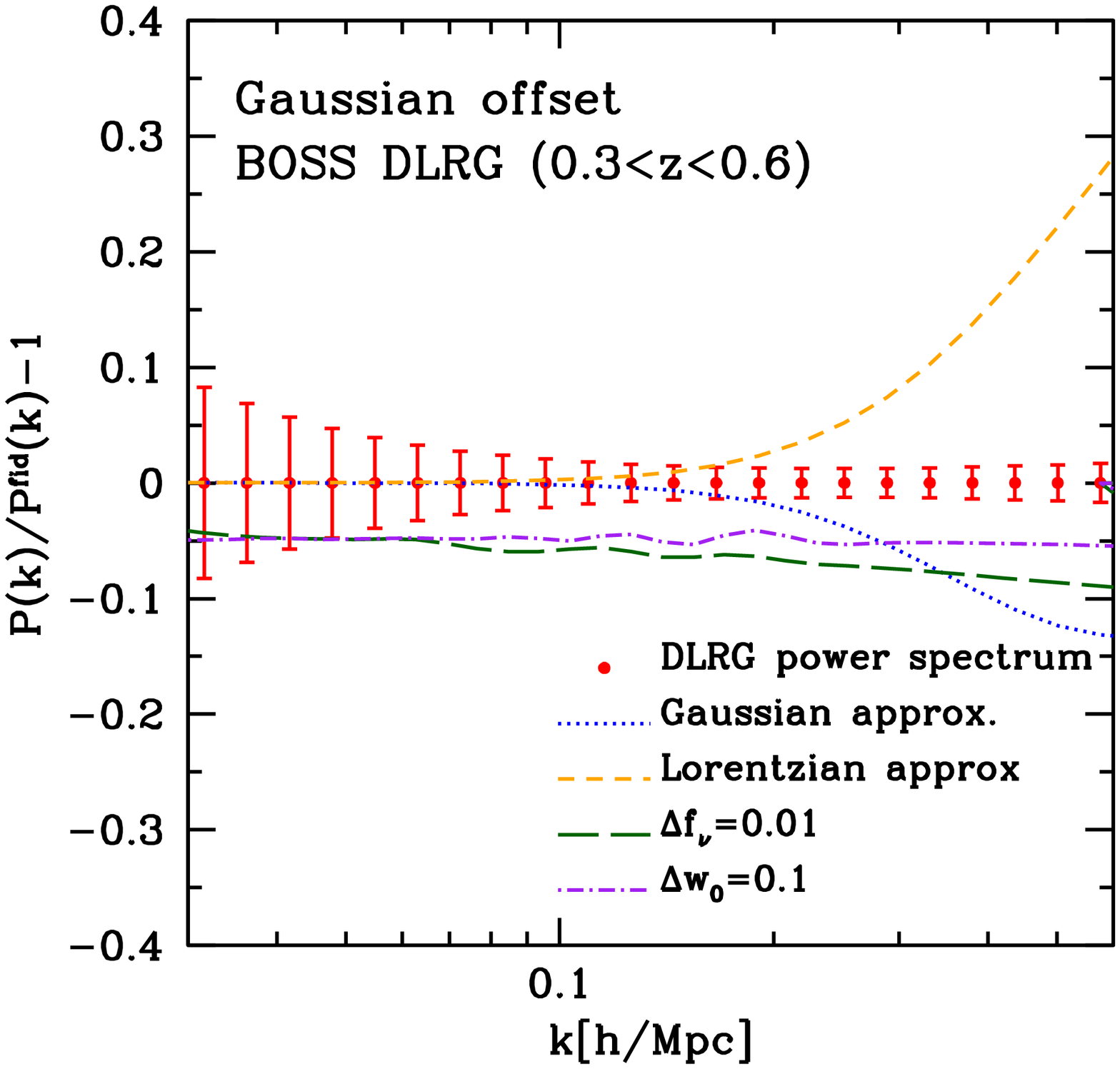}
\caption{{\it Left panel:} The relative difference between the
  monopole redshift-space power spectra of DLRGs with and without
  redshift distortion effect (the linear Kaiser effect plus the FoG
  effect). The data at each $k$ bins shows the halo model predictions
  we have developed in this paper, and the error bar at each $k$-bin
  shows the $1\sigma$ statistical uncertainties of the band power
  expected from the BOSS survey (see Table~\ref{tab:survey}).  For
  comparison, the dotted and dashed curves show the model predictions
  assuming a Gaussian or Lorentzian FoG form (see
  Eq.~[\ref{eq:FoG_form}]), which is specified with a single parameter
  $\overline{\sigma}_{v,{\rm off}}$ (Eq.~[\ref{eq:sigv}]). The full
  halo model prediction and the approximations show differences
  greater than the statistical error, more significantly at higher
  $k$'s.  {\it Right panel:} The difference ratio of the power spectra
  between the full FoG effect and the approximated FoG effects
  (Gaussian or Lorentzian FoG models): $P^{\rm app}_{s, {\rm
      DLRG}}/P^{\rm full}_{s,{\rm DLRG}}-1$.  Also shown is how the
  redshift-space power spectrum changes with changing the dark energy
  equation of state parameter $w_0$ and the neutrino mass
  $f_\nu(\equiv \Omega_{\nu 0}/\Omega_{\rm m0})$: here we consider
  variations to $w_0=-0.9$ and $f_\nu=0.01 $ (corresponding to the
  total neutrino mass $m_{\nu, {\rm tot}}=0.104~{\rm eV}$) from the
  fiducial values $w_0=-1$ and $f_\nu=0$, respectively.  }
\label{fig:fog}
\end{center}
\end{figure*}

To be more quantitative, Fig.~\ref{fig:fog} shows the ratio between
the monopole spectra of redshift- and real-space DLRG power spectra as
a function of $k$.  Since the linear Kaiser redshift distortion causes
an overall offset in the monopole power spectrum amplitudes, the
scale-dependent effect in the figure is solely due to the FoG
effect. The data points show the DLRG power spectrum assuming the
Gaussian radial profile model. The error bars are the expected
1$\sigma$ statistical uncertainties in measuring band powers of the
power spectrum at each $k$ bin for the BOSS survey, which include the
sampling variance and shot noise contamination. The plot implies that
the FoG effect can be very significant even at very large length
scales, $k\simeq 0.1~h{\rm Mpc}^{-1}$, relevant for BAO scales, and
the FoG suppression is more significant at larger $k$'s.  The FoG
effect suppresses the power spectrum amplitudes by 10 (30\%) around
$k= 0.1 (0.2)~h{\rm Mpc}^{-1}$.

The following empirical models of the FoG effect are often assumed in
the literature:
%%%%%%%%%%%%%%%%%%%%%%%%%%%%%%%%%%%%%%%%%%%%%%%%%%%%%%%%
\begin{eqnarray}
&{\rm Gaussian: }~~~~~&\exp[-(k\mu
\overline{\sigma}_{ v, {\rm
 off}}/aH(z))^2
%\sigma_{\rm FoG})^2
], 
\nonumber \\
&{\rm Lorentzian: }~~&\frac{1}
{1+(k\mu
%\sigma_{\rm FoG})^2
\overline{\sigma}_{ v, {\rm off}}/aH(z))^2
},
\label{eq:FoG_form}
\end{eqnarray}
%%%%%%%%%%%%%%%%%%%%%%%%%%%%%%%%%%%%%%%%%%%%%%%%%%%%%%%%
which are called the Gaussian and Lorentzian FoG models. These two
models are given by a single parameter $\overline{\sigma}_{v,{\rm
off}}$ and have the same form given as $ 1-(k
\mu\overline{\sigma}_{v,{\rm off}}/aH)^2$ at the small $k$ limit. Note
that we use the same notation $\overline{\sigma}_{v,{\rm off}}$ as
that of the radial profile model for notational simplicity, but keep
in mind that $\overline{\sigma}_{v,{\rm off}}$ in the equation above
is a free model parameter.

The dashed and dotted curves in Fig.~\ref{fig:pk} show the results for
these FoG models, which can be compared with the FoG effect due to
off-centered DLRGs we have developed in this paper. Here the model
parameter $\overline{\sigma}_{v,{\rm off}}$ is taken from
Eq.~(\ref{eq:sigv}) assuming the input Gaussian radial profile model.
At scales $k\simlt 0.15h$Mpc$^{-1}$, all the results well agree with
each other because the FoG effect can be well approximated as
$1-(k\mu\overline{\sigma}_{v, {\rm off}}/aH)^2$ at such small $k$'s.
At the larger $k$'s the model differences become significant: the
Gaussian approximation underestimates the power, while the Lorentzian
overestimates the power. The differences arise because the FoG effect
in our model arises after the integration of halo mass function and
the DLRG radial profile (see Eqs.~[\ref{eq:psoff}] and
[\ref{eq:ps_lrg_2}]), and therefore the resulting FoG effect does not
exactly follow a Gaussian form, even if we assume that the velocity
probability distribution of DLRGs is Gaussian for each halos.

The right panel of Fig.~\ref{fig:fog} more explicitly shows the
relative differences between the FoG effect and the FoG approximations
(Eq.~[\ref{eq:FoG_form}]). The plot shows that either Gaussian or
Lorentzian FoG model cannot be accurate enough compared to the
accuracies of upcoming galaxy surveys. The plot also shows how the
monopole power spectrum changes with changing the dark energy equation
of state parameter to $w_0=-0.9$ from $w_0=-1$ or the neutrino mass to
$f_\nu=0.01 (m_{\nu, {\rm tot}}=0.104~{\rm eV})$ from $f_\nu=0$. These
parameters are both sensitive to the power spectrum amplitudes and
therefore most affected by the FoG uncertainty as we will study below
more extensively.  The plot shows that the upcoming surveys have a
much higher statistical precision than the effects due to these
parameter changes, and also shows that a correction/calibration of the
FoG effect is very important in order not to have a biased estimate on
these parameters. Finally, for the FoG effect due to the NFW radial
profile model, we have also found similar results to the results in
Fig.~\ref{fig:fog}.
%%%%%%%%%%%%%%%%%%%%%%%%%%%%%%%%%%%%%%%%%%%%%%%%%%%%%%%%
%%%%%%%%%%%%%%%%%%%%%%%%%%%%%%%%%%%%%%%%%%%%%%%%%%%%%%%

The amount of the FoG suppression effect shown in Fig.~\ref{fig:fog}
obviously depends on the fiducial parameters of our FoG model. One of
the model uncertainties is the velocity distribution of DLRGs within a
halo. If we use the model given in Appendix~\ref{app:smallv}, the FoG
suppression effect is roughly half of the amplitude shown in
Fig.~\ref{fig:fog}.  A full analysis of both simulations and
observations will be essential to quantify the amplitude of this
suppression. In the following analysis, for simplicity we will
continue to assume the FoG effect computed using our fiducial Gaussian
and NFW radial profile models.

\subsection{The Impact of FoG Effect on Cosmological Parameter Estimations}
\label{sec:fisher}

How does an imperfect modeling of the FoG effect affect cosmological
parameter estimation? In particular how can adding the DLRG-galaxy
lensing information helps correct for the FoG effect on the
redshift-space DLRG power spectrum? In this subsection, we address
these questions.

We perform a Fisher analysis to estimate expected accuracies of
parameters for upcoming galaxy surveys, and possible biases due to an
imperfect modeling of the FoG effect.  As the observable we use the
redshift-space DLRG power spectrum. In the galaxy clustering analysis
one needs to assume a reference cosmological model to infer the
spatial position of each galaxy from the observed redshift and angular
position. However, the assumed cosmology generally differs from the
underlying true cosmology, which causes cosmological distortion effect
on the observed power spectrum.  Hence the observed redshift-space
power spectrum can be given as
\citep{AlcockPaczynski:79,SeoEisenstein:03}:
%%%%%%%%%%%%%%%%%%%%%%%%%%%%%%%%%%%%%%%%%%%%%%%%%%%%%%%
\begin{equation}
P_{s,{\rm DLRG}}^{\rm obs}(k_\parallel^{\rm fid},k_\perp^{\rm fid})=
\frac{D_A(z)^2_{\rm fid}}{D_A(z)^2}\frac{H(z)}{H(z)_{\rm fid}}
P_{s,{\rm DLRG}}(k_\parallel,k_\perp), 
\end{equation}
%%%%%%%%%%%%%%%%%%%%%%%%%%%%%%%%%%%%%%%%%%%%%%%%%%%%%%%
where $k_{\parallel}^{\rm fid}\equiv (D_{A}/D_{A,{\rm
fid}})k_{\parallel}$, $k_{\perp}^{\rm fid}\equiv (H_{\rm
fid}/H)k_\perp$, and $D_A(z)$ and $H(z)$ are the angular diameter
distance and the Hubble expansion rate at redshift $z$. The quantities
with subscript ``fid'' denote the quantities of the assumed reference
cosmology. We use Eq.~(\ref{eq:ps_lrg_2}) to compute the
redshift-space power spectrum.

Using the power spectrum covariance (Eq.~[\ref{eq:cov_pk}]), the
Fisher information matrix of the DLRG power spectrum measurement is
computed as
%%%%%%%%%%%%%%%%%%%%%%%%%%%%%%%%%%%%%%%%%%%%%%%%%%%%%%%
\begin{equation}
F_{\alpha\beta}^{\rm DLRG}\simeq\sum_{z_i}\frac{V_{z_i}}{8\pi^2}
\int_{-1}^1 d\mu \int^{k_{\rm max}}_{k_{\rm min}}k^2dk
\frac{\partial\ln P_{s,{\rm DLRG}}^{\rm obs}(k,\mu;z_i)}{\partial p_\alpha}
\frac{\partial\ln P_{s,{\rm DLRG}}^{\rm obs}(k,\mu;z_i)}{\partial p_\beta}
\left[1+\frac{1}{\bar{n}_{\rm DLRG}(z_i)P_{s,{\rm DLRG}}^{\rm obs}(k,\mu;z_i)}\right]^{-2}.
\label{eq:fisher}
\end{equation}
%%%%%%%%%%%%%%%%%%%%%%%%%%%%%%%%%%%%%%%%%%%%%%%%%%%%%%%
where the summation is over redshift slices, $V_{z_i}$ is the comoving
volume of the $i$-th redshift slice, and $p_\alpha$ denotes a set of
model parameters.  The redshift ranges of each slice are given in
Table~\ref{tab:survey}.  For each redshift slice, the minimum
wavelength ${k_{\rm min}}$ is set to be $2\pi/V_{z_i}^{1/3}$ and the
maximum wavelength $k_{\rm max}$ are chosen using the criterion given
by Eq.~(\ref{eq:Ckmax}) as listed in Table~\ref{tab:survey}. As we
discuss around Eq.~(\ref{eq:Ckmax}), the impact of the FoG effect
becomes more significant with including the power spectrum information
up to the higher $k_{\rm max}$.  We use a publicly available code CAMB
\citep{Lewisetal:00} to compute the input linear-mass power spectrum.

The galaxy power spectrum alone cannot determine all the cosmological
parameters due to severe parameter degeneracies
\citep[][]{Takadaetal:06}. Therefore, we combine CMB information with
the galaxy power spectrum, which helps efficiently in breaking the
parameter degeneracies. We use the CMB information expected from the
Planck experiments: the temperature spectrum, $E$-mode polarization
spectrum and the cross-spectrum over multipole range of $2\le l\le
1500$. We compute the CMB Fisher matrix $F^{\rm Planck}_{\alpha\beta}$
assuming the instrument noise and beam size of each frequency channel
in the Planck
website\footnote{http://www.sciops.esa.int/index.php?project=PLANCK}.
The Fisher matrix for a joint experiment of the CMB and galaxy power
spectra is simply estimated as $F_{\alpha\beta}=F_{\alpha\beta}^{\rm
Planck}+F_{\alpha\beta}^{\rm DLRG}$. The marginalized error of the
$\alpha$-th parameter, $\sigma(p_\alpha)$, is estimated as
$(\bmf{F}^{-1})_{\alpha\alpha}^{1/2}$, where $\bmf{F}^{-1}$ is the
inverse matrix of the Fisher matrix $\bmf{F}$.

The parameter estimation is sensitive to a set of model parameters as
well as a choice of the fiducial model. We include a fairly broad
range of model parameters:
%%%%%%%%%%%%%%%%%%%%%%%%%%%%%%%%%%%%%%%%%%%%%%%%%%%%%%%%
\begin{eqnarray}
p_\alpha\equiv \{ \Omega_{\rm b}h^2, \Omega_{\rm cdm}h^2, \Omega_K,
\Omega_{\rm DE}, \tau, A, n_s, 1/\bar{n}(z_i), \bar{b}(z_i), ~~r_{\rm
off}(z_i) ~{\rm or}~ c_{\rm off}(z_i) ~{\rm or}~ \overline{\sigma}_{v,
{\rm off}}(z_i),~~ f_\nu ~{\rm and/or}~ w_0 ~{\rm or}~ f_g(z=z_i) \}.
\label{eq:paras}
\end{eqnarray}
%%%%%%%%%%%%%%%%%%%%%%%%%%%%%%%%%%%%%%%%%%%%%%%%%%%%%%%%
The first 7 parameters are cosmological parameters, and the fiducial
values are given at the end of Section~\ref{sec:intro}. Following
\cite{SeoEisenstein:03} \citep[also see][]{Saitoetal:11}, we include
$1/\bar{n}(z_i)$ and $\bar{b}(z_i)$ as free parameters in order to
model uncertainties due to the residual shot noise and galaxy bias in
each redshift slice. The FoG effect fully computed based on the halo
model is parametrized by the off-centering parameter in each redshift
(Eq.~[\ref{eq:model_off}]): $r_{\rm off}(z_i)$ for the Gaussian radial
profile model, or $c_{\rm off}(z_i)$ for the NFW model.  When we
approximate the FoG effect with the Gaussian or Lorentzian form
(Eq.~[\ref{eq:FoG_form}]), we instead use $\overline{\sigma}_{v,{\rm
    off}}(z_i)$ (Eq.~[\ref{eq:ps_lrg_2}]) as a parameter. The fiducial
value of $\overline{\sigma}_{v,{\rm off}}(z_i)$ is computed based on
the halo model (eq.[\ref{eq:sigv}]).

In Eq.~(\ref{eq:paras}), we further include additional parameter(s):
the dark energy equation-of-state parameter $w_0(=0)$ and/or the
neutrino energy-density fraction
$f_\nu\equiv\Omega_\nu/\Omega_m(=0.01)$\footnote{The fiducial value
$f_\nu=0.01$ corresponds to the total neutrino mass $m_{\nu,{\rm
tot}}=0.104$eV for our fiducial $\Lambda$CDM model. This value is
close to the lower limit if the neutrinos obey the inverted mass
hierarchy.}, or the growth rate of each redshift slice
$f_{g}(z_i)\equiv \left. d\ln D/\ln a\right|_{z_i}$. The values in the
parenthesis denote the fiducial values, and the fiducial values of
$f_g$ in each redshift slice are taken from those of our fiducial
$\Lambda$CDM model. These parameters are all sensitive to the
small-scale amplitudes of galaxy power spectrum, and therefore
degenerate with the FoG suppression effect. We will pay special
attention to an issue of how a knowledge of the FoG effect helps
constrain these parameters and minimize a possible bias in the
parameter caused by the FoG uncertainty. The dimension of our Fisher
matrix is at most $19\times 19$, $23\times 23$, and $47\times 47$ for
BOSS, PFS and Euclid surveys, respectively.

In the parameter forecast we will consider the following four cases
for the treatment of the FoG effect:
%%%%%%%%%%%%%%%%%%%%%%%%%%%%%%%%%%%%%%%%%%%%%%%%%%%%%%%%%
\begin{itemize}
\item[(i)] The redshift-space power spectrum of DLRGs, measured from a
  hypothetical survey (BOSS, PFS or Euclid), is compared to the model
  power spectrum, which is given by the non-linear redshift-space
  matter power spectrum (Eq.~[\ref{eq:Kaiser}]) multiplied with either
  Gaussian or Lorentzian FoG model (see Eq.~[\ref{eq:FoG_form}]). In
  this fitting, the parameter $\overline{\sigma}_{v, {\rm off}}$ of
  the FoG model is treated as a free parameter (we will hereafter call
  ``w/o offset'').
\item[(ii)] The similar approach to the case (i), but we employ the
  external information on $\overline{\sigma}_{v, {\rm off}}$ from the
  DLRG-galaxy lensing measurements assuming combined imaging and
  spectroscopic surveys: the Subaru HSC survey combined with either
  BOSS or PFS survey or the combined Euclid imaging and spectroscopic
  surveys (we will call ``with offset''). To be more precise we use
  the statistical error on $\overline{\sigma}_{v,{\rm off}}$ in
  Table~\ref{tab:lens_error} as the prior of the Fisher analysis.
\item[(iii)] This is fully based on the halo model approach: the FoG
  effect is computed based on the halo model (Eq.~[\ref{eq:ps_lrg_2}])
  assuming either Gaussian or NFW radial profile model.  We use the
  lensing-derived constraints on $r_{\rm off}$ (Gaussian) or $c_{\rm
    off}$ (NFW) listed in Table~\ref{tab:lens_error} as the prior of
  the Fisher analysis (we will call ``FoG shape known'').
\item[(iv)] This is the extreme case that we neglect the FoG effect in
  the parameter estimation by setting $\overline{\sigma}_{v,{\rm
      off}}$ to be $=0$ (we will call ``FoG neglected'')..
\end{itemize}
%%%%%%%%%%%%%%%%%%%%%%%%%%%%%%%%%%%%%%%%%%%%%%%%%%%%%%%%

Based on the Fisher matrix formalism, we estimate the bias due to an
imperfect modeling of the FoG effect, $\delta p_\alpha (\equiv
p_{\alpha,{\rm est}}-p_{\alpha,{\rm true}})$, the difference between
the estimated and input values of the $\alpha$-th parameter
$p_\alpha$. The amount of the bias is estimated as
%%%%%%%%%%%%%%%%%%%%%%%%%%%%%%%%%%%%%%%%%%%%%%%%%%%%%%%
\begin{eqnarray}
\delta p_\alpha&=&\sum_\beta ({\bmf{F}}^{-1})_{\alpha\beta}b_\beta,
\end{eqnarray}
%%%%%%%%%%%%%%%%%%%%%%%%%%%%%%%%%%%%%%%%%%%%%%%%%%%%%%%
where
%%%%%%%%%%%%%%%%%%%%%%%%%%%%%%%%%%%%%%%%%%%%%%%%%%%%%%%
\begin{eqnarray}
b_\beta&\equiv &\sum_i\frac{V_{z_i}}{8\pi^2}
\int_{-1}^1 d\mu \int^{k_{\rm max}}_{k_{\rm min}}k^2 dk
\left[\frac{
P_{s, {\rm DLRG}}(k,\mu;z_i)}
{P^{\rm app}_{s,{\rm  DLRG}}(k,\mu; z_i)}
 - 1
\right]
\frac{\partial\ln P^{\rm app}_{s, {\rm DLRG}}(k,\mu;z_i)}{\partial p_\beta}
\left[1+\frac{1}{\bar{n}_{\rm DLRG}P_{s,
 {\rm DLRG}}(k,\mu;z_i)}\right]^{-2}. 
\end{eqnarray}
%%%%%%%%%%%%%%%%%%%%%%%%%%%%%%%%%%%%%%%%%%%%%%%%%%%%%%%
Here $P_{s, {\rm DLRG}}$ denotes the true power spectrum where the FoG
effect is computed based on the halo model. On the other hand, $P_{s,
  {\rm DLRG}}^{\rm app}$ is the approximated spectrum using Gaussian
or Lorentzian FoG model.
%Since the spectra $P_{s, {\rm DLRG}}$ and $P_{s,
%{\rm DLRG}}^{\rm app}$ differ as demonstrated in Fig.~\ref{fig:fog},
%the use of $P^{\rm app}_{s, {\rm DLRG}}$ in the model fitting may
%cause a bias in the best-fit parameter, even after marginalization
%over other parameters' uncertainties. 
The total error including both statistical and systematic errors is
estimated as
%%%%%%%%%%%%%%%%%%%%%%%%%%%%%%%%%%%%%%%%%%%%%%%%%%%%%%%
\begin{equation}
\label{eq:toterr}
\left[\Delta p_\alpha\right]^2
=[\sigma_{\rm stat}(p_\alpha)]^2 + (\delta p_\alpha)^2.
\end{equation}
%%%%%%%%%%%%%%%%%%%%%%%%%%%%%%%%%%%%%%%%%%%%%%%%%%%%%%%

\begin{figure*}
\begin{center}
\includegraphics[width=8cm]{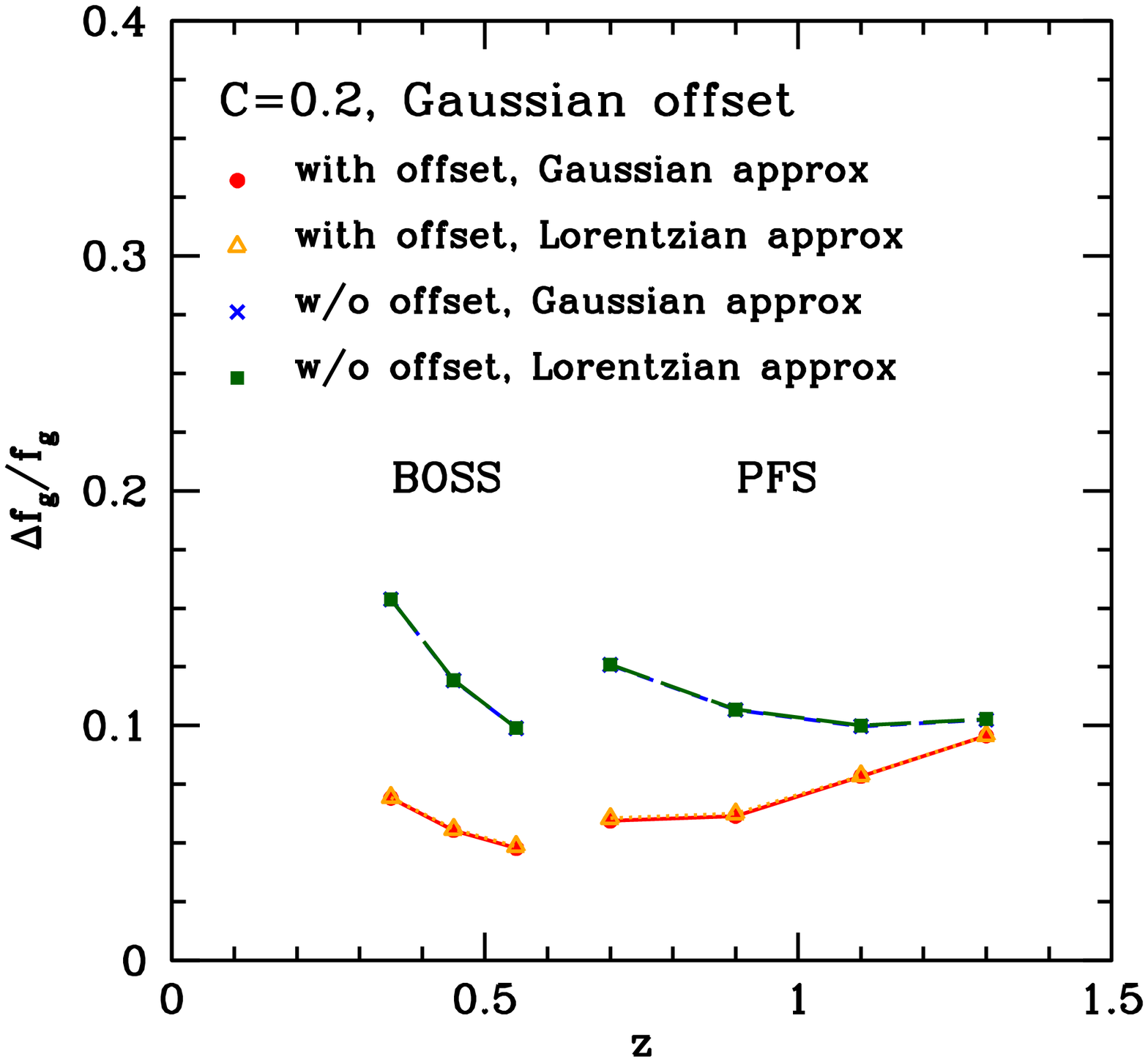}
\includegraphics[width=8cm]{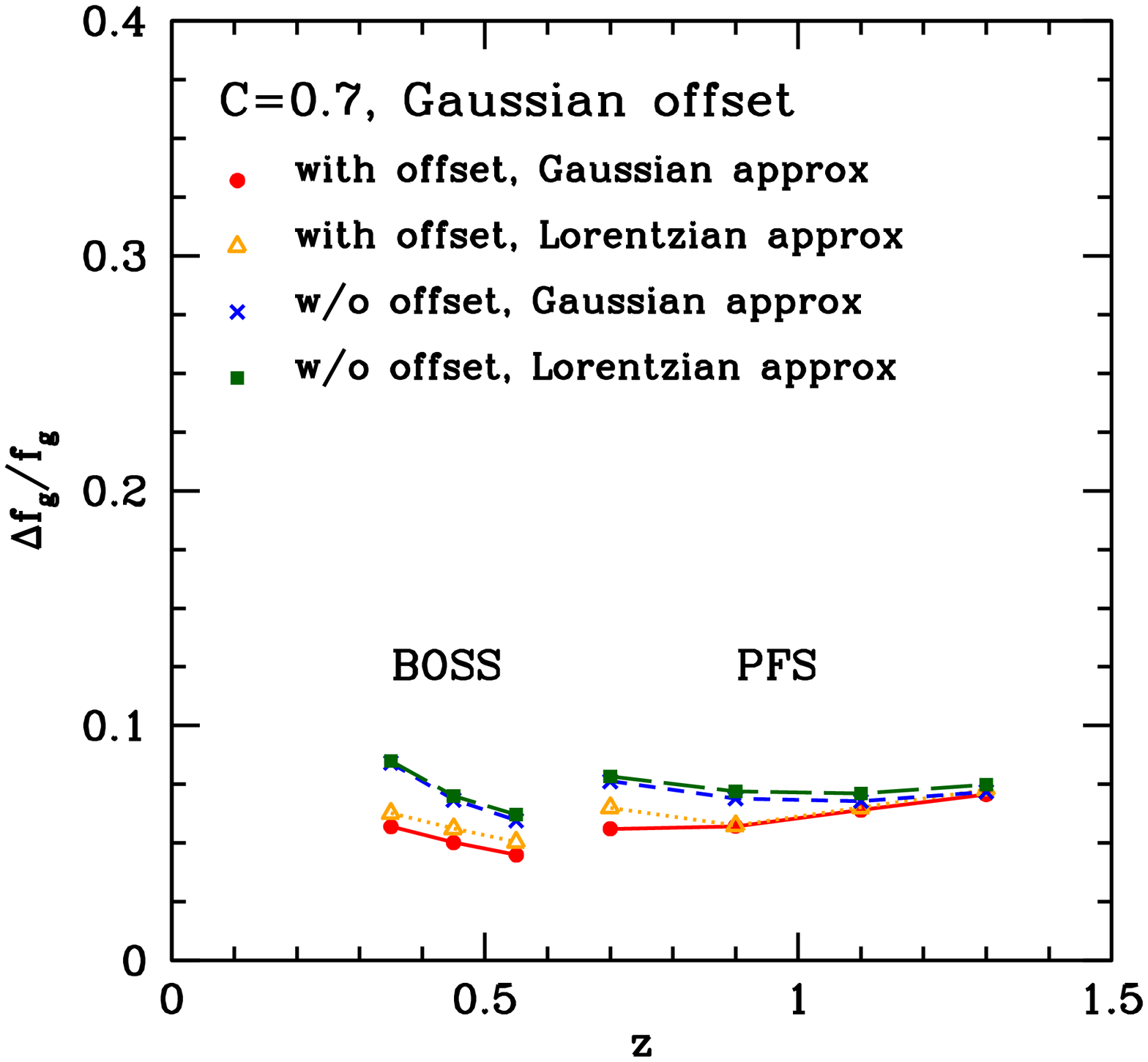}
\includegraphics[width=8cm]{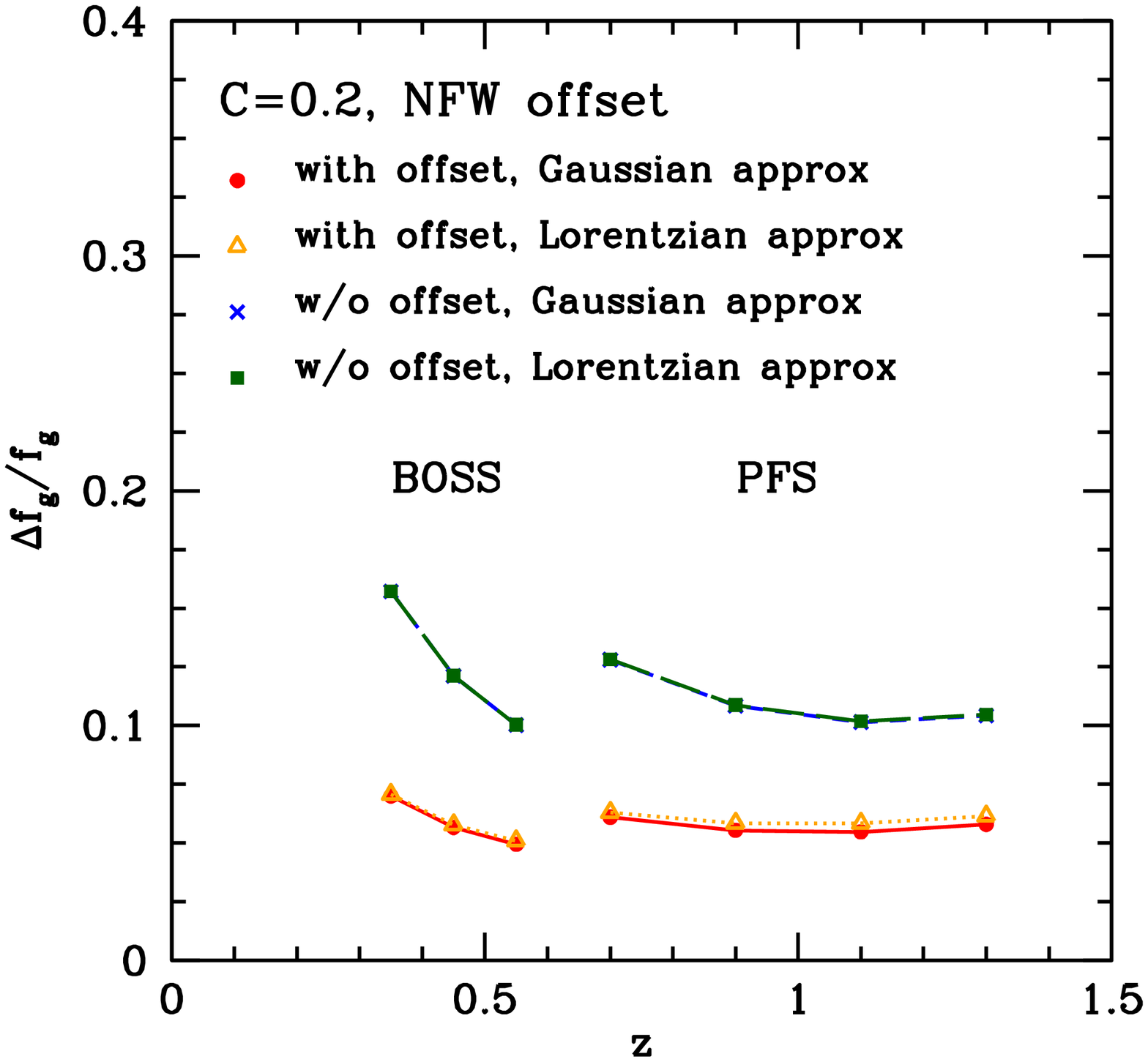}
\includegraphics[width=8cm]{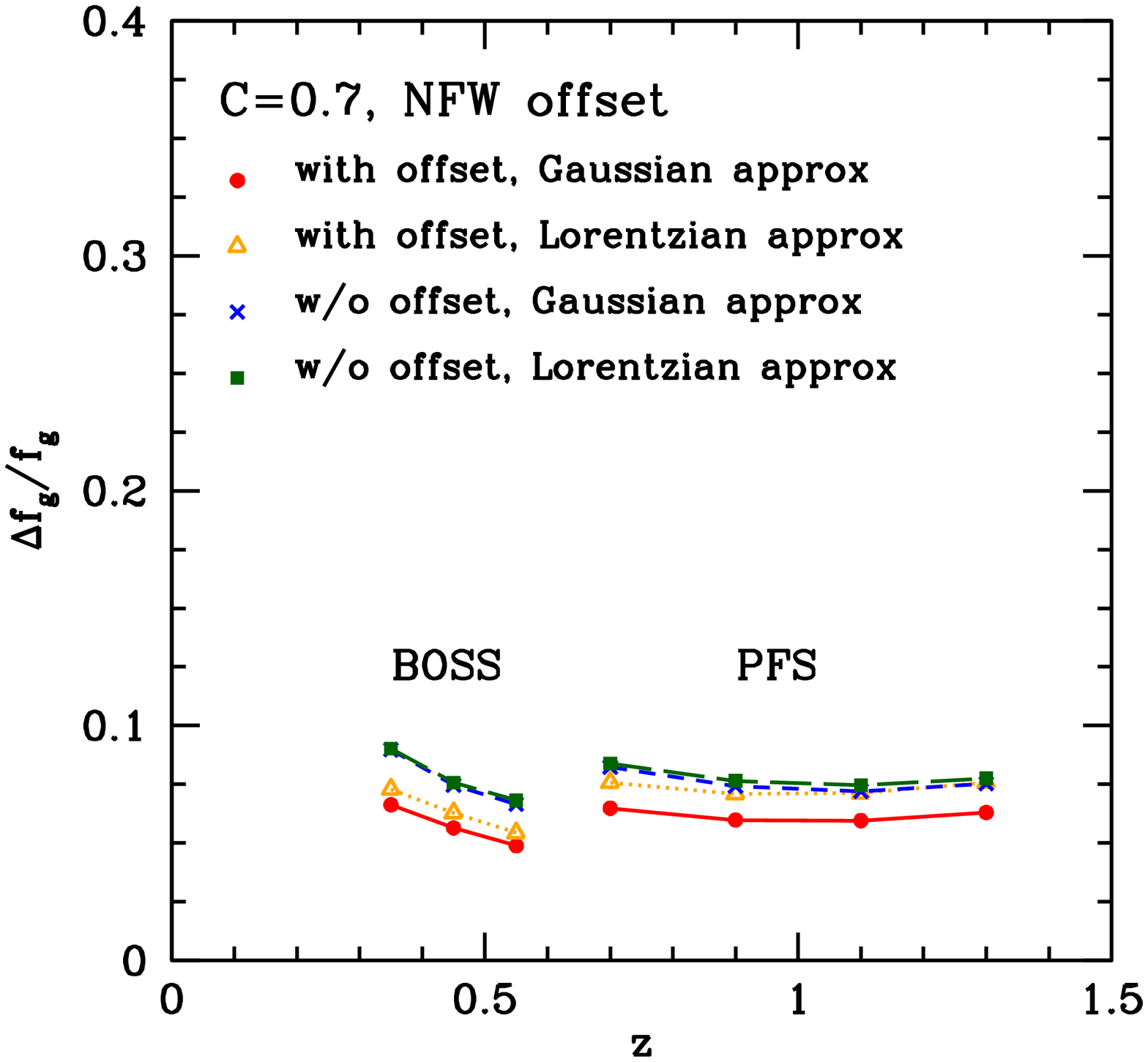} 
\caption{The marginalized error on the growth rate $f_g(\equiv d\ln
  D/d\ln a)$ at each redshift slices of BOSS and PFS surveys, expected
  when either of the two surveys is combined with the Planck CMB
  information. Note that the error shown here includes both the
  statistical and systematic error contributions (see
  Eq.~[\ref{eq:toterr}]). The left- and right-side panels differ in
  the maximum wavenumber $k_{\rm max}$ that is determined by the
  criterion of either $C=0.2$ or $0.7$, respectively (see
  Eqs.~[\ref{eq:fisher}] and [\ref{eq:Ckmax}] and
  Table~\ref{tab:survey}).  For the upper- and lower-side panels, we
%CH
  assume Gaussian and NFW radial profiles of DLRGs respectively, as in
  Fig.~\ref{fig:profile}.
%  assume the FoG models computed assuming our fiducial Gaussian and
%  NFW DLRG radial profiles, respectively, as in Fig.~\ref{fig:profile}.  
%  The different curves in each panel are the
%  results for different treatments of the FoG effect.  
  The long- and short-dashed curves show the results when the FoG
  effect is modeled by either Gaussian or Lorentzian form and the
  model parameter $\sigma_{v,{\rm off}}$ (Eq.~[\ref{eq:FoG_form}]) is
  treated as a free parameter in the model fitting. The solid and
  dotted curves show the results including the DLRG-galaxy weak
  lensing information on $\sigma_{v,{\rm off}}$ in
  Table~\ref{tab:lens_error}.
% to correct the FoG effect assuming either Gaussian or Lorentzian FoG form. 
  An inaccuracy of the approximated FoG form biases the growth rate
  estimation
%, which contributes to the error budget, 
  more significantly when including up to the higher $k$'s, as implied
  in the right-side panels.}
\label{fig:fz}
\end{center}
\end{figure*}

\begin{figure}
\begin{center}
\includegraphics[width=8cm]{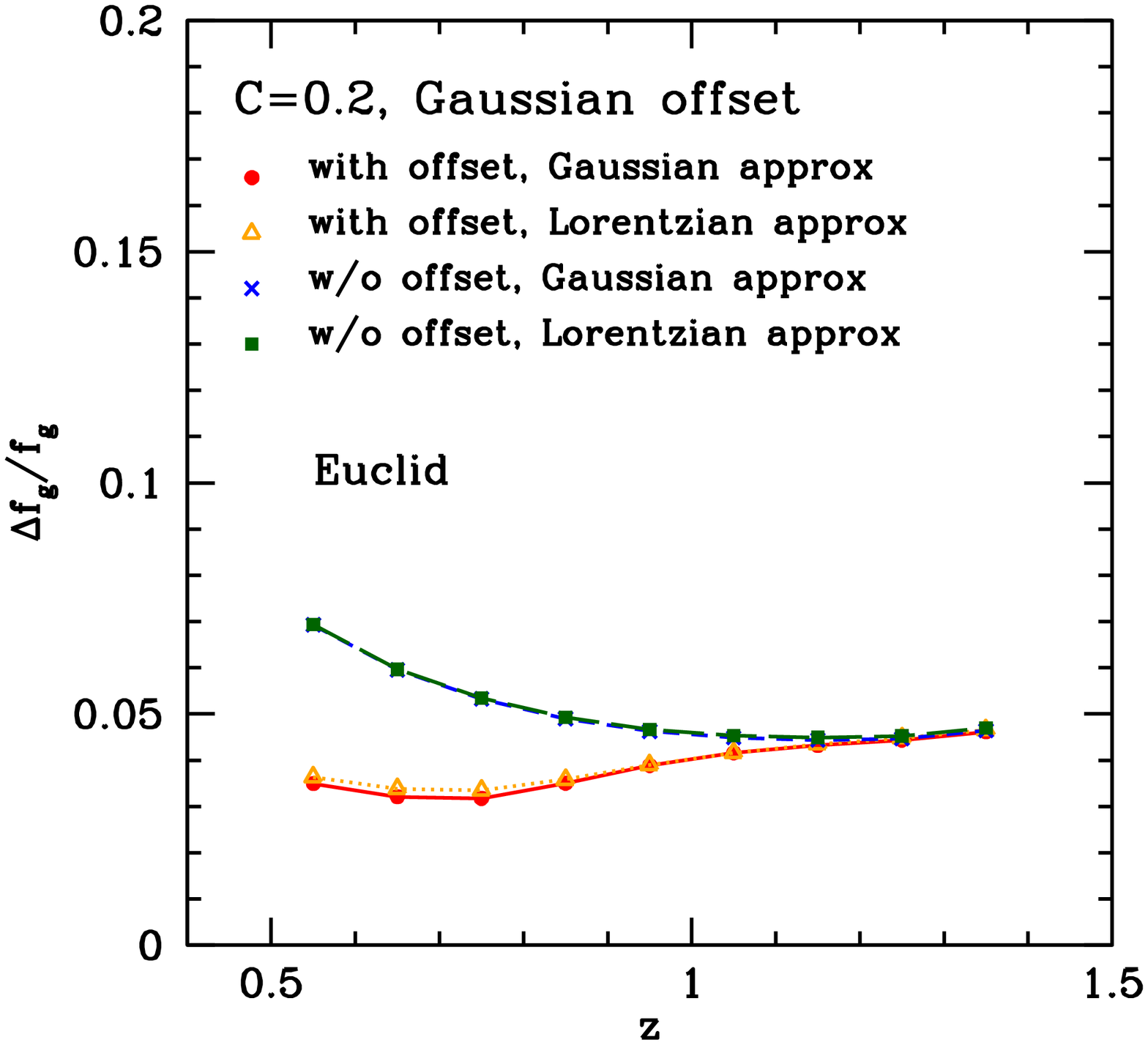}
\includegraphics[width=8cm]{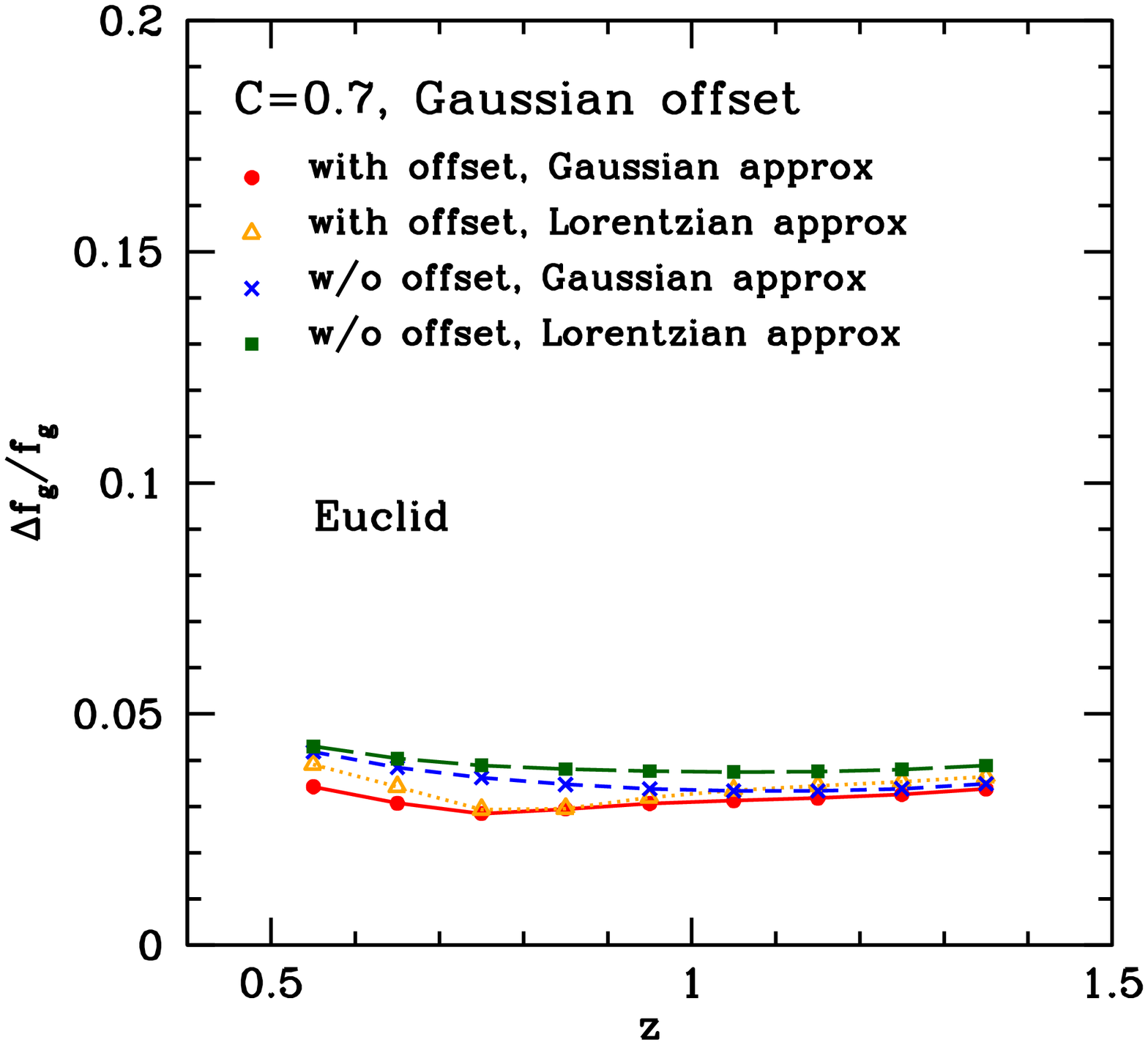}
\includegraphics[width=8cm]{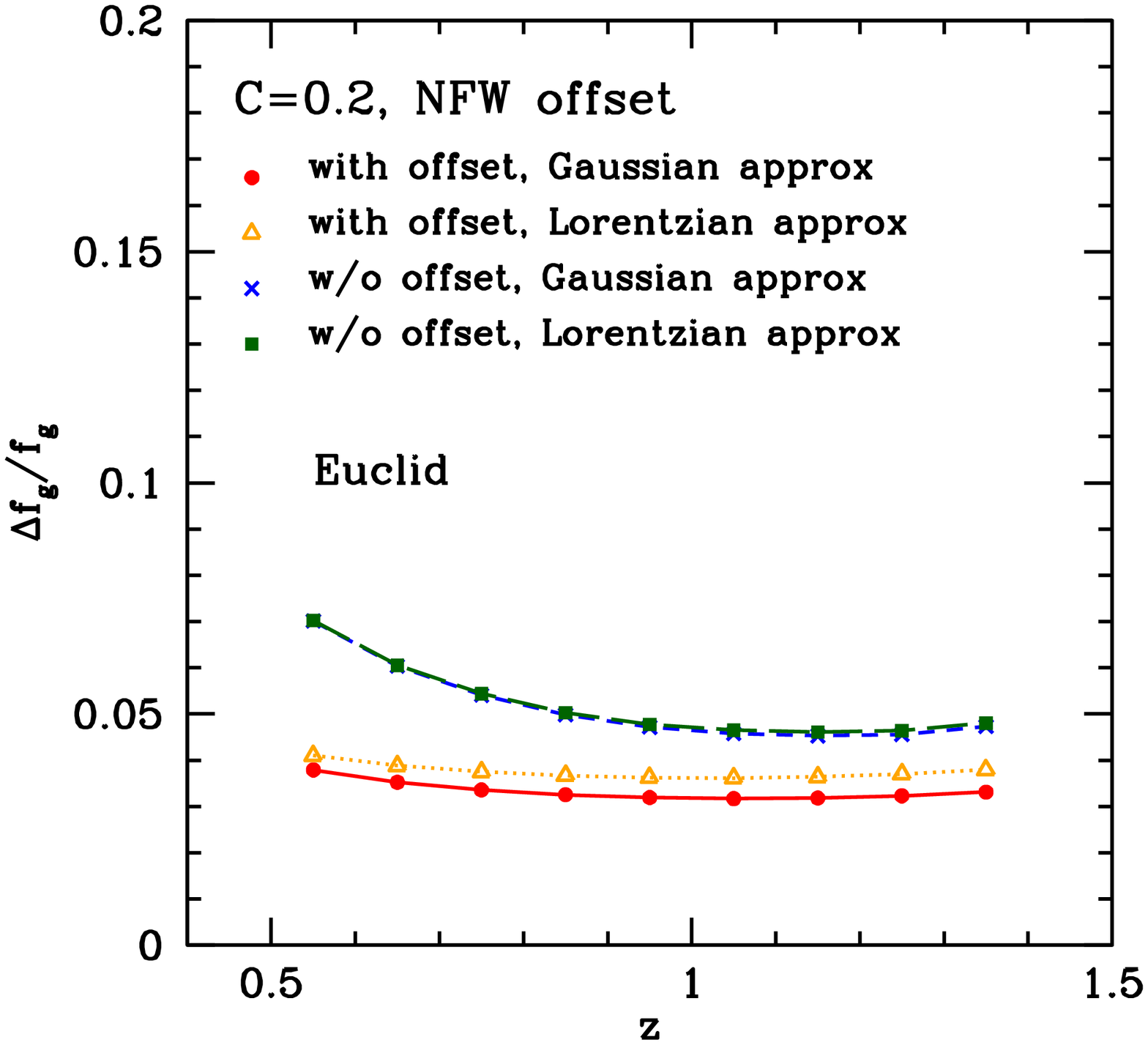}
\includegraphics[width=8cm]{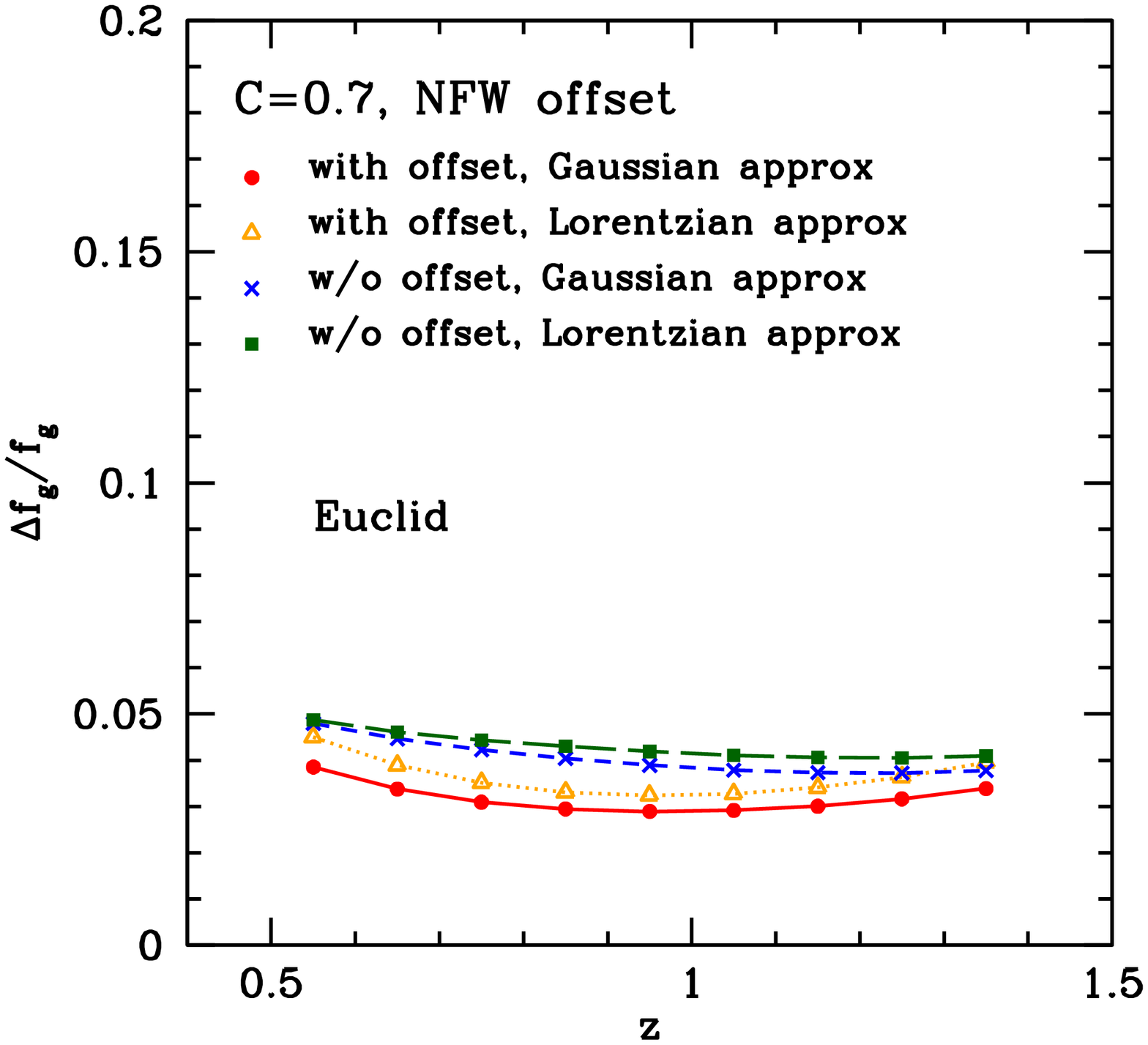}
\caption{Similar to Fig \ref{fig:fz}, but for the Euclid survey.}
\label{fig:fz_euclid}
\end{center}
\end{figure}

We estimate the impact of the FoG effect on the measurement of the
growth rate in each redshift slice. Figs.~\ref{fig:fz} and
\ref{fig:fz_euclid} compare the marginalized, fractional errors of
$f_g(z_i)$ for the cases (i) and (ii), the cases with and without the
FoG effect correction using the DLRG-galaxy lensing information of the
DLRG radial profile (see Table~\ref{tab:lens_error}). The upper panels
show the results for the Gaussian radial profile model, while the
lower panels show the results for the NFW radial profile model. The
left- and right-side panels are different in the maximum wavenumber
$k_{\rm max}$, which is determined by $C=0.2$ (left panels) and $C=0.7
$ (right) in Eq.~(\ref{eq:Ckmax}) roughly corresponding to $k_{\rm
max}\simeq 0.1$ and $0.2~h{\rm Mpc}^{-1}$ respectively (see
Table~\ref{tab:lens_error}).  Note that the plotted errors include the
statistical and systematic errors (Eq.~[\ref{eq:toterr}]).  It is
found that the lensing information significantly improves the
constraint on $f_g$ over a wide range of redshifts $z$: the
improvement is up to a factor of 2 for $C=0.2$ (the lower $k_{\rm
max}$), but less significant at higher $z$. This result can be
understood as follows. The DLRG-galaxy lensing measurements are more
accurate for lower $z$ DLRGs because of higher number densities of
background galaxies, which reduces the shot noise contamination.  (see
Table~\ref{tab:lens_error}). The error of $f_g$ improves less at
higher $k$ where the systematic error due to the FoG model inaccuracy
is dominated over the statistical error.
%In addition, if including the
%information up to the higher $k_{\rm max}$, the Gaussian or Lorentzian
%FoG model is not accurate enough as implied in
%Fig.~\ref{fig:fog}. Therefore a less improvement in the error of $f_g$
%in the right-side panels is ascribed to the fact that the systematic
%error due to the model inaccuracy becomes more significant at the
%higher $k$, compared to the statistical precision. 
The Gaussian FoG model gives a better improvement than the Lorentzian
model because we assume the Gaussian velocity distribution of DLRGs
within halos.  Finally, comparing the upper- and lower-side panels
shows that the NFW radial profile model gives a better performance of
the lensing FoG correction, as implied by the accuracies of the DLRG
velocity dispersion reconstruction in Table~\ref{tab:lens_error}.  As
a result these spectroscopic surveys allow for constraints on the
growth rate to 5\% precision or even better at each redshift when the
lensing information is combined.

%%%%%%%%%%%%%%%%%%%%%%%%%%%%%%%%%%%%%%%%%%%%%%%%%%%%%%%%
%%%%%%%%%%%%%%%%%%%%%%%%%%%%%%%%%%%%%%%%%%%%%%%%%%%%%%%

%%%%%%%%%%%%%%%%%%%%%%%%%%%%%%%%%%%%%%%%%%%%%%%%%%%%%%%
%%%%%%%%%%%%%%%%%%%%%%%%%%%%%%%%%%%%%%%%%%%%%%%%%%%%%%%

Tables~\ref{tab:fnu} and \ref{tab:w0} study the impact of FoG effect
on the parameter $f_\nu$ and $w_0$, respectively.  The tables show the
marginalized error on each parameter and the amount of bias in the
best-fit parameter if the FoG effect is corrected imperfectly or
ignored. Note that in this analysis, either $w_0$ or $f_\nu $ is fixed
to the fiducial value: for the table of $w_0$, $f_\nu$ is fixed to
$f_\nu=0$ and vice versa.  Although the dark energy equation-of-state
parameter $w_0$ can be more robustly constrained by the BAO peak
location \citep[][]{Eisensteinetal:05}, adding the amplitude and shape
information of power spectrum can significantly improve the error
because dark energy is sensitive to the power spectrum amplitude via
the growth rate \cite[][]{Saitoetal:11}. First of all, the table shows
that the FoG effect is very significant even at such large length
scales ($k_{\rm max}\simeq 0.1~h{\rm Mpc}^{-1}$ for $C=0.2$).  For
example, the column ``FoG neglected'' indicates that the parameters
are significantly biased if the FoG effect ignored. The bias can be
larger than the statistical error by more than a factor of 5. That is,
if the FoG effect is ignored, a non-zero neutrino mass or $w_0\ne -1$
may incorrectly be claimed.

The table also shows that the parameter accuracy of $f_\nu$ or $w_0$
can be improved by correcting for the FoG effect based on the lensing
information of DLRGs radial profile.  For $C=0.2$ ($k_{\rm max}\simeq
0.1~h{\rm Mpc}^{-1}$), comparing the columns labeled as ``w/o offset''
and ``with offset'' shows that the lensing FoG correction improves the
marginalized error of $f_\nu$ or $w_0$ by about 5-25\% for BOSS, PFS
or Euclid.  However, as the power spectrum information up to the
higher $k_{\rm max}$ is included, the bias due to the imperfect
modeling of FoG effect becomes more significant, as shown in the rows
denoted by ``$C=0.7$''. Thus even if the statistical precision is
apparently improved in such high-$k$ regime, a more accurate modeling
of the FoG effect is needed in order not to have any significant
parameter bias.  For BOSS or PFS surveys, the amount of the bias on
$f_\nu$ or $w_0$ is still smaller than the (marginalized) statistical
error, while the bias becomes significant for Euclid survey.  Finally
we remark on the results that directions in the parameter bias are
opposite for the results with and without the lensing priors on the
off-centering parameters. For example, for a Gaussian FoG model case,
we have found that, if including the lensing prior on the FoG
suppression scale (i.e. the case with ``with offset''), the model
prediction tends to overestimate the FoG suppression as implied in
Fig.~\ref{fig:fog}, and thus prefers a smaller $f_\nu$, i.e. $\delta
f_\nu<0$. On the other hand, if the FoG suppression scale is treated
as a free parameter (i.e. the case ``w/o offset''), the fitting tends
to prefer a smaller FoG suppression, and in turn prefer a larger
neutrino mass (larger $f_\nu$); $\delta f_\nu>0$.

\begin{figure*}
\begin{center}
\includegraphics[width=8cm]{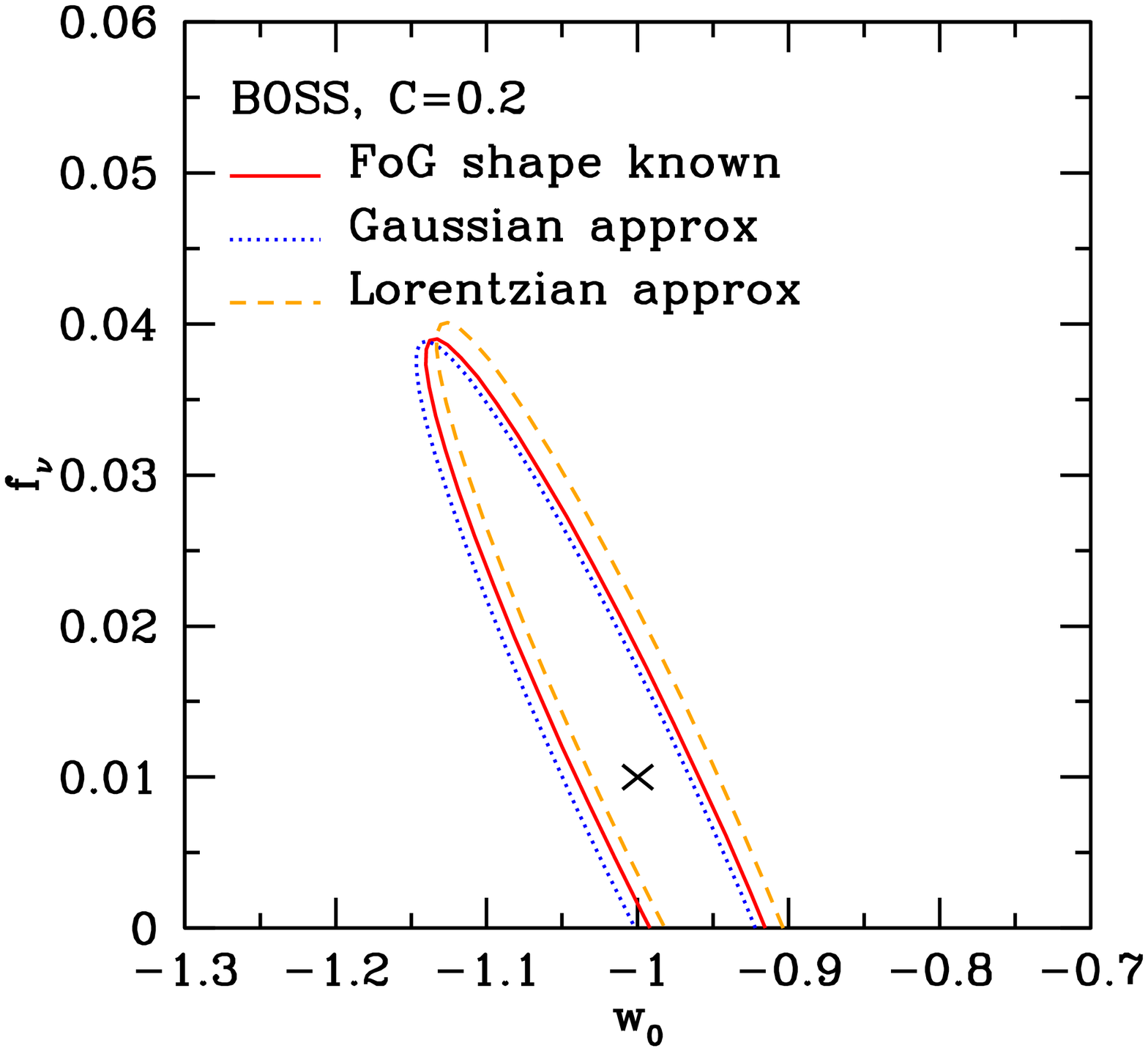}
\includegraphics[width=8cm]{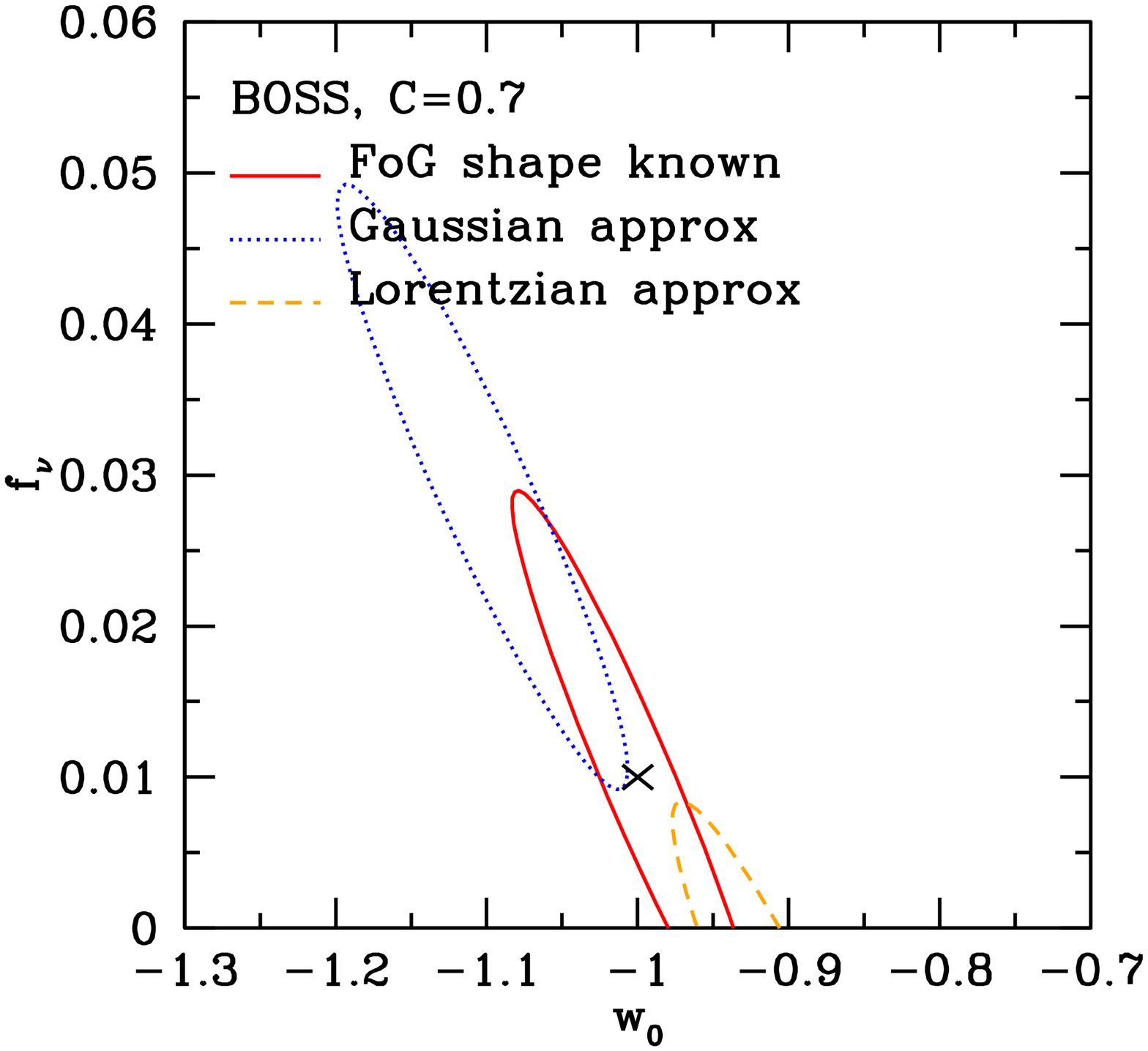}
\includegraphics[width=8cm]{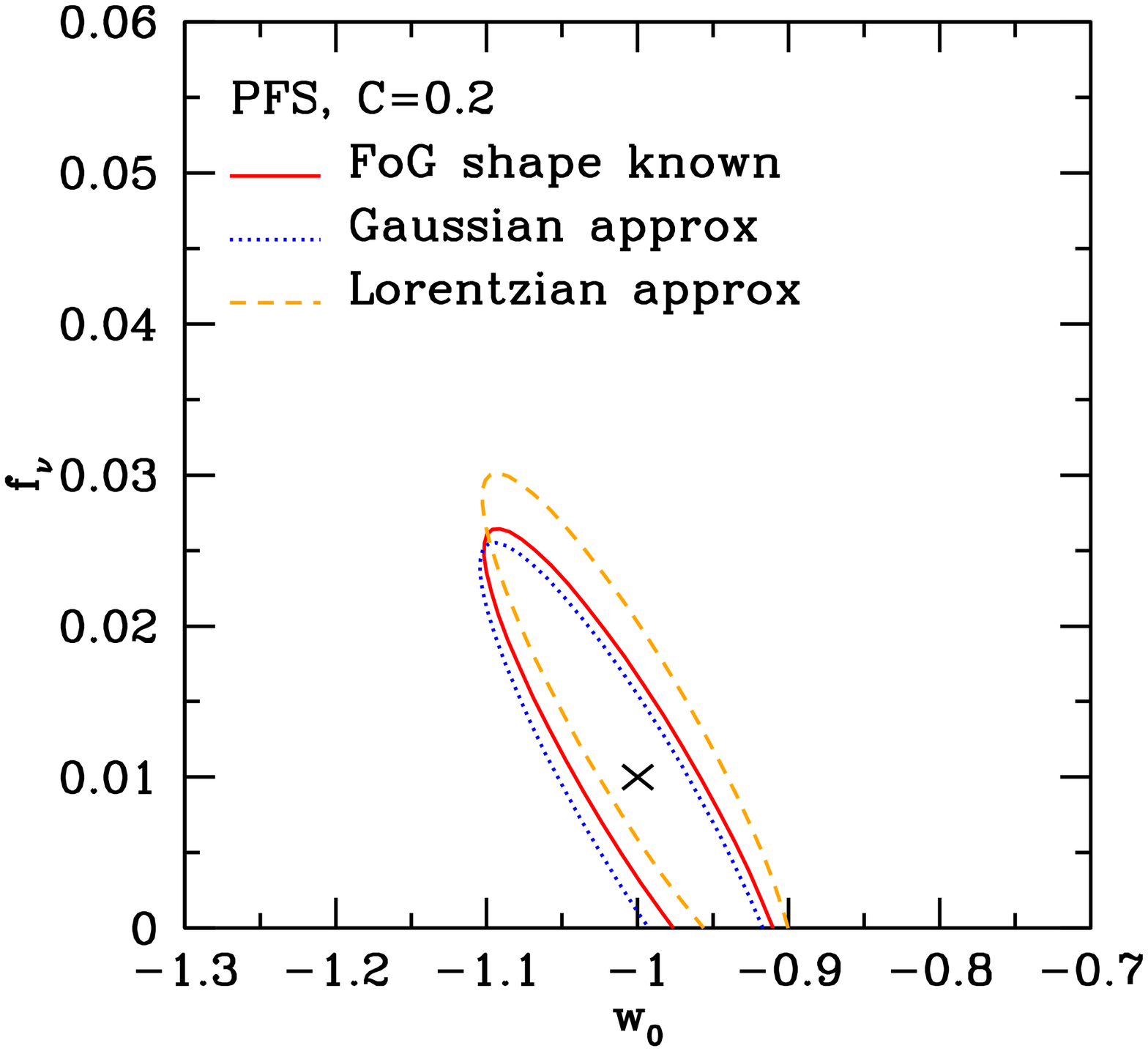}
\includegraphics[width=8cm]{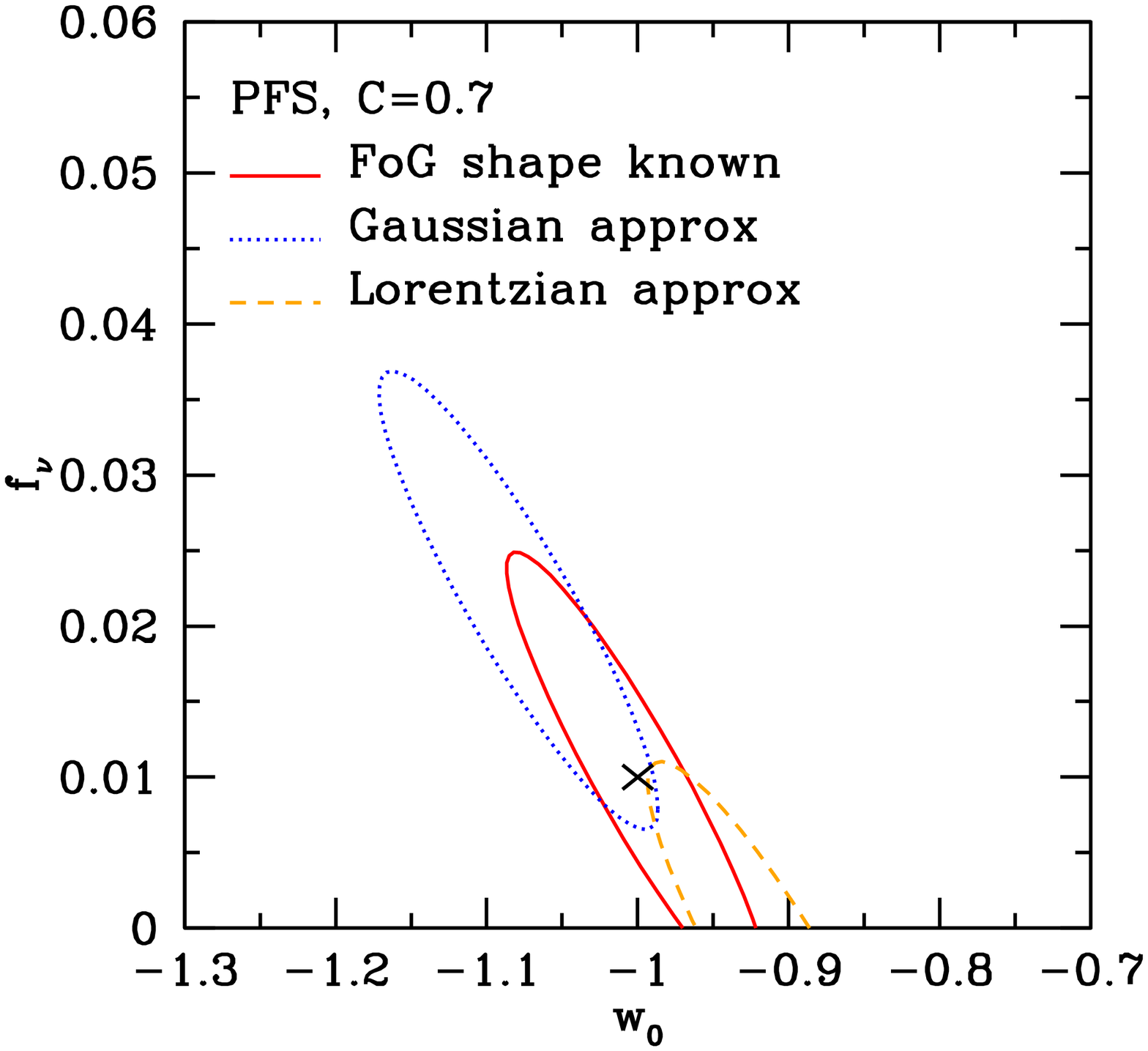} 
\caption{The marginalized error ellipses in a sub-space of
  $(w_0,f_\nu)$, where both the parameters are included in the Fisher
  analysis. Solid ellipse represents the error when the FoG shape is
  completely known (i.e. ``FoG shape known''), while the other
  ellipses show the errors when the FoG effect is approximated with
  Gaussian or Lorentzian forms ``with offset'' (dotted and dashed
  respectively). The imperfect modeling of the FoG effect biases the
  results different from the input values denoted by cross
  symbols. The lensing information is added for all plots. The
  different panels differ in the hypothetical spectroscopic survey
  (BOSS or PFS) and the maximum wavenumber $k_{\rm max}$ used in the
  Fisher analysis ($C=0.2$ or 0.7).}
\label{fig:covcont}
\end{center}
\end{figure*}

The Gaussian and Lorentzian approximations are not sufficiently
accurate to fully describe the FoG shape at $k\simgt
0.15h$/Mpc. Fig.~\ref{fig:covcont} shows the systematic impact on the
marginalized errors of $(f_\nu, w_0)$ for BOSS and PFS surveys with
different $k_{\rm max}$ values.  In this Fisher analysis both $f_\nu$
and $w_0$ are treated as free parameters in the model fitting (the
growth rate $f_g(z_i)$ is fixed to the fiducial value). The error
ellipse in each panel shows that the parameter $w_0$ and $f_\nu$ are
correlated with each other in the measured DLRG power spectrum.  The
figure shows that an imperfect modeling of the FoG effect may bias the
parameters.  However, the amount of the bias is still at 1-$\sigma$
confidence level in the two-dimensional space for these surveys.

\begin{figure}
\begin{center}
\includegraphics[width=8cm]{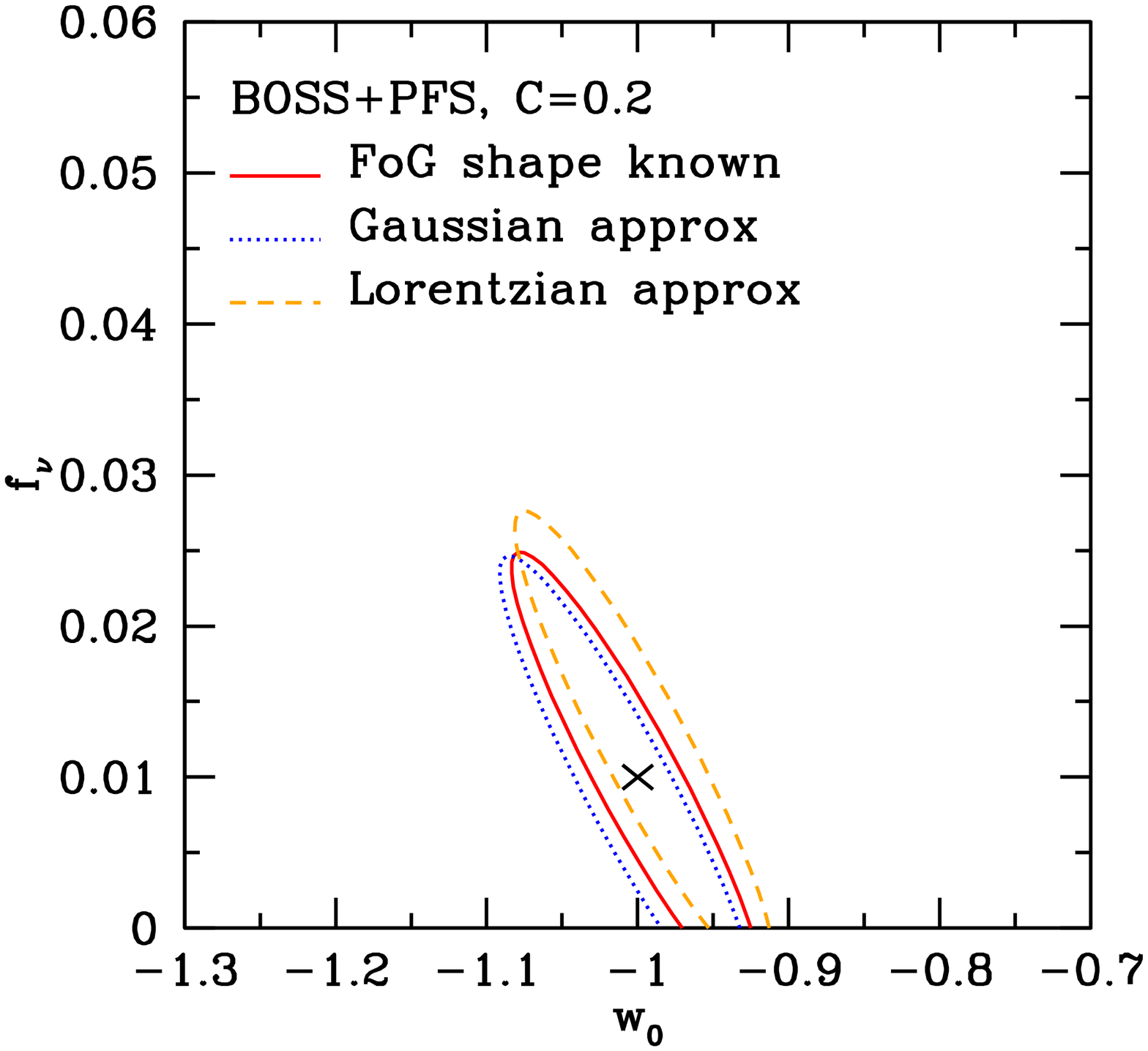}
\includegraphics[width=8cm]{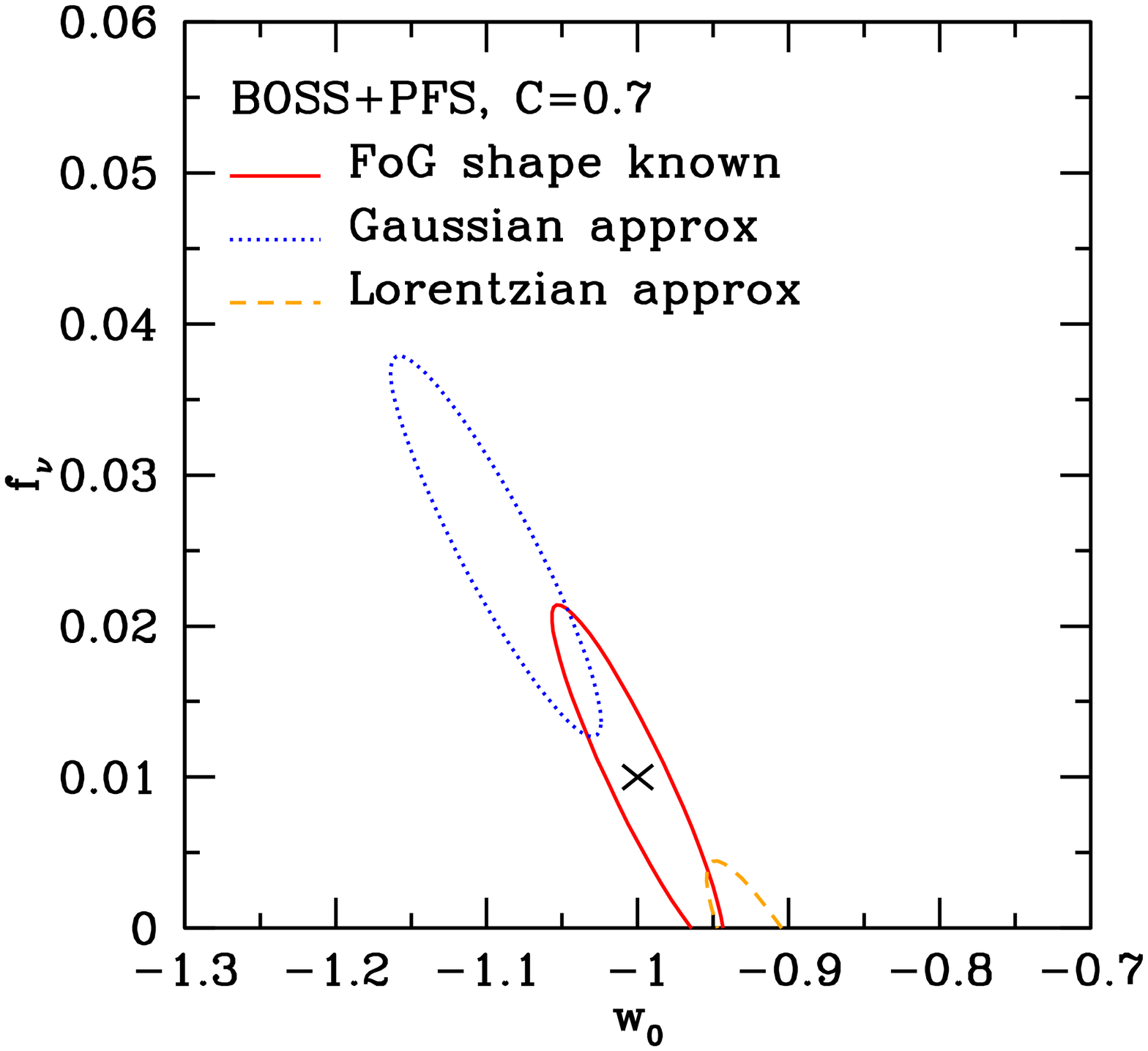}
\includegraphics[width=8cm]{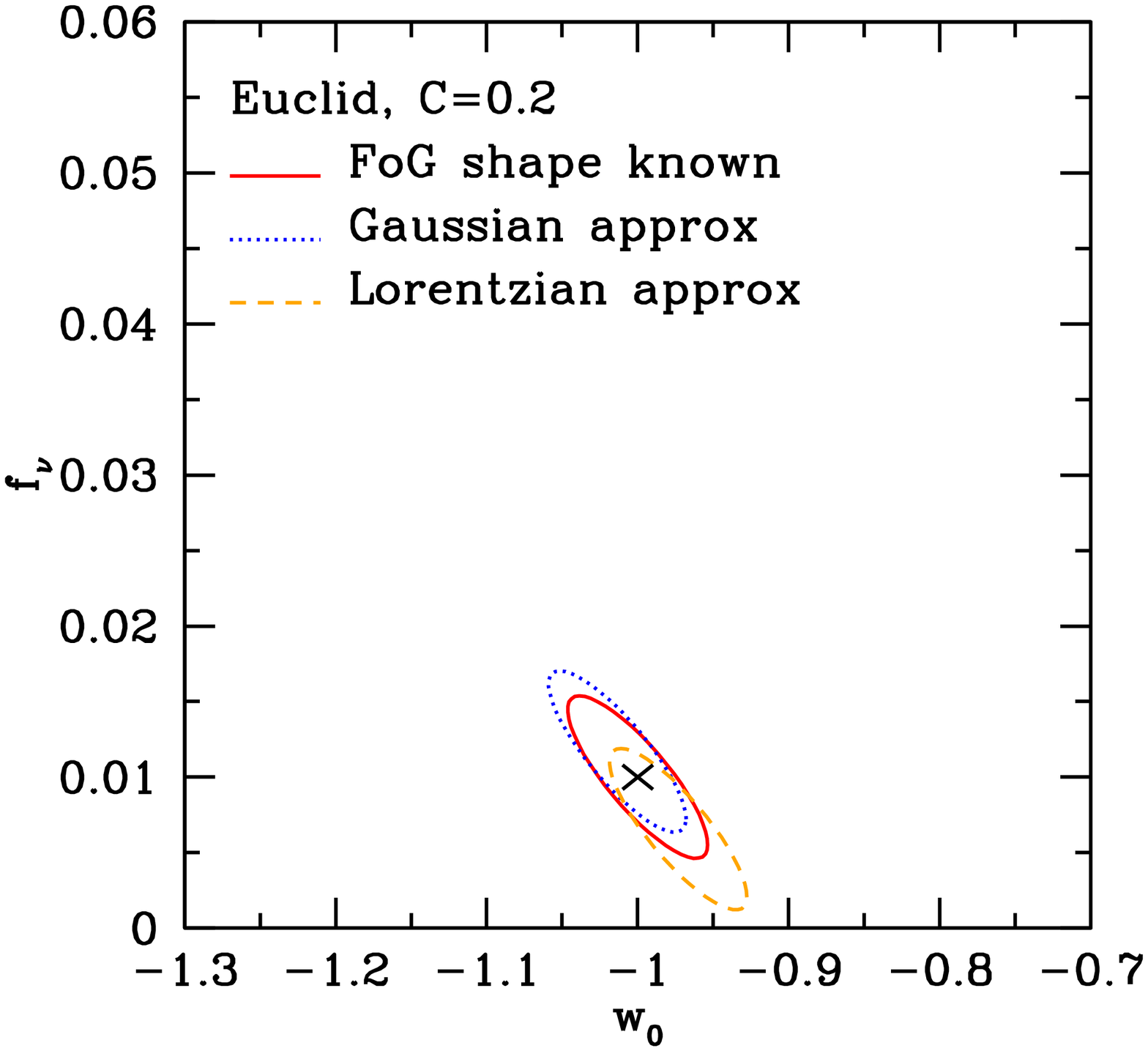}
\includegraphics[width=8cm]{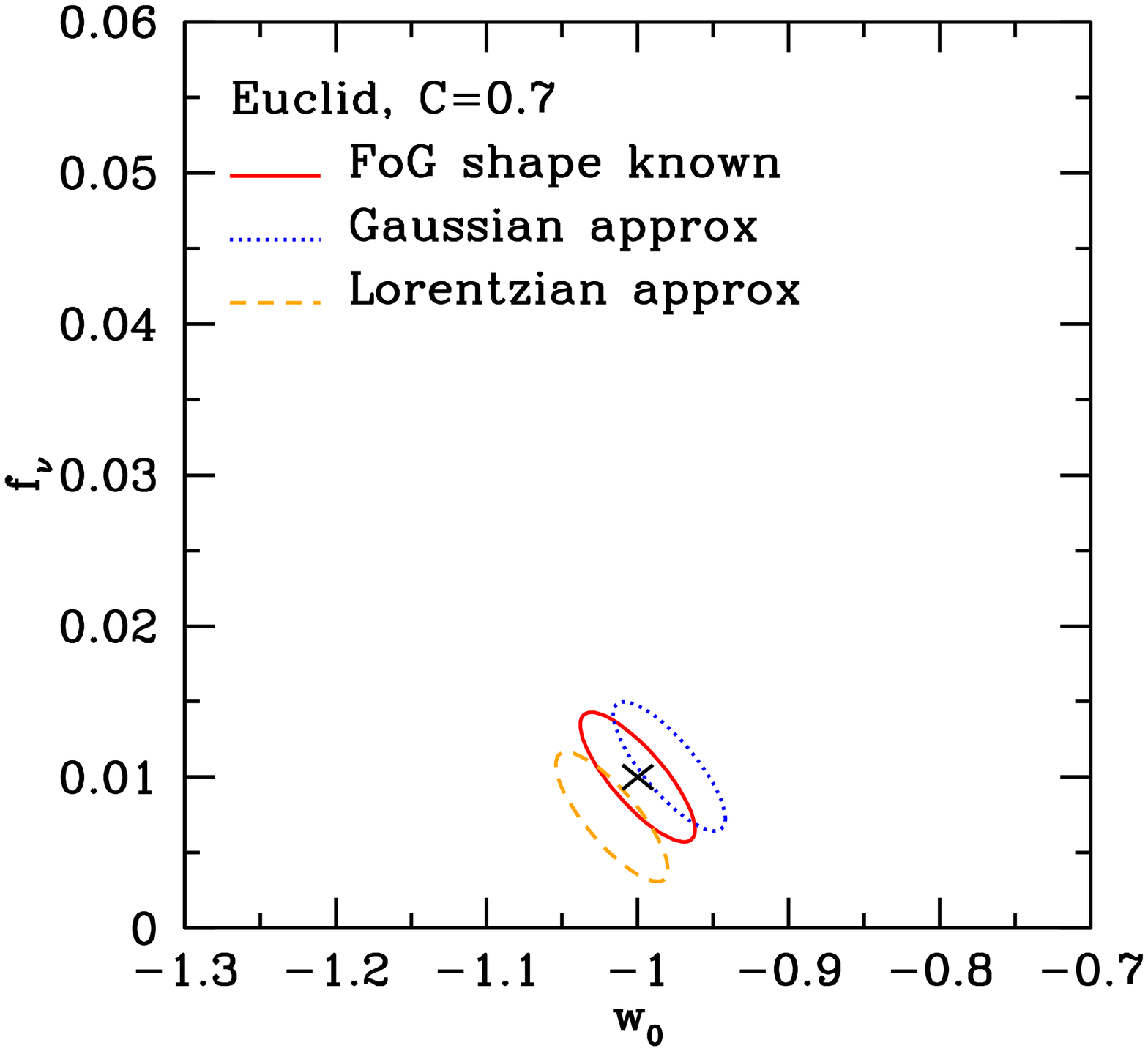}
\caption{Similar to the previous figure, but for the joint experiments
  of HSC+BOSS+PFS+Planck ({\em upper panels}) and Euclid+Planck ({\em
  lower}).
\label{fig:fnu2}}
\end{center}
\end{figure}

Fig.~\ref{fig:fnu2} shows the forecasts for the combined experiments
of Planck CMB combined with BOSS, PFS and HSC (upper panels) and
Planck CMB with Euclid (lower), respectively. The figure shows that
the error ellipses are further shrunk by having the power spectrum
information over a wider range of redshifts. As a result, the
marginalized errors of $\sigma(f_\nu)\simeq 0.011 (0.009)$ [or
  equivalently $\sigma(m_{\nu,{\rm tot}})\simeq 0.11 (0.093)~{\rm
    eV}$] and $\sigma(w_0)\simeq 0.063 (0.051)$ can be achieved at
$C=0.2 (0.7)$ for the BOSS+PFS+HSC survey.  In Euclid survey, the
error decreases to $\sigma (f_\nu)\simeq 0.0035 (0.0028)$
[$\sigma(m_{\nu,{\rm tot}})=0.035 (0.028)$~eV] and $\sigma(w_0)\simeq
0.029 (0.024)$ at $C=0.2(0.7)$.  In particular the Euclid survey
allows for a detection of the non-zero neutrino masses at more than
$1\sigma$ significance, because the lower mass bound for the normal or
inverted mass hierarchy is about $0.05~$ or $0.1$~eV, respectively.
Thus such a high-precision determination of these parameters is
potentially feasible for these surveys {\em if} we can use a
sufficiently accurate model of the FoG in order to include the power
spectrum amplitude information up to such higher $k$'s.

%%%%%%%%%%%%%%%%%%%%%%%%%%%%%%%%%%%%%%%%%%%%%%%%%%%%%%%%
%C.H. add ``FoG shape known'' result
\begin{table*}
\begin{center}
Marginalized Error and Bias of Neutrino Mass $(f_\nu)$\\
\begin{tabular}{ccccccccccc}
  \hline\hline
& & \multicolumn{3}{c}{w/o offset} & & \multicolumn{3}{c}{with offset} & FoG shape known & FoG neglected \\
\cline{3-5}\cline{7-9} 
Survey & $k_{\rm max}$ &
\raisebox{-1ex}{$\sigma(f_\nu)$} & \raisebox{-1ex}{$\delta f_\nu$(Gauss)} & \raisebox{-1ex}{$\delta f_\nu$(Lorentz)} & &
\raisebox{-1ex}{$\sigma(f_\nu)$} & \raisebox{-1ex}{$\delta f_\nu$(Gauss)} & \raisebox{-1ex}{$\delta f_\nu$(Lorentz)} &
\raisebox{-1ex}{$\sigma(f_\nu)$} & \raisebox{-1ex}{$\delta f_\nu$} 
\\ \hline
BOSS   & C=0.2 & $ 0.0080$ & $ 0.0001$ & $-0.0002$ & & $ 0.0058(27\%)$ & $-0.0009$ & $ 0.0014$ & $ 0.0063 $& $0.03$ \\
       & C=0.7 & $ 0.0052$ & $ 0.0022$ & $-0.0027$ & & $ 0.0046(12\%)$ & $-0.0022$ & $ 0.0031$ & $ 0.0050 $& $0.08$ \\
PFS    & C=0.2 & $ 0.0068$ & $ 0.0004$ & $-0.0007$ & & $ 0.0057(16\%)$ & $-0.0006$ & $ 0.0010$ & $ 0.0063 $& $0.04$ \\
       & C=0.7 & $ 0.0051$ & $ 0.0022$ & $-0.0032$ & & $ 0.0049( 4\%)$ & $ 0.0003$ & $-0.0003$ & $ 0.0051 $& $0.08$ \\
EUCLID & C=0.2 & $ 0.0025$ & $ 0.0005$ & $-0.0010$ & & $ 0.0023( 9\%)$ & $-0.0002$ & $ 0.0003$ & $ 0.0024 $& $0.04$ \\
       & C=0.7 & $ 0.0020$ & $ 0.0025$ & $-0.0035$ & & $ 0.0020( 1\%)$ & $ 0.0017$ & $-0.0025$ & $ 0.0020 $& $0.07$ \\
\hline
\end{tabular}
\end{center}
\caption{Marginalized error and bias of $f_\nu$ for either BOSS, PFS,
  or Euclid survey combined with the Planck CMB information. Here we
  assume the FoG effect due to our fiducial Gaussian DLRG radial
  profile model, as in Fig.~\ref{fig:fz}, and assume $f_\nu=0.01
  (m_{\nu, {\rm tot}}=0.104~{\rm eV})$ as the fiducial values.  Each
  column shows the statistical error ($\sigma(f_\nu)$) and the amount
  of bias ($\delta f_\nu\equiv f_{\nu,{\rm est}}- f_{\nu, {\rm
      true}}$) due to the difference of the assumed Gaussian (Gauss)
  or Lorentzian (Lorentz) FoG model from the input FoG effect that is
  computed based on our halo model. The columns labeled as ``with
  offset'' and ``w/o offset'' show the results with and without the
  lensing information being used to correct the FoG effect. The
  percentage numbers in parentheses in the $\sigma(f_\nu )$ column
  denote an improvement in the errors due to the lensing information.
  The column ``FoG shape known'' shows the statistical error when the
  same lensing information is added but assuming the shape of the FoG
  effect is known based on the halo model (see text for details).  The
  column ``FoG neglected'' shows the parameter bias where the FoG
  effect is completely ignored (i.e. $\overline{\sigma}_{v, {\rm
      off}}=0$ is set). These results show a significant bias, more
  than 100\% bias compared to the input value $f_\nu=0.01$.
\label{tab:fnu}
}
\end{table*}
%%%%%%%%%%%%%%%%%%%%%%%%%%%%%%%%%%%%%%%%%%%%%%%%%%%%%%%%
\begin{table*}
\begin{center}
Marginalized Error and Bias of $w_0$\\
\begin{tabular}{ccccccccccc}
  \hline\hline
& & \multicolumn{3}{c}{w/o offset} & & \multicolumn{3}{c}{with offset} & FoG shape known & FoG neglected \\
\cline{3-5}\cline{7-9}
Survey & $k_{\rm max}$ &
\raisebox{-1ex}{$\sigma(w_0)$} & \raisebox{-1ex}{$\delta w_0$(Gauss)} & \raisebox{-1ex}{$\delta w_0$(Lorentz)} & &
\raisebox{-1ex}{$\sigma(w_0)$} & \raisebox{-1ex}{$\delta w_0$(Gauss)} & \raisebox{-1ex}{$\delta w_0$(Lorentz)} &
 \raisebox{-1ex}{$\sigma(w_0)$} & \raisebox{-1ex}{$\delta w_0$} 
\\ \hline
BOSS   & C=0.2 & $ 0.038$ & $ 0.000$ & $ 0.000$ & & $ 0.029(23\%)$ & $-0.004$ & $ 0.007$ & $ 0.031 $&$ 0.15$ \\
       & C=0.7 & $ 0.027$ & $ 0.007$ & $-0.009$ & & $ 0.021(20\%)$ & $-0.018$ & $ 0.024$ & $ 0.024 $&$ 0.41$ \\
PFS    & C=0.2 & $ 0.039$ & $ 0.001$ & $-0.001$ & & $ 0.034(13\%)$ & $-0.004$ & $ 0.007$ & $ 0.037 $&$ 0.20$ \\
       & C=0.7 & $ 0.032$ & $ 0.007$ & $-0.009$ & & $ 0.030( 9\%)$ & $-0.009$ & $ 0.015$ & $ 0.032 $&$ 0.53$ \\
EUCLID & C=0.2 & $ 0.017$ & $ 0.003$ & $-0.005$ & & $ 0.014(17\%)$ & $-0.003$ & $ 0.006$ & $ 0.016$ & $ 0.25$ \\
       & C=0.7 & $ 0.014$ & $ 0.016$ & $-0.022$ & & $ 0.013( 5\%)$ & $ 0.004$ & $-0.004$ & $ 0.014$ & $ 0.64$ \\
\hline
\end{tabular}
\end{center}
\caption{Similar to Table~ \ref{tab:fnu}, but for the dark energy
 equation-of-state parameter $w_0$. Note that for these results the
 neutrino mass is fixed to the fiducial value $f_\nu=0$.
\label{tab:w0}
}
\end{table*}
%%%%%%%%%%%%%%%%%%%%%%%%%%%%%%%%%%%%%%%%%%%%%%%%%%%%%%%%

%%%%%%%%%%%%%%%%%%%%%%%%%%%%%%%%%%%%%%%%%%%%%%%%%%%%%%%%
%%%%%%%%%%%%%%%%%%%%%%%%%%%%%%%%%%%%%%%%%%%%%%%%%%%%%%%

%%%%%%%%%%%%%%%%%%%%%%%%%%%%%%%%%%%%%%%%%%%%%%%%%%%%%%%
%%%%%%%%%%%%%%%%%%%%%%%%%%%%%%%%%%%%%%%%%%%%%%%%%%%%%%%

\section{Summary and Discussion}
\label{sec:summary}

Our lack of understanding of the relationship between the positions of
galaxies in redshift space and the underlying distribution of matter
is the largest systematic source of errors in the analysis of redshift
surveys.  If the dominant galaxy (e.g., the brightest cluster galaxy)
in each halo was at rest in the center of the halo, then there would
be minimal Finger-of-God (FoG) effects in a properly selected sample,
nor any suppression of the power spectrum amplitude due to finite size
of galaxy halos.  We have quantified the deviations from this simple
universe in terms of galaxy radial distribution, the probability of
finding a galaxy at a given distance from the center of the halo.
Since galaxies not in the centers of the gravitational potential are
moving relative to the potential, thus the amplitude of this
displacement is directly related to the amplitude of the FoG effect.

In this paper, we have showed how galaxy-galaxy lensing observations,
which measure the relationship between galaxy positions and the
distribution of dark matter in a statistical sense, can be used to
measure this profile and reduce the uncertainties in analyses of
galaxy power spectrum measurements.  We have studied a case of
luminous red galaxies (LRG), especially the dominant LRG in each halo
(DLRG), as a working example for demonstrating an expected performance
of our proposing method.  We have assumed DLRG off-centered profiles
motivated by the weak lensing study \citep{Ogurietal:10} and the
analysis \citep{Hoetal:09} studying offsets of LRGs relative to the
peak in X-ray brightness.  Our DLRG radial profile model predicts that
the FoG effect suppresses the DLRG power spectrum amplitudes at $k\sim
0.1$ and 0.2$h$/Mpc by 10 and 30\%, respectively, much greater than
the statistical precision of power spectrum measurements for ongoing
and upcoming spectroscopic surveys (see Fig. 8). The halos with masses
$\sim 10^{14}M_\odot$ make the dominant contribution to the FoG effect
on the DLRG power spectrum.

If we can measure the radial distribution, then we can model this
effect and remove (or at least significantly reduce) this source of
systematic uncertainty. We have illustrated this correction for
ongoing and upcoming surveys: the BOSS survey, the Subaru imaging
(HSC) and spectroscopic (PFS) surveys and ESA's Euclid mission (see
Table \ref{tab:survey}). For example, the Subaru HSC imaging survey
covering an area of 2000 square degrees can accurately measure the
DLRG-galaxy lensing profile down to scales of an arcminute enabling an
accurate characterization of the DLRG off-centered profile (see
Fig. \ref{fig:lrglens} and Table \ref{tab:lens_error}). As we have
showed, this radial distribution can be used to correct the FoG effect
and significantly improve parameter estimation from the redshift-space
power spectrum measurements. We have illustrated this effect by
discussing the impact of FoG effect on the parameters $w_0$ and
$f_\nu$ as well as the growth rate $f_g\equiv d\ln D/d\ln a$ at each
redshift slice.  We have found that combining the lensing measurements
with the redshift-space power spectrum can improve the constraints on
$f_g$ up to a factor of 2, compared to the case that the FoG effect is
considered unknown, if the power spectrum information down to $k_{\rm
max}\simeq 0.1$ or $0.2h$/Mpc is included (see Fig.\ref{fig:fz}).
Including the lensing correction also improves measurements of $w_0$
and $f_\nu$ by 5-25\%. Such a factor of 2 improvement in the parameter
corresponds to a factor of 4 larger survey volume, and therefore the
combined imaging and spectroscopic surveys can bring a huge beneficial
synergy.

The analyses in this paper have made a number of simplifying
assumptions that would have to be improved in a realistic analysis of
a large spectroscopic survey.  We have assumed a linear model for
redshift distortion effect on halo power spectrum (the Kaiser
formulation in Eq.~[\ref{eq:Kaiser}]). The quality of the rapidly
improving redshift data requires more refined model based on extended
perturbation theory as well as simulations
\citep{Scoccimarro:04,Taruyaetal:10,Tangetal:11,SatoMatsubara:11}.  We
have also assumed a linear bias model. \cite{Saitoetal:11} found that
non-linear bias corrections are important for the SDSS DLRGs.

In this paper, for simplicity, we have assumed that the velocity
dispersion of DLRGs is determined by an enclosed mass (mostly dark
matter) at a given radius and that the velocity distribution of DLRGs
obeys a Gaussian distribution in a statistical average sense (see
Eqs.~[\ref{eq:sigv_rM}] and [\ref{eq:vel_gauss}]). However, the
velocity distribution should be by nature sensitive to various
complicated physics and galaxy formation physics, so it is still
difficult to understand based on theoretical studies. Therefore an
observational approach to tackling this issue is rather more
adequate. For example, one may be able to use spectroscopic redshifts
of galaxies, from a survey data itself or from a dedicated survey with
optimized fiber positioning, to explore the velocity structures of
DLRGs around massive halos based on the stacking analysis
\citep[][]{Beckeretal:07,WhiteCohn:10,Skibbaetal:11} (also see Lam et
al. in preparation). Or an alternative approach on the analysis side
is to use an empirical model of the FoG effect which is given by a
multiplicative function of the Gaussian FoG form and a perturbative
functional form [$1+a_0(k\mu)^2 +a_1(k\mu)^4+\cdots$], where the
coefficients $a_0, a_1$ are treated as nuisance parameters. The model
defined in this way can have more degrees of freedom to describe
possible complicated scale-dependences in genuine FoG effects
\citep[][for a similar approach]{Tangetal:11}.  Then, by using the
generalized FoG form, we may be able to derive more robust, unbiased
cosmological constraints including marginalization over the nuisance
parameters, but still using the lensing information of off-centered
DLRGs to constrain the Gaussian FoG part as we did in this paper.

There is more work needed to explore the effects of DLRG offsets on
galaxy power spectra and lensing.  Since galaxies of a given type lies
in a range of halo masses, it is important to trace the dependence of
the DLRG clustering and the FoG effect on halo mass scale: for the
clustering, halos with typical mass scale of $\sim 10^{13}M_\odot$
gives a dominant contribution, while the FoG effect arises mainly from
more massive halos with masses around $10^{14}M_\odot$.  It would be
useful to select a sample of massive halos focused on the high mass
range to more accurately explore the offset of the DLRG from the halo
centers as defined by optical, X-ray and SZ data.  In this approach,
weak lensing can play an important role, as various halo center
indicators can be used to monitor weak lensing signals at small
angular scales in both individual cluster lensing and stacked analysis
bases
\citep[][]{Johnstonetal:07,Leauthaudetal:10,Okabeetal:10,Ogurietal:10,OguriTakada:10}.

Besides the cosmological use, the off-centering information of DLRGs
should be very useful to develop a more physical understanding of the
dynamical processes of DLRGs within the main halo. DLRGs are likely to
reside on one of the most massive sub-halos within the main host halo
with $\simgt 10^{13}M_\odot$.  Therefore DLRG within each halo tends
to sink towards the halo center due to dynamical friction, which is
one of basic explanations for a more centrally concentrated
distribution of DLRGs within halos compared to the dark matter
distribution.  The dynamical processes and assembly histories of
DLRGs, within a given time scale of cluster-scale halos, are a key
information to understanding the nature of DLRGs in the context of CDM
dominated structure formation scenario. Again the galaxy-galaxy
lensing can offer a new method of tackling these issues.

\bigskip
 
\section*{Acknowledgments}
We thank Joanne Cohn, Daniel Eisenstein, Issha Kayo, Eiichiro Komatsu,
Rachel Mandelbaum, Nikhil Padmanabhan, John Peacock, Beth Reid and
Martin White for useful discussion and valuable comments.  CH
acknowledges support from a Japan Society for Promotion of Science
(JSPS) fellowship.  MT thanks Department of Astrophysical Sciences,
Princeton University for its warm hospitality during his visit, where
this work was initiated.  DNS and CH acknowledge support from NSF
grant AST-0707731 and the NASA AST theory program.  DNS thanks the
IPMU for its warm hospitality during his visit, where the work was
completed.  This work is in part supported in part by JSPS
Core-to-Core Program ``International Research Network for Dark
Energy'', by Grant-in-Aid for Scientific Research from the JSPS
Promotion of Science, by Grant-in-Aid for Scientific Research on
Priority Areas No. 467 ``Probing the Dark Energy through an Extremely
Wide \& Deep Survey with Subaru Telescope'', by World Premier
International Research Center Initiative (WPI Initiative), MEXT,
Japan, and by the FIRST program ``Subaru Measurements of Images and
Redshifts (SuMIRe)'', CSTP, Japan.

%\bibliography{mn,refs}
\bibliography{mn-jour,refs}

\appendix
\section{Isothermal Velocity Dispersion Model}
\label{app:smallv}

In this appendix, we show an alternative model of the velocity
dispersion of DLRGs following the theory in \cite{BinneyTremaine},
which differs from the model in Section~\ref{sec:sigv}.

Let us begin our discussion with assuming that the phase space density
of DLRGs, which reside in host halo of mass $M$, obeys an isothermal
distribution (Eq.~[4-116] in \cite{BinneyTremaine}):
%%%%%%%%%%%%%%%%%%%%%%%%%%%%%%%%%%%%%%%%%%%%%%%%%%%%%%%
\begin{equation}
\label{eq:DLRGphasedensity}
f(r,\mathbf{v};M)=\frac{\rho_1}{(2\pi\sigma_{\rm DLRG}^2(r;M))^{3/2}}
\exp\left(\frac{\Psi(r)-|\mathbf{v}|^2/2}{\sigma_{v,{\rm iso}}^2(r;M)}\right)
\end{equation}
%%%%%%%%%%%%%%%%%%%%%%%%%%%%%%%%%%%%%%%%%%%%%%%%%%%%%%%
where $\Psi(r)$ is the gravitational potential determined by dark
matter distribution, $\sigma_{v,{\rm iso}}^2(r;M)$ is the 1D velocity
dispersion of DLRGs and $\rho_1$ is the normalization constant.  Here
we assume a spherically symmetric distribution for the DLRG
distribution in a statistical average sense.  Although DLRGs obey such
an isothermal distribution in the potential well of dark matter, this
is an alternative model to estimate the DLRG velocity dispersion, so
let us continue our discussion.

Integrating the equation above over velocities yields the radial
profile of DLRGs, which should be equivalent to the off-centered
profile in our language:
%%%%%%%%%%%%%%%%%%%%%%%%%%%%%%%%%%%%%%%%%%%%%%%%%%%%%%%
\begin{equation}
\label{eq:DLRGprofile}
p_{\rm off}(r;M)=\rho_1
\exp\left(\frac{\Psi(r)}{\sigma^2_{v, {\rm iso}}(r;M)}\right).
\end{equation}
%%%%%%%%%%%%%%%%%%%%%%%%%%%%%%%%%%%%%%%%%%%%%%%%%%%%%%%
This equation gives us an inverse problem: once the off-centered
profile and the gravitational potential are given, the velocity
dispersion $\sigma_{v, {\rm iso}}$is determined.

Assuming an NFW profile for the total mass profile within the halo,
the gravitational potential is given via Poisson's equation as
%%%%%%%%%%%%%%%%%%%%%%%%%%%%%%%%%%%%%%%%%%%%%%%%%%%%%%%
\begin{equation}
\label{eq:poisson}
\frac{1}{r^2}\frac{d}{dr}\left(r^2\frac{d\Psi(r)}{dr}\right)
=-4\pi G\rho_{\rm DM}(r) ,
\end{equation}
%%%%%%%%%%%%%%%%%%%%%%%%%%%%%%%%%%%%%%%%%%%%%%%%%%%%%%%
Hence the potential is given as 
%%%%%%%%%%%%%%%%%%%%%%%%%%%%%%%%%%%%%%%%%%%%%%%%%%%%%%%
\begin{eqnarray}
\label{eq:potential}
\Psi(r)&=& -\int_0^r\! dr'~ \frac{GM(<r')}{r'^2}, \nonumber \\
&=&2\sigma_{v}^2(r; M)\frac{I(r/r_s)}{m_{\rm nfw}(c)/c},
\end{eqnarray}
%%%%%%%%%%%%%%%%%%%%%%%%%%%%%%%%%%%%%%%%%%%%%%%%%%%%%%%
with
%%%%%%%%%%%%%%%%%%%%%%%%%%%%%%%%%%%%%%%%%%%%%%%%%%%%%%%
\begin{eqnarray}
&&m_{\rm nfw}(x)\equiv \ln(1+x) -\frac{x}{1+x},\nonumber\\
&&I(x)\equiv \int_0^x dx'\frac{m_{\rm nfw}(x')}{x'^2}=1-\frac{\ln(1+x)}{x} .
\label{eq:Ix}
\end{eqnarray}
%%%%%%%%%%%%%%%%%%%%%%%%%%%%%%%%%%%%%%%%%%%%%%%%%%%%%%%
Here we have rewritten the equation above in terms of the virial
velocity dispersion $\sigma_{v}(r; M)$ defined in
Eq.~(\ref{eq:sigv_rM}).

Inserting Eq.~(\ref{eq:potential}) into Eq.~(\ref{eq:DLRGprofile})
yields
%%%%%%%%%%%%%%%%%%%%%%%%%%%%%%%%%%%%%%%%%%%%%%%%%%%%%%%
\begin{equation}
\label{eq:DLRGprofile2}
p_{\rm off}(r;M)=\rho_1\exp\left(
-\frac{2\sigma_{v}^2(r; M)}{\sigma_{v,{\rm iso}}^2(r;M)}
\frac{I(r/r_s)}{m_{\rm nfw}(c)/c}
\right).
\end{equation}
%%%%%%%%%%%%%%%%%%%%%%%%%%%%%%%%%%%%%%%%%%%%%%%%%%%%%%%
The normalization constant $\rho_1$ is determined by imposing that the
velocity dispersion $\sigma_{v,{\rm iso}}$ is the same as $\sigma_{v}$
at the limit $r\rightarrow 0$:
%%%%%%%%%%%%%%%%%%%%%%%%%%%%%%%%%%%%%%%%%%%%%%%%%%%%%%%
\begin{eqnarray}
\rho_1&=&p_{\rm off}(r_{\rm piv})\exp\left(
\frac{2I(r_{\rm piv}/r_s)}{m(r_{\rm piv}/r_s)/r_{\rm piv}/r_s}
\right) \nonumber \\
&\simeq &p_{\rm off}(r_{\rm piv})\exp(2)~~~~~~(r_{\rm piv}/r_s \ll 1)
\end{eqnarray}
%%%%%%%%%%%%%%%%%%%%%%%%%%%%%%%%%%%%%%%%%%%%%%%%%%%%%%%
where $r_{\rm piv}$ is a pivot radius we choose to impose the
conditions that $r_{\rm piv}$ is very small as well as $\sigma_{v,
{\rm iso}}=\sigma_{v}$ at $r=r_{\rm piv}$.  Hence, the velocity
dispersion for an isothermal distribution model can be given in terms
of the off-centered profile as
%%%%%%%%%%%%%%%%%%%%%%%%%%%%%%%%%%%%%%%%%%%%%%%%%%%%%%%
\begin{equation}
\label{eq:sigv_iso}
\sigma_{v, {\rm iso}}^2(r;M)
=\sigma_{\rm vir}^2(M)\frac{I(r/r_s)}{m_{\rm nfw}(c)/c}
\left[1-\frac{1}{2}
\ln\left(\frac{p_{\rm off}(r;M)}{p_{\rm off}(r_{\rm piv};M)}\right)
\right]^{-1}.
\end{equation}
%%%%%%%%%%%%%%%%%%%%%%%%%%%%%%%%%%%%%%%%%%%%%%%%%%%%%%%

%%%%%%%%%%%%%%%%%%%%%%%%%%%%%%%%%%%%%%%%%%%%%%%%%%%%%%%
\begin{figure}
\begin{center}
\includegraphics[width=8cm]{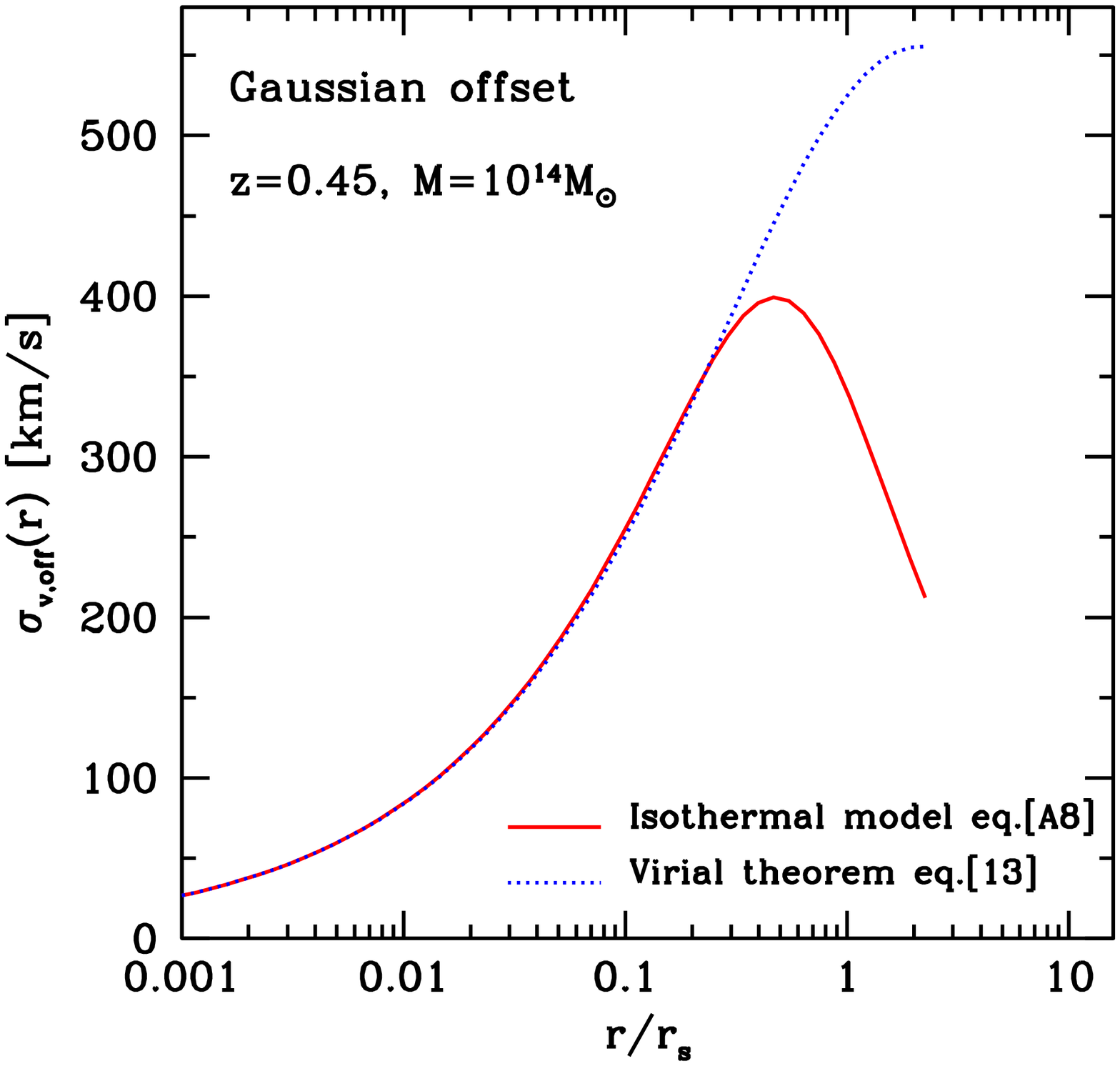}
\includegraphics[width=8cm]{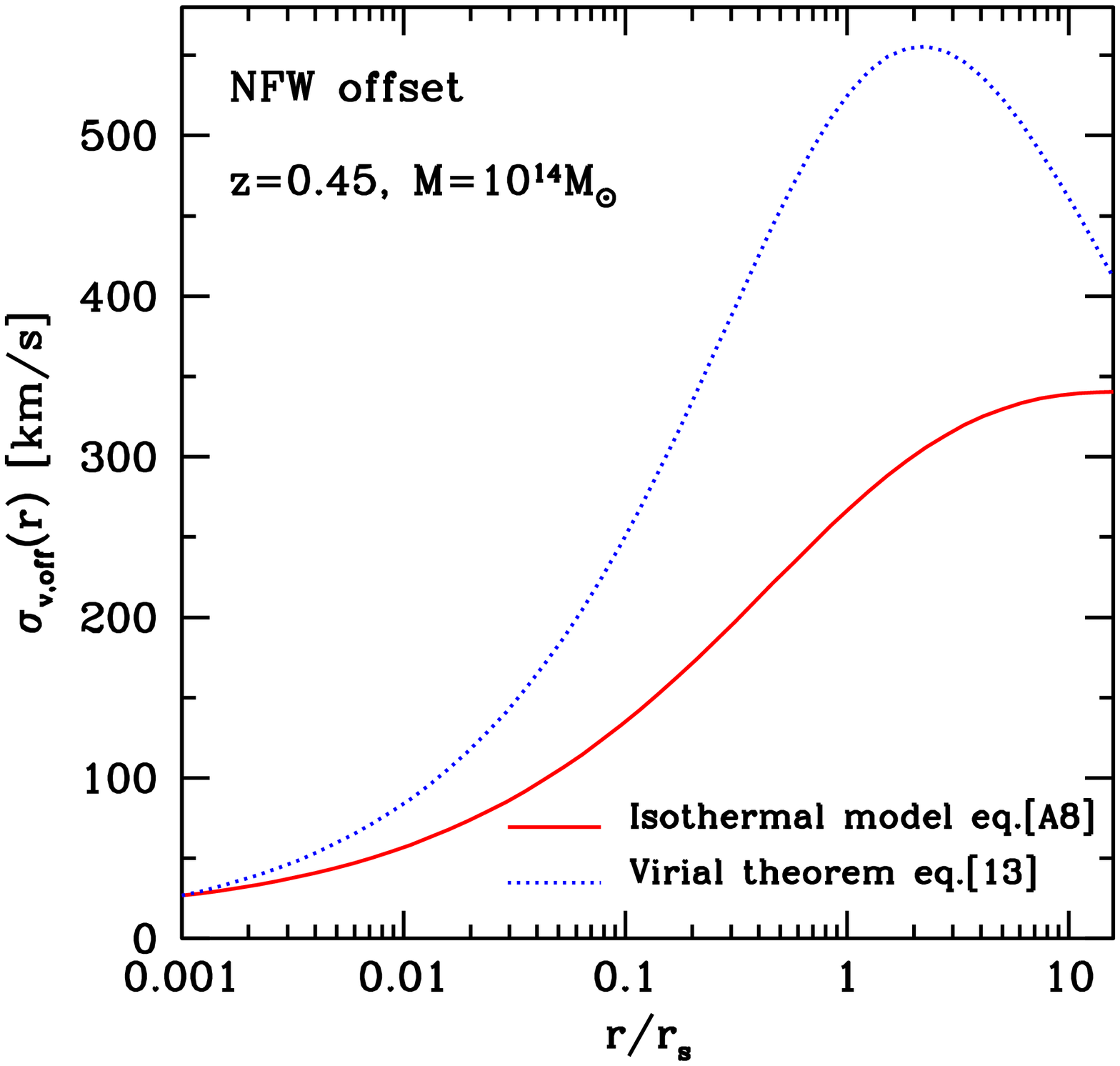} 
\caption{Comparing the models of DLRG velocity dispersion within a
  halo with $10^{14}M_\odot$ and at redshift $z=0.45$. The two models
  to be compared are a model based on the virial theorem
  (Eq.~[\ref{eq:sigv_rM}]) and a model derived assuming an isothermal
  velocity distribution given the off-centered profile of DLRGs and
  the dark matter mass profile. The left- and right panels show the
  results assuming the Gaussian and NFW off-centered profiles,
  respectively. The model given in Appendix~\ref{app:smallv} gives a
  smaller velocity dispersion than our fiducial model.
\label{fig:iso}}
\end{center}
\end{figure}
%%%%%%%%%%%%%%%%%%%%%%%%%%%%%%%%%%%%%%%%%%%%%%%%%%%%%%%

Fig.~\ref{fig:iso} shows the velocity dispersion for this isothermal
distribution model, comparing with our fiducial model based on a
simple virial theorem (Eq.~[\ref{eq:sigv_rM}]). To compute this
velocity profile, we use $r_{\rm piv}/r_s=10^{-4}$. However note that
the dependence of $r_{\rm piv}$ is small because the dependence is
logarithmic as implied in Eq.~(\ref{eq:sigv_iso}). The figure shows
that the isothermal model gives a smaller velocity dispersion than the
fiducial model, and therefore the resulting FoG effect is smaller than
the results shown in the main text, by about factor of 2 in the power
spectrum amplitudes.

\bibliographystyle{mn}

\end{document}